\documentclass[aps,superscriptaddress,showpacs,showkeys,nofootinbib,floatfix]{revtex4}
\usepackage{graphicx,epsfig,wrapfig,amssymb,amsmath,multirow,amsfonts,latexsym}
\usepackage[normalem]{ulem} 
\renewcommand\sout{\bgroup \color{blue} \ULdepth=-.5ex \ULset}

\usepackage{color}
\usepackage{pdfpages}

\newcommand{\be}{\begin{equation}}
\newcommand{\ee}{\end{equation}}
\newcommand{\ba}{\begin{eqnarray}}
\newcommand{\ea}{\end{eqnarray}}
\newcommand{\la}{\langle}
\newcommand{\ra}{\rangle}
\newcommand{\di}{ {\rm d} }
\newcommand{\pTsqx}[3]{{\la k_{\perp,\rm #2}^{2\;\,#3}(#1)\ra}}

\def \to {\rightarrow}

\newcommand{\bea}{\begin{eqnarray}}
\newcommand{\eea}{\end{eqnarray}}
\newcommand{\nn}{\nonumber}
\def\ket#1{\hbox{$\vert #1\rangle$}}   
\begin{document}
%
%
\newcommand*{\Pavia}{Dipartimento di Fisica, 
  Universit\`a degli Studi di Pavia, Pavia, Italy}\affiliation{\Pavia}
\newcommand*{\INFN}{Istituto Nazionale di Fisica Nucleare, 
  Sezione di Pavia, Pavia, Italy}\affiliation{\INFN}
\newcommand*{\UConn}{Department of Physics, University of Connecticut, 
  Storrs, CT 06269, USA}\affiliation{\UConn}
\title{
   Pion TMDs in light-front constituent approach, and\\ 
   Boer-Mulders effect in the pion-induced Drell-Yan process}
\author{B.~Pasquini}\affiliation{\Pavia}\affiliation{\INFN}
\author{P.~Schweitzer}\affiliation{\UConn}
\vspace{0.5in}
\begin{abstract}{
    At leading twist the transverse-momentum dependent parton 
    distributions of the pion consist of two functions, the unpolarized 
    $f_{1,\pi}(x,\boldsymbol k^{\, 2}_\perp)$ and the Boer-Mulders function 
    $h_{1,\pi}^{\perp }(x,\boldsymbol k^{\, 2}_\perp)$.
    We study both functions within a  light-front constituent model 
    of the pion, comparing the results with different pion models 
    and the corresponding nucleon distributions from a light-front 
    constituent model. 
    After evolution from the model scale  to the relevant experimental scales, 
    the results for the collinear pion valence parton distribution function 
    $f_{1,\pi}(x)$ 
    are in very good agreement with available parameterizations.
    Using the light-front constituent model results for the 
    Boer-Mulders functions of the pion and nucleon, we calculate  
    the coefficient $\nu$ in the angular distribution of Drell-Yan 
    dileptons produced in pion-nucleus scattering, which is 
    responsible for the violation of the Lam-Tung relation.
    We find a good agreement with data, and carefully discuss the 
    range of applicability of our approach.}
\end{abstract}
\maketitle

\section{Introduction}

Transverse momentum dependent distribution functions 
(TMDs) \cite{Collins:1981uw,Collins:2003fm,Collins-book}
provide unique insights in the 3D hadronic structure 
\cite{Tangerman:1994eh,Kotzinian:1994dv,Mulders:1995dh,Boer:1997nt,
Bacchetta:2006tn}, by taking into account the transverse motion of 
partons and spin-orbit correlations. The Drell-Yan process (DY) 
\cite{Christenson:1970um,Drell:1970wh} is 
basically the only source for this type of information for hadrons other 
than the nucleon, that are available as secondary beams in high energy 
experiments, such as the pion which is the main focus of this work. 
DY experiments with pions were reported in
Refs.~\cite{Badier:1981ti,Palestini:1985zc,
Falciano:1986wk,Guanziroli:1987rp,Conway:1989fs,Bordalo:1987cs}, see 
\cite{Stirling:1993gc} for a compilation of DY data till 1993 and 
\cite{McGaughey:1999mq,Reimer:2007iy,Arnold:2008kf,Chang:2013opa,
Peng:2014hta} for reviews of later data and theoretical progress.
TMDs describe hard processes like DY on the basis of factorization 
theorems \cite{Collins:1981uk,Ji:2004wu,Collins:2004nx,Echevarria:2011rb}.
The QCD evolution properties of some of the TMDs were studied in 
Refs.~\cite{Collins-book} and 
\cite{Collins:1984kg,Aybat:2011zv,Aybat:2011ge,Cherednikov:2007tw,
Bacchetta:2013pqa,Echevarria:2012pw,Vladimirov:2014aja}.

This work is devoted to the study of leading-twist TMDs of the pion.
At leading twist the pion structure is described in terms of two TMDs,
$f_{1,\pi}(x,\boldsymbol{k}^{2}_\perp)$ and 
$h_{1,\pi}^{\perp }(x,\boldsymbol{k}^2_{\perp})$. The unpolarized TMD 
$f_{1,\pi}(x,\boldsymbol{k}^{2}_\perp)$ describes the distribution of unpolarized 
partons carrying the longitudinal momentum fraction $x$ of the 
pion, and the transverse momentum $k_\perp$. 
The so-called Boer-Mulders function $h_{1,\pi}^{\perp }(x,\boldsymbol{k}^{2}_\perp)$ 
\cite{Boer:1997nt,Boer:1999mm} describes a spin-orbit correlation of 
transversely polarized partons, which is chiraly and (``naively'') 
time-reversal odd.
``Chiraly odd'' means that the operator structure defining 
$h_{1,\pi}^{\perp }(x,\boldsymbol{k}^{2}_\perp)$ flips the chirality of the partons, implying that 
this function can enter the description of a process only in combination 
with another chiral odd function. ``Time reversal odd'' (T-odd) means 
that under time reversal transformations the correlation flips sign, 
while the Wilson lines inherent in the TMD operator definitions are 
transformed from future- to past-pointing or vice versa.
This implies that T-odd functions appear with different signs 
in deep-inelastic scattering (DIS) and DY process
\cite{Brodsky:2002cx,Collins:2002kn,Ji:2002aa,Brodsky:2002rv,Boer:2002ju,
Belitsky:2002sm,Boer:2003cm}. The different signs of T-odd TMDs in 
different processes can be tested experimentally in the case of 
the nucleon, though this is not feasible for the pion.

However, the T-odd correlations as described by the Boer-Mulders 
functions in pion and nucleon may be responsible for the violation 
of the Lam-Tung relation, which connects the coefficients in the 
angular distribution of the DY lepton pairs 
\cite{Lam:1978pu,Collins:1978yt,Mirkes:1994dp}. 
The $\pi$-nucleus DY data
\cite{Falciano:1986wk,Guanziroli:1987rp,Conway:1989fs} show
a significant violation of this relation, which calls for
a nonperturbative leading-twist mechanism beyond  collinear 
factorization. The Boer-Mulders effect provides 
such a mechanism within the TMD factorization framework \cite{Boer:1999mm}, 
though  alternative mechanisms have also been proposed 
\cite{Brandenburg:1993cj,Brandenburg:1994wf,
Boer:2004mv,Brandenburg:2006xu,Nachtmann:2014qta}.
Indications for the violation of the Lam-Tung relation were 
also observed in $pp$- and $pd$-induced DY \cite{Zhu:2006gx}.

In order to perform the nonperturbative calculations of the
pion TMDs $f_{1,\pi}(x,\boldsymbol{k}^{2}_\perp)$ and 
$h_{1,\pi}^{\perp }(x,\boldsymbol{k}^{2}_\perp)$ we use
the light-front formalism, where hadrons are described in terms of
light-front wave functions (LFWFs).
The latter are expressed as an expansion of various quark, antiquark and
gluon Fock components. In principle, there is
an infinite number of LFWFs in such an expansion. 
However, there are many situations
   where one can successfully model hadronic wave functions by confining
   oneself to the contribution of the minimal Fock-space configuration with
a few partons.
We will refer to this approach as the light-front constituent model (LFCM).
The LFCM was successfully applied to describe many nucleon
properties \cite{Boffi:2002yy,Boffi:2003yj,Pasquini:2004gc,
Pasquini:2005dk,Pasquini:2006iv,Pasquini:2007xz,Pasquini:2007iz,
Boffi:2007yc,Pasquini:2009ki,Lorce:2011dv} including TMDs
\cite{Pasquini:2008ax,Boffi:2009sh,Pasquini:2010af,Pasquini:2011tk}.
For the pion, the specific model we will adopt for the minimal Fock-space
components of the LFWF
 has been originally proposed in
Refs.~\cite{Schlumpf:1994bc,Chung:1988mu}, and has been applied to study
some partonic properties of pion in
Refs.~\cite{Frederico:2009fk,Salme':2012rv}.
However, the present work is the first application to study the TMDs in
the pion.

The description of nucleon TMDs within the LFCM was shown to agree
with phenomenology within (10-30)$\,\%$ in the valence-$x$ region
after evolution from the low initial scale of the model to 
experimentally relevant scales \cite{Boffi:2009sh,Pasquini:2011tk}.
This is in particular the case for the Boer-Mulders function of
the nucleon \cite{Pasquini:2011tk}. 

In this work we derive and calculate the unpolarized TMD 
and Boer-Mulders function of the pion, $f_{1,\pi}(x,\boldsymbol{k}^{2}_\perp)$ 
and $h_{1,\pi}^{\perp }(x,\boldsymbol{k}^{2}_\perp)$, and compute the coefficient 
$\nu$ in $\pi$-nucleus induced DY. 
We find that the valence distribution function $f_{1,\pi}(x)$ 
of  the pion obtained from the LFCM agrees well with available
parameterizations. 
We compare our results for the pion Boer-Mulder function with previous
results from spectator and bag models 
Refs.~\cite{Lu:2004hu,Burkardt:2007xm,Gamberg:2009uk,Lu:2012hh} 
as well as with lattice QCD \cite{Engelhardt:2013nba}.
We show that $h_{1,\pi}^{\perp }(x,\boldsymbol{k}^{2}_\perp)$, in combination with 
the nucleon Boer-Mulders function $h_{1,N}^{\perp }(x,\boldsymbol{k}^{2}_\perp)$ 
from the LFCM of Ref.~\cite{Pasquini:2010af}, gives a good description 
of the DY data on the coefficient $\nu$.
For other model studies of nucleon Boer-Mulders function and 
phenomenological work related to the violation of the Lam-Tung 
relation we refer to Refs.~\cite{Boer:1999mm} and 
\cite{Bianconi:2006hc,Lu:2005rq,Gamberg:2005ip,Sissakian:2005yp,Barone:2006ws,Zhang:2008nu,Lu:2009ip,Barone:2010gk,Lu:2011mz,Liu:2012fha,Liu:2012vn,Chang:2013pba,Chen:2013zpy}.

The outline of this work is as follows.
In Sec.~\ref{Sec-2:initial-scale} 
we determine the initial scale of the pion LFCM approach.
In Sec.~\ref{Sec-3:lcwf} 
we review the classification of the pion light-front wave 
function, in the minimal ($q\bar q$) Fock space configuration, 
in terms of light-front amplitudes describing the different $\bar q q$
orbital angular momentum components in the pion state.
In Sec.~\ref{Sec-4:TMDs-in-pion} 
we derive the representation of the leading-twist pion TMDs 
as overlap of light-front amplitudes.
In Sec.~\ref{Sec-5:results-from-LFCM} 
we use a specific model for the pion LFWFs to obtain numerical results 
for pion TMDs at the initial hadronic scale. We then evolve and discuss 
our results. 
Sec.~\ref{Sec-6:DY}
gives a brief review of the DY formalism at a leading-order parton-model 
level.
Sec.~\ref{Sec-7:f1-in-DY} 
is dedicated to a discussion of the unpolarized TMD in the DY process,
and establishes the range of applicability of our approach.
In Sec.~\ref{Sec-8:BM-in-DY}
we discuss the Boer-Mulders effect in $\pi$-nucleus induced DY.
Finally, in Sec.~\ref{Sec-9:summary}
we summarize our results and give an outlook.

\section{Initial scale of pion constituent approach}
\label{Sec-2:initial-scale}

Parton distribution functions are defined within a certain
regularization scheme at a given renormalization scale.
The results from a constituent approach refer to an assumed
low initial scale $\mu_0$, at which a pion is thought 
to consist of a ``valence'' quark-antiquark pair only, while
a nucleon is similarly assumed to consist of 3 ``valence'' quarks only.
The value of $\mu_0$ is not known a priori, but it can be determined
in a way independent of the details of the constituent model.
Therefore we shall first address this point, before embarking with 
the actual study of TMDs in LFCM.

It is crucial to determine $\mu_0$ for two reasons.
First, the model parton distributions have to be evolved from a
well-defined initial scale $\mu_0$ to experimentally relevant 
scales $Q\sim$ few GeV before they can be confronted with data.
Second, the initial scale $\mu_0$ determines the value of the running 
coupling constant $\alpha_s(\mu_0^2)$ which enters the overall 
normalization of the Boer-Mulders function, when the initial (final) 
state interaction effects are taken into account via the one-gluon 
exchange mechanism (see the discussion in in Sec.~\ref{sec:bm}).

To determine $\mu_0$ we use the following standard procedure
\cite{Pasquini:2004gc,Broniowski:2007si,Davidson:2001cc,Courtoy:2008nf}.
At the initial scale the entire pion momentum must
be carried by valence $\bar q$, $q$ degrees of freedom,
$\langle x\rangle_v=1$, while sea-quark and gluon contributions 
are set to zero. Similarly to Ref.~\cite{Broniowski:2007si} we 
then require the initial scale $\mu_0$ to be such that after 
evolution from $\mu_0$ to say $Q^2=4\,{\rm GeV}^2$ the 
phenomenological value for the pion momentum fraction carried 
by valence quarks is reproduced. We take
    \begin{eqnarray}
      \langle x\rangle_v
      &=& \int_0^1\di x\;x\biggl[
          (f_{1,\pi^+}^{u}-f_{1,\pi^+}^{\bar u})(x)+
          (f_{1,\pi^+}^{\bar d}-f_{1,\pi^+}^{d})(x)\biggr] \nonumber\\
      &=& 0.47 \pm 0.02 \;\;\mbox{at $\;Q^2=4$ GeV$^2$}
      \label{eq:xv-init}
    \end{eqnarray} 
in leading (LO) and next-to-leading order (NLO) from the parameterizations
\cite{Sutton:1991ay,Gluck:1999xe}. We use
$\alpha_{\rm  LO}(M^2_Z)=0.13939$ and  
$\alpha_{\rm NLO}(M_Z^2)=0.12018$ in $\overline{\rm MS}$ scheme 
for the strong coupling constant at the $Z_0$ mass $M_Z=91.1$ 
GeV from the global fit of parton distribution functions (PDFs) to 
hard-scattering data from Ref.~\cite{Martin:2009iq}.
    These values correspond to 
    $\Lambda_{\rm LO} = 359,\,322,\,255$ MeV and
    $\Lambda_{\rm NLO}= 402,\,341,\,239$ MeV for respectively $N_F=3,\,4,\,5$ 
    flavors in the variable flavor-number scheme with heavy-quark mass 
    thresholds at $m_c=1.4$ GeV, $m_b=4.75$ GeV, $m_t=175$ GeV 
    \cite{Martin:2009iq}.

With these parameters the above described procedure yields
at LO and NLO for the initial scale
    \ba
    \mu_{0,{\rm LO}} &=& 460\,\mbox{MeV}, \;\;\;\;\; 
    \frac{\alpha_{{\rm LO}}(\mu_0^2)}{4\pi}=0.225, 
    \label{eq:fixing-initial-scale-LO}
    \\
    \mu_{0,{\rm NLO}} &=& 555\, \mbox{MeV}, \;\;\;\;\; 
    \frac{\alpha_{{\rm NLO}}(\mu_0^2)}{4\pi}=0.0938.
    \label{eq:fixing-initial-scale-NLO}
    \ea
In a different NLO scheme the numerical values in 
Eq.~(\ref{eq:fixing-initial-scale-NLO}) would be somewhat different,
but it is beyond the scope of this work to study scheme-dependence effects.
In the following we shall assume that theoretical uncertainties 
due to scheme dependence are smaller than the generic accuracy
of (light-front) constituent model approaches.

At this point it is instructive to compare the LO and NLO initial scales 
for pion distribution functions, 
Eqs.~\eqref{eq:fixing-initial-scale-LO}-\eqref{eq:fixing-initial-scale-NLO},
with those obtained in the case of the nucleon \cite{Pasquini:2011tk}.
In principle, the constituent model approaches for pion 
and nucleon can be viewed as unrelated models.
Nevertheless, the underlying physical assumption is the same.
At some low ``hadronic scale'' one deals with constituent 
(``valence'') degrees of freedom carrying the total hadron momentum:
a constituent quark-antiquark pair in the pion case, or
3 constituent quarks in the nucleon case.
For the underlying physical picture to be successful one should 
expect the initial scale to be ``universal,'' i.e.\ 
independent of the considered hadron. 
It is therefore gratifying to observe how close numerically the results 
are for the nucleon case (in \cite{Pasquini:2011tk} it was obtained
$\mu_{0,\rm LO}=420\,\mbox{MeV}$ and $\mu_{0,{\rm NLO}}=508\,\mbox{MeV}$ for 
the nucleon) as compared to the pion case in
Eqs.~\eqref{eq:fixing-initial-scale-LO}-\eqref{eq:fixing-initial-scale-NLO}.
This is an encouraging indication for the usefulness of the constituent
model picture.

\section{Light-front amplitudes in pion constituent approach}
\label{Sec-3:lcwf}

In this section, we review  the classification of the light-front wave 
function for the pion, considering the minimal Fock space configuration,
i.e.\ $q\bar q$. According to the total quark orbital angular momentum 
projection, the $q\bar q$ LFWF of the pion can be written in terms of two 
light-front amplitudes carrying the total quark orbital angular momentum  
$l_z=0$ and $|l_z=1|$, i.e.    
   \begin{equation}
   \label{OAMwf}   
   |\pi(p)\rangle_{q\bar q} = 
   |\pi(p)\rangle_{q\bar q}^{l_z=0}  +  |\pi(p)\rangle_{q\bar q}^{|l_z=1|}  .
   \end{equation}
The different angular-momentum components of the state
in Eq.~\eqref{OAMwf} are given by~\cite{Burkardt:2002uc,Ji:2003yj}
\begin{eqnarray}
   |\pi(p)\rangle_{q\bar q}^{l_z=0} &=&T_\pi\,
\int d[1]d[2] \psi^{(1)}(1,2)
 \frac{\delta_{ij}}{\sqrt{3}}
   \left[q^{\dagger}_{i\uparrow}(1)\bar q^{\dagger}_{j\downarrow}(2)-q^{\dagger}_{i\downarrow}(1)\bar q^{\dagger}_{j\uparrow}(2)\right]            |0\rangle \ ,
\label{lca1}\\
|\pi(p)\rangle_{q\bar q}^{|l_z|=1} &=&T_\pi\,
\int d[1]d[2] \psi^{(2)}(1,2)
\frac{\delta_{ij}}{\sqrt{3}}
\left[k_{1\perp}^-q^{\dagger}_{i\uparrow}(1)\bar q^{\dagger}_{j\uparrow}(2)
+k_{1\perp}^+q^{\dagger}_{i\downarrow}(1)\bar q^{\dagger}_{j\downarrow}(2)\right]      |0\rangle  \ ,
\label{lca2}
\end{eqnarray}    where
$k^\pm_{i\perp}=k^x_i\pm k^y_i$, and $q^{\dagger}_{i\lambda}$ and    $\bar q^{\dagger}_{i\lambda}$   are creation operators of a quark and antiquark    with flavor $q$, helicity $\lambda$ and color $i$, respectively.
   In Eqs.~(\ref{lca1}) and (\ref{lca2}), $T_\pi$ is the isospin factor which
   projects on the different members of the isotriplet of the pion,    and is defined as    $T_\pi=\sum_{\tau_q\tau_{\bar q}}\langle    1/2 \tau_q 1/2 \tau_{\bar q}| 1 \tau_\pi \rangle$ with $\tau_q, \tau_{\bar q}$ and $\tau_\pi$ the isospin of the quark, antiquark and pion state, respectively.
   Furthermore, the amplitudes $\psi^{(1,2)}(1,2)$    are functions of quark momenta with arguments 1 representing $x_1$    and $\boldsymbol k_{1\perp}$ and so on.
   They depend on the transverse momenta only through scalar products, e.g.   $\boldsymbol k_{i\perp}\cdot \boldsymbol k_{j\perp}$.   Since momentum conservation implies $\boldsymbol k_{1\perp}+\boldsymbol k_{2\perp}=0$ and $x_1+x_2=1$,  $\psi^{(1,2)}_{q\bar q}(1,2)$  depend only
   on the variables $\bar x=x_1$ and $\boldsymbol{\kappa}_{\perp}^2$, with  $\boldsymbol{\kappa}_{\perp}=\boldsymbol k_{1\perp}$.   The integration measure in Eqs.~(\ref{lca1}) and (\ref{lca2}) is defined as
\begin{equation}
\label{eq:7}
d[1]d[2]=
   \frac{dx_1dx_2 }{\sqrt{x_1x_2}}\delta\left(1-\sum_{i=1}^2 x_i\right)
\frac{d^2 \boldsymbol{k}_{1\perp}d^2\boldsymbol{k}_{2\perp}}{[2(2\pi^3)]} \delta^2\left(\sum_{i=1}^2 \boldsymbol{k}_{i\perp}\right)=   \frac{d\bar x}{\sqrt{\bar x(1-\bar x)}}   \frac{d^2 \boldsymbol{\kappa}_{\perp}}{[2(2\pi^3)]}.
   \end{equation}
In the following, we will describe the above LFWF  amplitudes in a light-front constituent  model which was already successfully applied for describing the charge form factor and decay constant of the pion~\cite{Chung:1988mu,Schlumpf:1994bc} and the  generalized parton distributions~\cite{Frederico:2009fk,Salme':2012rv}. 

The $q\bar q$ component of the light-front state of the pion  can be written as   \be\label{eq:12}
\ket{\pi(p)}_{q\bar q} = T_\pi \,\sum_{\lambda_i,c_i}
   \int d[1]d[2]   \Psi^{[f]}_{q\bar q}(\{x_i,\boldsymbol{ k}_{i\perp };\lambda_i\})   \frac{\delta_{ij}}{\sqrt{3}}   q^{\dagger}_{i  \lambda_1}(1)   \bar q^{\dagger}_{j\lambda_2}(2)   |0\rangle\, .
\ee
In Eq.~(\ref{eq:12}),
   the LFWF $\Psi^{[f]}_{q\bar q}(\{x_i,\boldsymbol{ k}_{i\perp };\lambda_i\})$   satisfies Poincar\`e covariance and    is an eigenstate of the total angular momentum operator in the light-front dynamics. These properties can be fulfilled by constructing the wave function as the product of a momentum wave function, which is spherically symmetric and invariant under permutation of the two constituent partons,
   and a spin wave function, which is uniquely determined by symmetry requirements, i.e.,
\begin{eqnarray}
   \label{eq:13}
\Psi^{[f]}_{q\bar q}(\{x_i,\boldsymbol{ k}_{i\perp }; \lambda_i\})   &=&   \tilde \psi_\pi(\bar x,\boldsymbol{ \kappa}_{\perp })   \tilde\Phi(\lambda_1,\lambda_2).   \end{eqnarray}   In the above equation,    the spin-dependent part  is given by   \begin{eqnarray}
   \tilde\Phi(\lambda_1,\lambda_2)   &=&\sum_{\mu_1\mu_2}   \langle 1/2,\mu_1; 1/2, \mu_2|0, 0 \rangle   D_{\mu_1\lambda_1}^{1/2*}(R_{M}(\boldsymbol{ \kappa}))   D_{\mu_2\lambda_2}^{1/2*}(R_{M}(-\boldsymbol{ \kappa})),   \label{eq:15}
\end{eqnarray}
where $\boldsymbol{ \kappa}=\{\boldsymbol{\kappa}_{\perp},\kappa_{z}\}$,
   with
\begin{eqnarray}
\kappa_{z}=M_0(\bar x,\boldsymbol{\kappa}_{\perp})(\bar x-\frac{1}{2}),
\end{eqnarray}   and the free mass defined as
\begin{eqnarray}
M_0^2(\bar x,\boldsymbol{\kappa}_{\perp})=\frac{m^2+|\boldsymbol{\kappa}_{\perp}|^2}{\bar x(1-\bar x)},
\end{eqnarray}
with $m$ the quark mass.
In Eq.~(\ref{eq:15}), $D_{\lambda\mu}^{1/2}(R_{M}(\bar x,\boldsymbol{\kappa}_\perp))$ is   the matrix element of the Melosh rotation $R_{M}$~\cite{Melosh:1974cu}
\begin{eqnarray}
D_{\lambda\mu}^{1/2}(R_{M}(\boldsymbol{\kappa})) &=&
   \langle\lambda|R_{M}(\boldsymbol{\kappa})|\mu\rangle\nonumber\\
&=&   \langle\lambda|\frac{m + \bar xM_0 -   i\boldsymbol{\sigma}\cdot(\hat{\boldsymbol{z}}\times\boldsymbol{\kappa}_\perp)}{\sqrt{(m   + \bar xM_0)^2 + \boldsymbol{\kappa}^{\, 2}_\perp}}|\mu\rangle.
   \label{eq:16}
\end{eqnarray}
The Melosh rotation corresponds to the unitary transformation which converts    the Pauli spinors of the quark and antiquark in the pion rest-frame to the light-front spinor.    Making explicit the dependence on the quark and antiquark helicities, the spin wave function of Eq.~(\ref{eq:15}) takes the following values:
   \begin{eqnarray}
\label{eq:17}
\tilde \Phi\left(\uparrow,\uparrow\right)   &=&   \prod_i\frac{1}{\sqrt{N(x_i,\boldsymbol{\kappa}_{\perp })}}
\kappa_{\perp}^-(-a_1+a_2),   \\\label{eq:18}
\tilde \Phi\left(\uparrow,\downarrow\right)
&=&\prod_i\frac{1}{\sqrt{N(x_i,\boldsymbol{\kappa}_{\perp })}}
   (   a_1a_2-\kappa_{\perp}^+\kappa^-_{\perp}),   \\\label{eq:19}
\tilde \Phi\left(\downarrow,\uparrow\right)   &=&
\prod_i\frac{1}{\sqrt{N(x_i,\boldsymbol{\kappa}_{\perp })}}
   (-a_1a_2+   \kappa_{\perp}^+\kappa^-_{\perp}),  \\\label{eq:20}   \tilde \Phi_\uparrow\left(\downarrow,\downarrow\right)   &=&   \prod_i\frac{1}{\sqrt{N(x_i,\boldsymbol{\kappa}_{\perp })}}
\kappa_{\perp}^+(-a_1-a_2),
\end{eqnarray}
where $a_i=(m+x_i M_0)$,    and $N(x_i,\boldsymbol{\kappa}_{\perp })=   [(m+x_i M_0)^2+ \boldsymbol{\kappa}^2_{\perp}]$.      Taking into account the quark-helicity dependence in Eqs.~(\ref{eq:17})-(\ref{eq:20}), the pion state can be mapped out into the different angular momentum components. As a result, the pion wave function amplitudes in the light-front CQM
   read
\begin{eqnarray}
\psi^{(1)}(1,2)&=&\tilde \psi(x,\boldsymbol{\kappa}_{\perp })
   \prod_i\frac{1}{\sqrt{N(x_i,\boldsymbol{\kappa}_{\perp })}}   
   \frac{1}{\sqrt{2}}(   a_1 a_2 -\kappa_{\perp}^-\kappa_{\perp}^+),
   \label{eq:24}\\   \psi^{(2)}(1,2)&=&\tilde \psi(x,\boldsymbol{\kappa}_{\perp })   \prod_i\frac{1}{\sqrt{N(x_i,\boldsymbol{\kappa}_{\perp })}}\frac{1}{\sqrt{2}}   
   (-a_1 - a_2).
\label{eq:25}
\end{eqnarray}
\section{Twist-2 TMD\lowercase{s} in pion constituent approach}
\label{Sec-4:TMDs-in-pion}

The quark TMDs  are defined through the following correlation function~\cite{Mulders:1995dh,Boer:1997nt,Bacchetta:2006tn}
\begin{eqnarray}
\label{correlator}
\Phi_{ij}(x,\boldsymbol k_\perp^2)=  \int \frac{d\xi^-d^2\boldsymbol \xi_\perp}{(2\pi)^3}e^{i\xi\cdot k  }  
\langle p |\bar\psi_j(0){\cal L}^\dagger( 0,\boldsymbol 0_\perp |n ) {\cal L} ( \xi^-,\boldsymbol\xi_\perp | n )\psi_i(\xi)| p\rangle|_{\boldsymbol\xi^+=0} \ ,
\end{eqnarray}
where $x=k^+/p^+$ and for  a generic four-vector $a^\mu=(a^+,a^-,\boldsymbol{a}_\perp)$ we used the light-front components $a^\pm=(a^0\pm a^3)/\sqrt{2}$.
The Wilson lines ${\cal L}$ connecting the two quarks fields ensure the color gauge invariance of the correlator in 
Eq.~\eqref{correlator} and are defined as~\cite{Mulders:1995dh,Boer:1997nt,Bacchetta:2006tn}
\begin{equation}  
\label{gauge-link}
{\cal L}(\xi^-,{\boldsymbol \xi}_\perp| n)={\cal P}\exp\left(-ig\int_{\boldsymbol \xi^-}^{n\cdot \infty}   {\rm d} \eta^-\cdot A^+(\eta^-,\boldsymbol\xi_\perp)\right){\cal P}\exp\left(-ig\int_{\boldsymbol \xi_\perp}^{\infty}   {\rm d}^2\boldsymbol \eta_\perp\cdot \boldsymbol A_\perp(\xi^-=n\cdot\infty,\boldsymbol\eta_\perp)\right) \ ,
\end{equation}
where the   vector $n$ depends on the process under consideration. For instance,  the future-pointing Wilson lines with  $n=(0,+1,0)$ are appropriate for defining TMDs in semi-inclusive DIS, whereas in the Drell-Yan process the Wilson line are necessarily past-pointing with $n=(0,-1,0).$
In particular, this reverses the sign of all T-odd distributions functions entering  the correlator.

For a pion target, the information content of the correlator~\eqref{correlator} is summarized at leading twist by two 
TMDs that can be projected out from the correlator as follows
\begin{eqnarray}
\label{f1}
\frac{1}{2}{\rm Tr} [\Phi\gamma^+]&=&f_{1,\pi}(x,\boldsymbol k_\perp^2),\\
\label{h1}
\frac{1}{2}{\rm Tr} [\Phi i\sigma^{i+}\gamma_5]&=&\frac{\varepsilon^{ij}k_\perp^j}{M_\pi} h_{1,\pi}^\perp(x,\boldsymbol k_\perp^2).
\end{eqnarray}
The function $f_{1,\pi}(x,\boldsymbol k^2_\perp)$ is the unpolarized quark 
distribution, which integrated over $\boldsymbol k_\perp$ gives the familiar 
light-front momentum distribution $f_{1,\pi}(x)$
(in the parton model; in the TMD factorization framework
the relation is more subtle \cite{Collins-book}), and $h_{1,\pi}^\perp(x,\boldsymbol k^2_\perp)$ is the Boer-Mulders TMD~\cite{Boer:1997nt}, which is a T-odd function, i.e.  it changes sign under "naive time reversal",  defined as usual
time reversal, but without interchange of initial and final states. In the following, we will present the model calculation of the Boer-Mulders function for the SIDIS process, denoted as $h_{1,\pi}^{\perp}(x,\boldsymbol k^2_\perp)_{{\rm DIS}}$, while
for the Boer-Mulders function in the Drell-Yan process we will use $h_{1\,\pi}^{\perp}(x,\boldsymbol k^2_\perp)_{{\rm DY}}= - h_{1,\pi}^{\perp}(x,\boldsymbol k^2_\perp)_{{\rm DIS}}$ .

 \subsection{Unpolarized parton distribution function in the pion}

From the definition~\eqref{f1}, the  $f_{1,\pi}$ TMD to leading order in the gauge field is given by
\begin{equation}
\label{f1-tmd}
f_{1,\pi}(x,\boldsymbol k^{\, 2}_\perp)=
   \int\frac{d\xi^-d^2\boldsymbol \xi_\perp}{(2\pi)^3}e^{i(\xi^- k^+-\boldsymbol\xi_\perp\cdot\boldsymbol k_\perp)}   \langle \pi ( p )|\bar\psi(0)\gamma^+\psi(\xi^-,\boldsymbol \xi_\perp)|\pi( p )\rangle \ .
   \end{equation}
By using  the canonical expansion of the quark fields in
terms of Fock operators and taking into account the $q\bar q$ component of the light-front state of the pion in Eq.~\eqref{eq:12}, we find the final result
\begin{eqnarray}
f_{1,\pi}^{q}(x,\boldsymbol k^{2}_\perp)=f_{1,\pi}^{\bar q}(x,\boldsymbol k^{2}_\perp)
&=&T^2_\pi\int d[1] d[2]\sqrt{x_1x_2}\delta(x-x_1)\delta^2(\boldsymbol k_\perp-\boldsymbol k_{1\perp} )|\tilde\psi_\pi(x_1,\boldsymbol k_{\perp 1})|^2\nonumber\\
&=&T^2_\pi\frac{1}{2(2\pi)^3}
|\tilde\psi_\pi(x,\boldsymbol k_{\perp })|^2.
\end{eqnarray}
The unpolarized TMD involves a matrix element which is diagonal in the quark orbital angular momentum. As a consequence, it takes the following expression
in terms of the wave function amplitudes in Eqs.~\eqref{lca1}-\eqref{lca2}
\begin{equation}
f_{1,\pi}^{q}(x,\boldsymbol k^{2}_\perp)
=T^2_\pi\frac{1}{2(2\pi)^3}\, 2\left[
|\tilde\psi^{(1)}(x,\boldsymbol k_\perp^2)|^2+\boldsymbol k_\perp^2|\tilde\psi^{(2)}(x,\boldsymbol k_\perp^2)|^2\right].
\end{equation}


\subsection{Boer-Mulders function of the pion}   \label{sec:bm}
\begin{figure}
\centerline{
\epsfig{file=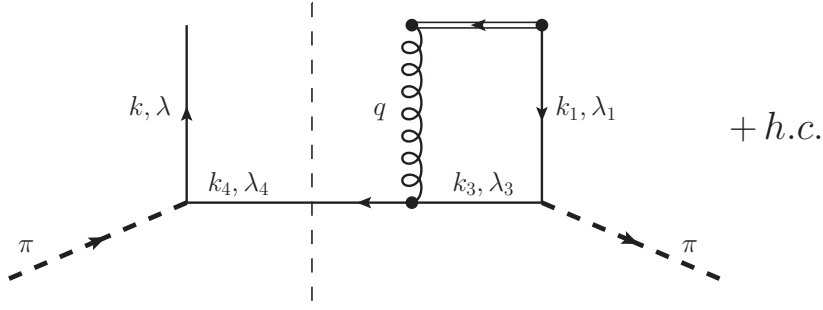,width=0.7\columnwidth}}
\vspace{-11.5 truecm}
\caption{\label{fig1}
The leading contribution from the one-gluon exchange mechanism to
the T-odd distribution function of the pion.}
\end{figure}

Using  the definitions~\eqref{correlator} and \eqref{h1}, the quark Boer-Mulders function of the pion is given by
\begin{equation}
\label{bm}
h_{1,\pi}^\perp(x,\boldsymbol k^{\, 2}_\perp)_{{\rm DIS}}=\epsilon^{ij} k^j\frac{M_\pi}{2\boldsymbol k^{\, 2}_\perp}
   \int\frac{d\xi^-d^2\boldsymbol \xi_\perp}{(2\pi)^3}e^{i(\xi^- k^+-\boldsymbol\xi_\perp\cdot\boldsymbol k_\perp)}   \langle \pi ( p )|\bar\psi(0){\cal L}^\dagger(0,\boldsymbol 0_\perp|\, n)  i\sigma^{i+}\gamma_5{\cal L}(\xi^-,\boldsymbol \xi_\perp|\, n)\psi(\xi^-,\boldsymbol \xi_\perp)|\pi( p )\rangle \ ,
   \end{equation}
with $n=(0,+1,0)$.
 The gauge link ${\cal L}$ is crucial  to obtain a non-zero Boer-Mulders  function.
In the light-front gauge, it  reduces to a transverse gauge-link  at $\xi^-=\infty$, given by the second term in Eq.~\eqref{gauge-link}.
Furthermore, we expand  the above gauge link to take into account the first order non-vanishing contribution corresponding to
    the one-gluon exchange diagram shown in Fig.~\ref{fig1}.   Following the procedure outlined in Ref.~\cite{Pasquini:2010af} for the analogous calculation   of the T-odd TMDs of the nucleon, we obtain the following result for the quark Boer-Mulders function of the pion
   \begin{eqnarray}
&&h_{1,\pi}^{\perp\,q}(x,\boldsymbol k^{\, 2}_\perp)_{{\rm DIS}}=-g^2\,  M_\pi   
  \frac{k^{x}_{\perp}-i k^{y}_{\perp}}{\boldsymbol k^{\, 2}_\perp}   \frac{1}{(2\pi)^{11}}\frac{1}{\sqrt{2k^+}}   \int\frac{{\rm d}k_3^+{\rm d}^2\boldsymbol k_{3\perp}}{\sqrt{(2k_3^+)(2k^+_4)}}
   \int\frac{{\rm d}^2\boldsymbol q_\perp}{2k^+_1}    \nonumber\\
&&\times   \Big\{\frac{1}{\boldsymbol{q}^{\,2}_\perp}\sum_{\lambda_3}   \sum_{\bar q}\sum_{i.j}\sum_{k,l}   T^a_{ij}T^b_{kl}\delta_{ab}   \langle \pi(p)|q^{\dagger}_{i\uparrow}(k_1)q_{j\downarrow}(k)
   \bar q_{k\lambda_3}(k_4)\bar q^{\dagger}_{l\lambda_3}(k_3)   |\pi(p)\rangle\Big\},   \label{eq:bm1}
\end{eqnarray}
where the parton momenta are  defined as $k_1=k-q$, $k_4=k_3-q$.   
The above equation corresponds to the diagram of Fig.~\ref{fig1}
with $\lambda=-\lambda_1$ and $\lambda_4=\lambda_3$ for the helicity 
of the interacting and spectator partons, respectively, i.e. the 
helicity is conserved at the antiquark-gluon vertex,  while the helicity 
of the struck quark  flips from the initial to the final state.   
For angular momentum conservation, the quark helicity flip  must be 
compensated by a transfer of one unit of orbital angular momentum.      
Inserting in Eq.~(\ref{eq:bm1}) the light-front wave function amplitude 
decomposition of the pion state introduced  in Sec.~\ref{Sec-3:lcwf}, 
one finds the following results in terms of the light-front 
amplitudes $\psi^{(i)}$
  \begin{eqnarray}
    h_{1,\pi}^{\perp\,q}(x,\boldsymbol k^{\, 2}_\perp)_{{\rm DIS}}&=&\frac{4}{3}g^2\, T^{2}_{\pi}\,  M_\pi   
    \frac{k^{x}_{\perp}-i k^{y}_{\perp}}{\boldsymbol k^{\, 2}_\perp}
    \int \frac{{\rm d^2}\boldsymbol q_\perp}{(2\pi)^5}   \frac{1}{\boldsymbol q^{\, 2}_\perp}   
    \,{\cal H}_{\pi}^{\perp\,q},  \label{eq:bm-overlap}
  \end{eqnarray}
where the function ${\cal H}^{\perp\, q}_\pi$ is
  \begin{eqnarray}
    {\cal H}_{\pi}^{\perp\, q}&=&   
    -k'^{+}_{\perp}\psi^{(1)}(1,2)\psi^{(2)*}(1',2')   +k^{+}_{\perp}\psi^{(1)*}(1,2)\psi^{(2)}(1',2')
    \label{eq:H-up}
  \end{eqnarray}
with $\boldsymbol{k}'_\perp=\boldsymbol{k}_\perp-\boldsymbol{q}_\perp$ 
and the parton coordinates 
$1=(x,\boldsymbol{k}_\perp)$,    
$2=(1-x,-\boldsymbol{k}_\perp)$, and   
$1'=(x,\boldsymbol{k}'_\perp)$, 
$2'=(1-x,-\boldsymbol{k}'_\perp)$.
In the model for the light-front amplitudes introduced in Sec.~\ref{Sec-3:lcwf}, we find the following explicit results
   \begin{eqnarray}
&&h_{1,\pi}^{\perp\,q}(x,\boldsymbol k^{\, 2}_\perp)_{{\rm DIS}}=\frac{4}{3}g^2\, T^{2}_\pi \, M_\pi   
\frac{k^{x}_{\perp}-i k^{y}_{\perp}}{\boldsymbol k^{\, 2}_\perp}  \int \frac{{\rm d}^2\boldsymbol q_\perp}{(2\pi)^5}\frac{1}{\boldsymbol q^{\, 2}_\perp}   \tilde\psi^*(\{x'_i\},\{\boldsymbol k'_{i\perp}\})
   \,\tilde\psi(\{x_i\},\{\boldsymbol k_{i\perp}\})\nonumber\\   &&\times \frac{1}{2}\prod_{i=1}^2 N^{-1}(\boldsymbol{k}\,'_i)    N^{-1}(\boldsymbol{k}_i)   \left[\tilde A_1 A_2+\boldsymbol{\tilde{B_1}}\cdot\boldsymbol B_2\right],
   \label{eq:bm-overlap2}
\end{eqnarray}
where  we introduced the definitions
\begin{eqnarray}   \tilde A_1&=&  (m+ x_1  M_0)(k'^x_1 +i k'^y_1)- (m+ x'_1M'_0)(k^x_1+i k^y_1),\nn   \\   \tilde B_1^x &=&  -i(m+ x'_1M'_0) (m+ x_1  M_0)   +i(k'^x_1+ik'^y_1)(k^x_1+ ik^y_1),\nn
   \\   \tilde B_1^y &=& (m+ x'_1M'_0) (m+ x_1  M_0)+(k'_{1,x}+ik'_{1y})(k_1^x+ ik_1^y),\nn   \\   \tilde B_1^z &=&  i(m+ x'_1M'_0)(k_1^x+ik_1^y) +i(m+ x_1  M_0)(k'^x_1+ik'^y_1),
\end{eqnarray}
for the spin-dependent contribution from the active quark, and
   \begin{eqnarray}
A_2&=&  (m+ x_2  M_0)(m+ x'_2M'_0)+k^x_2 k'^x_2+k^y_2 k'^y_2,\nn   \\   B_2^x &=&  -(m+ x'_2M'_0) k^x_2   +(m+ x_2  M_0)k'^y_2,\nn   \\   B_2^y &=&  (m+ x'_2M'_0) k^x_2   -(m+ x_2  M_0)k'^x_2,\nn   \\   B_2^z &=&  k'^x_2 k_2^y-k'^y_2k^x_2,
\end{eqnarray}
   for the spin-dependent contribution of the spectator antiquark.

The Boer-Mulders function for the valence antiquark can be obtained 
through a similar calculation by replacing the antiquark spectator 
with the quark spectator.   As a result, one finds 
$h_{1,\pi}^{\perp\,\bar q}=h_{1,\pi}^{\perp\,q}$.

\section{Results from a light-front constituent model}
\label{Sec-5:results-from-LFCM}

Up to this point we made only general assumptions.
We have chosen to work in a constituent approach of the pion,
and determined its initial scale in Sec.~\ref{Sec-2:initial-scale}.
We have then chosen to use the light-front formalism and presented in 
Sec.~\ref{Sec-3:lcwf} a general discussion of light-front amplitudes 
in the pion constituent approach. 
In Sec.~\ref{Sec-4:TMDs-in-pion} we derived a a model independent 
representation of the leading-twist pion TMDs as overlap of light-front 
amplitudes for the $q\bar q$ Fock-state of the pion.
In this Section we will apply the formalism from Secs.~\ref{Sec-3:lcwf}
and \ref{Sec-4:TMDs-in-pion} to obtain predictions for pion TMDs
using a specific model for the momentum-dependent part of the 
light-front wave function.

\subsection{ 
   Model for the momentum-dependent wave function}
   \label{Sec-5A:chosing-the-model}

The formalism described in the previous sections is applied 
to a specific choice for the LFCM, namely the model proposed in 
Refs.~\cite{Schlumpf:1994bc,Chung:1988mu}.
The model is specified by adopting the following exponential form 
for the momentum-dependent part of the pion wave function
   \be
   \label{eq:psifc2}    
  \tilde\psi_\pi(\bar x,{\boldsymbol \kappa}_{\perp })     
   = [2(2\pi)^3]^{1/2}  \left(\frac{M_0(\bar x,{\boldsymbol \kappa}_\perp)}{4~\bar x (1-\bar x)}\right)^{1/2}
   \frac{1}{\pi^{3/4}\beta^{3/2}}\exp{(-\boldsymbol{\kappa}^2/(2\beta^2))}.
   \ee
The wave function in Eq.~(\ref{eq:psifc2}) is normalized as 
  $$
  \int_0^1~d\bar x\int\frac{d{\boldsymbol \kappa}_\perp}{ 2(2\pi)^3}~
  |\tilde \psi_\pi(\bar x,{\boldsymbol \kappa}_{\perp })|^2=1
  $$  
(recalling   
that $d\kappa_z= d\bar x~ M_0(\bar x,{\boldsymbol \kappa}_\perp)/[4 \bar x (1-\bar x)]$), 
and depends on the free parameter $\beta$ and the quark mass $m$, 
which have been fitted to the pion charge radius and decay constant.         
 In particular, we take $m=0.250$ GeV and $\beta=0.3194$~\cite{Schlumpf:1994bc}.    
 As we are considering only the leading $q\bar q$ Fock-space component in the pion LFWF, the quark (antiquark) contribution to the pion distribution functions at the hadronic scale of the model coincides with the valence quark $q_v$ (antiquark $\bar q_v$) contribution, while the sea quark contribution is vanishing.
 Furthermore, isospin symmetry imposes
 $j_{\pi^{+}}^{u_{v}}=j_{\pi^{+}}^{\bar d_{v}}=j_{\pi^{-}}^{d_{v}}=j_{\pi^{-}}^{\bar u_{v}}=\tfrac{1}{2} \,j_{\pi^{0}}^{u_{v}} =\tfrac{1}{2} \,j_{\pi^{0}}^{\bar u_{v}} = \tfrac{1}{2} \,j_{\pi^{0}}^{d_{v}} = \tfrac{1}{2} \,j_{\pi^{0}}^{\bar d_{v}} $, with $j=f_{1}, \, h^{\perp}_{1}$.
 In the following, we will refer to 
distributions of valence quarks and antiquarks in charged pions,
using the notation  $j^{q_v}_{\pi}$ and $j^{\bar q_v}_{\pi}$, respectively.

\subsection{\boldmath 
  Results for $f^{q_v}_{1,\pi}(x,\boldsymbol{k}^{2}_\perp)$ at the hadronic scale}
  \label{Sec-5B:f1-at-low-scale}

\begin{figure}[b!]
\centering 
\begin{tabular}{ccc}
\epsfig{file=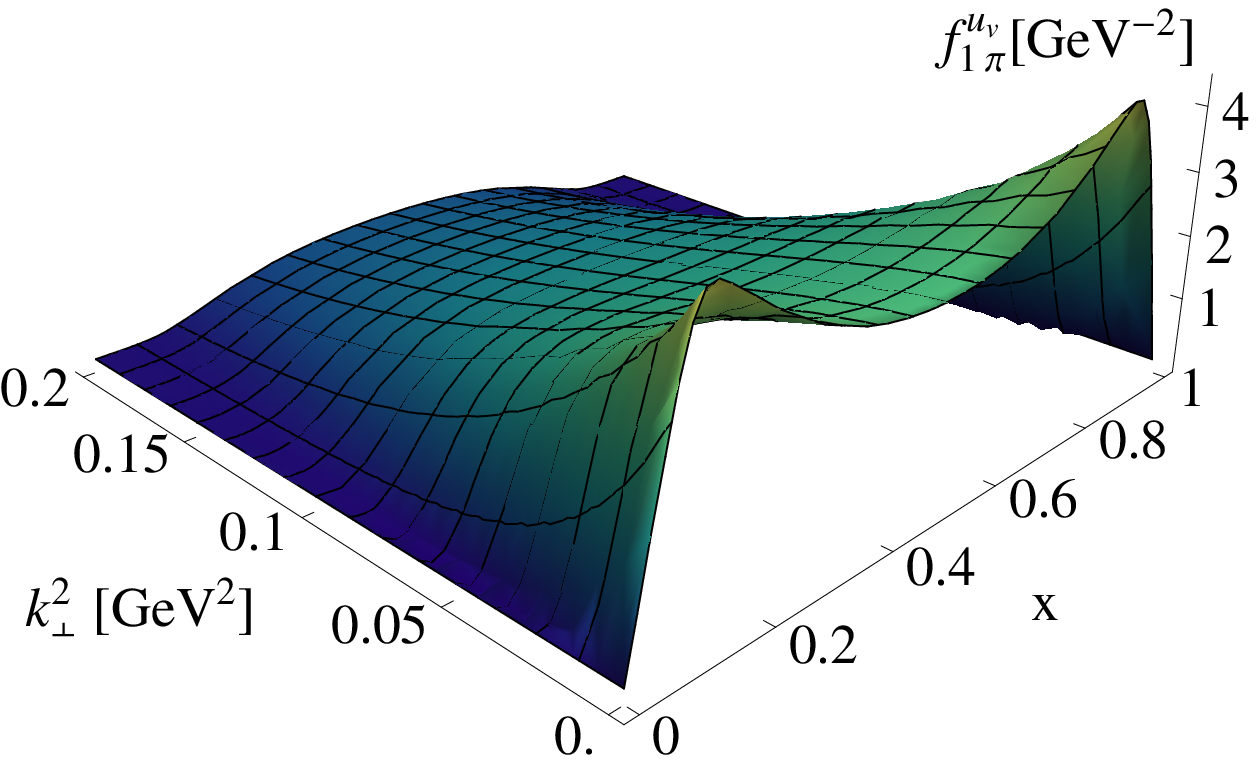,height=4.5cm} \hspace{1cm} \ & 
\epsfig{file=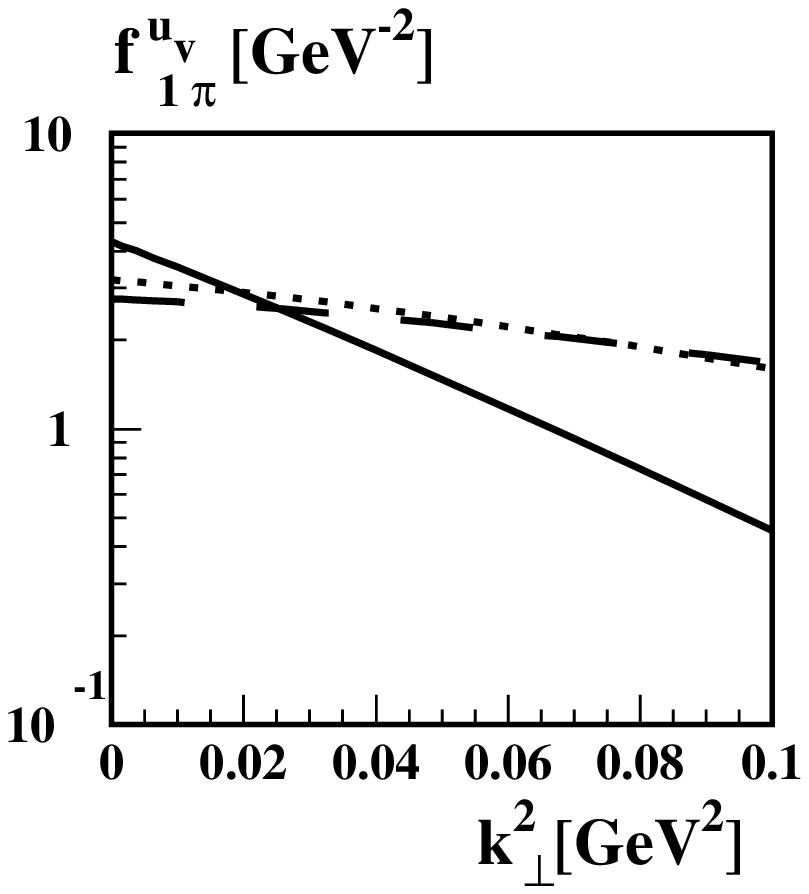,height=5cm} \ & 
\epsfig{file=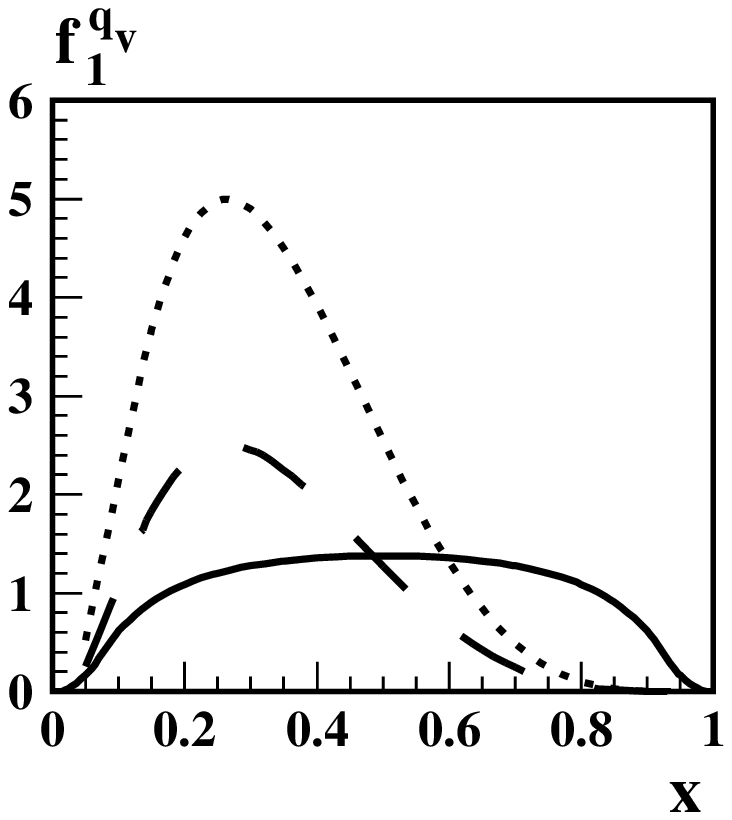,height=5cm}  
\\
(a) & (b) & (c) 
\end{tabular}
\caption{\label{Fig-2:tmd-f1-x-ksq}
  The valence-quark unpolarized TMD of the pion
  from the LFWF~\eqref{eq:psifc2} at the  hadronic scale.
  (a) $f_{1,\pi^{+}}^{u_v}(x,\boldsymbol k_\perp^2)$
      as function of $x$ and $\boldsymbol k^2_\perp$.
  (b) $f_{1,\pi}^{u_v}$ as function of $\boldsymbol k^{2}_{\perp}$ 
      for selected values of $x$ 
      ($x=0.1$ solid curve, $x=0.3$ dotted curve, $x=0.5$ dashed curve).
  (c) Comparison of the unpolarized PDFs as functions of $x$ in pion and 
      nucleon from LFCM approaches at their initial hadronic scales.
      Solid line: $f_{1,\pi}^{u_v}(x)$ in the pion, obtained in this work.  
      Dotted (dashed) curve:  $f_{1p}^{u_v}(x)$ ($f_{1p}^{d_v}(x)$) 
      in the proton from the light-front constituent quark model of 
      Ref.~\cite{Pasquini:2008ax}.}
\end{figure}

In Fig.~\ref{Fig-2:tmd-f1-x-ksq}, we show the model predictions for the 
valence-quark contribution to the unpolarized TMD as function of $x$ and 
$\boldsymbol k^{2}_{\perp}$. The results refer  to the low hadronic scale 
determined in Sec~\ref{Sec-2:initial-scale}. For the $q\bar q$ component 
of the pion state, the distribution of quark  with longitudinal momentum 
fraction $x$ is equal to the distribution of antiquark with longitudinal 
momentum fraction $1-x$, i.e.\ 
$f_{1,\pi}^{     q_v}(  x,\boldsymbol k^{2}_{\perp})
=f_{1,\pi}^{\bar q_v}(1-x,\boldsymbol k^{2}_{\perp})$. 
Furthermore, one has the relation 
$f_{1,\pi}^{     q_v}(x,\boldsymbol k^{2}_{\perp})
=f_{1,\pi}^{\bar q_v}(x,\boldsymbol k^{2}_{\perp})$,  
which gives as final result a momentum distribution symmetric 
with respect to $x=1/2$. We also observe a rapid fall off with 
$\boldsymbol k^{2}_{\perp}$, with a decreasing slope at larger $x$. 
This behavior can be better seen 
in Fig.~\ref{Fig-2:tmd-f1-x-ksq}b where we plot the $f_{1,\pi}^{u_v}$ 
TMD as function of  $\boldsymbol k^2_\perp$  at different values of $x$.
We notice that the $\boldsymbol k^{2}_{\perp}$ dependence is definitely 
not Gaussian, but it can be approximated by a Gaussian function with 
reasonable accuracy.
Upon integration over $\boldsymbol k_\perp$, we obtain the unpolarized PDF.
In Fig.~\ref{Fig-2:tmd-f1-x-ksq}c we compare the 
unpolarized quark distribution of the pion $f_{1,\pi}(x)$ with the results 
of the unpolarized quark  distribution of the proton $f_{1,p}(x)$ obtained 
from the three-quark LFWF of Ref.~\cite{Pasquini:2008ax}. The shape of 
the distributions for the pion and proton is quite different, reflecting 
the different valence-quark structure of the hadrons. For the proton, 
the momentum distribution of the valence-quark is peaked at $x \approx 1/3$.
Moreover, the SU(6) symmetry for the spin-flavor structure
of the LFWF in~\cite{Pasquini:2008ax} gives $f_{1,p}^{u_v}(x)=2f_{1,p}^{d_v}(x)$.

\subsection{\boldmath 
  Evolved results for $f^{q_v}_{1,\pi}(x)$ in comparison to parametrizations}
  \label{Subsec-5C:f1-evolution}

\begin{figure}[b!]
\vspace{-3mm}
\begin{center}\begin{tabular}{ccc}
  \epsfig{file=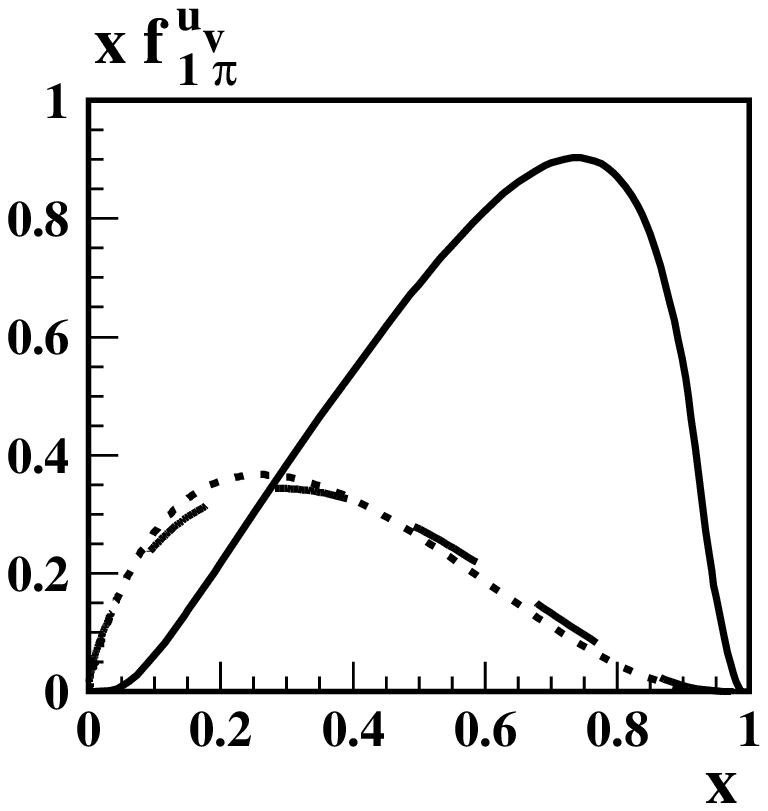,height=6cm} 
 &\epsfig{file=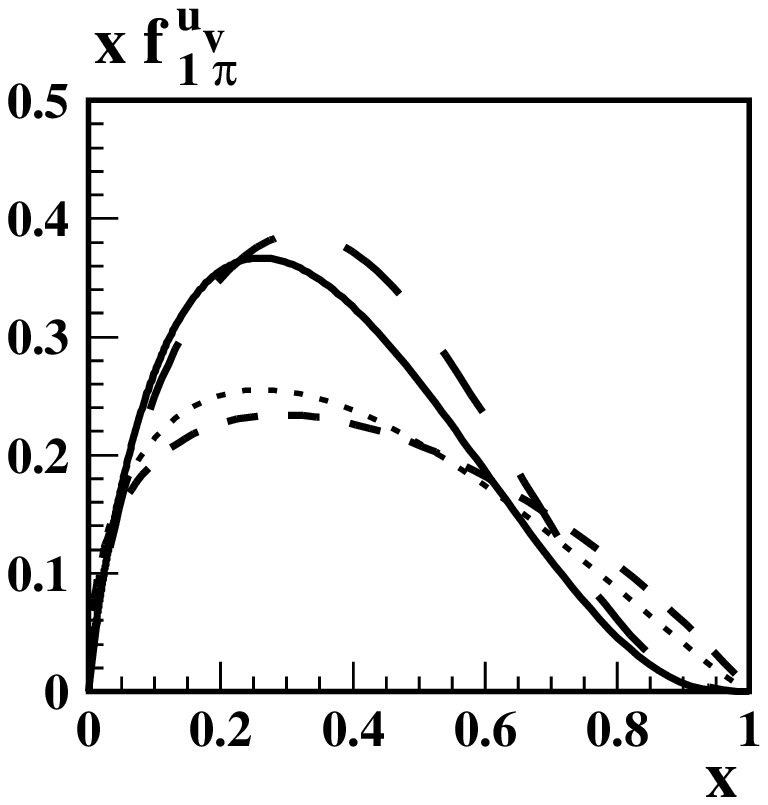,height=6cm}
 &\epsfig{file=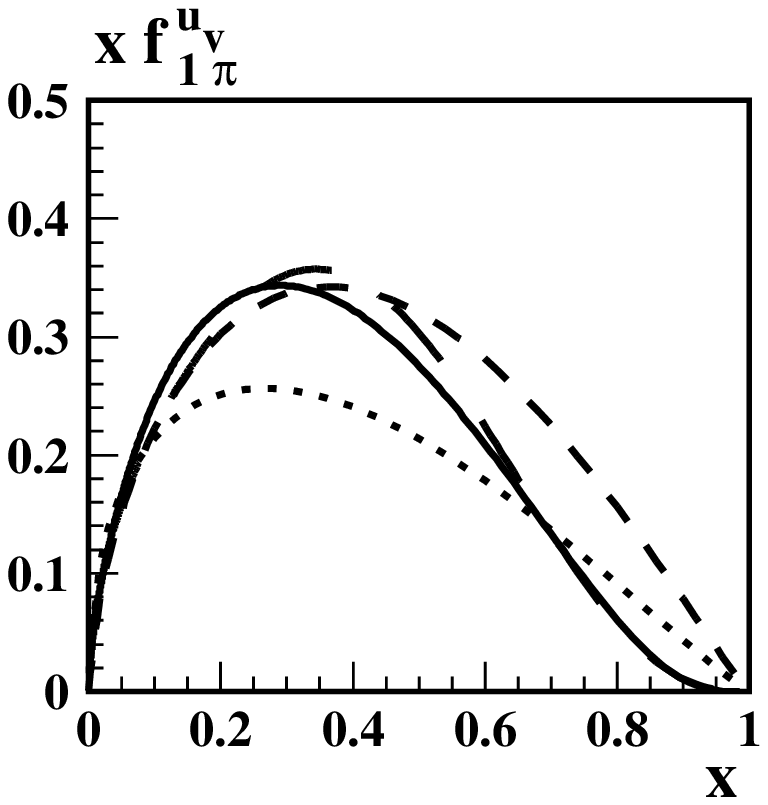,height=6cm} 
\vspace{-8mm}

\\
 (a) \ \ & (b) \ \ & (c) \ \  
\end{tabular}\end{center}
\vspace{-5mm}
\caption{  
  \label{Fig-3:f1-evolved}  
  (a) 
  $x f_{1,\pi}^{u_v}(x)$ as function of $x$.
  Solid line: at initial scale of the model.
  Dotted line: LO-evolved to $25\,{\rm GeV}^2$.
  Dashed line: NLO-evolved to $25\,{\rm GeV}^2$.
  (b) 
  $x f_{1,\pi}^{u_v}(x)$ as function of $x$ after LO-evolution to $Q^2=25$ GeV$^2$
  in comparison to the LO-parameterizations from
  \cite{Owens:1984zj}  (dashed  curve) and 
  \cite{Gluck:1991ey}  (dotted curve), and the calculation of 
  \cite{Hecht:2000xa} (long-dashed curve).
  (c) 
  $x f_{1,\pi}^{u_v}(x)$ as function of $x$ after NLO-evolution to $Q^2=25$ GeV$^2$
  in comparison to the NLO-parameterizations from
  \cite{Aicher:2010cb} (long-dashed curve), 
  \cite{Sutton:1991ay} (dashed curve), and 
  \cite{Gluck:1991ey,Gluck:1999xe} (dotted curve).}
\end{figure}

As a first test of the applicability of the LFCM  to the description of partonic properties of the 
pion, we compare the results for $f_{1,\pi}^{u_v}(x)$, evolved from the 
initial scale of the model to $Q^2=25\,{\rm GeV}^2$, with available 
parameterizations ~\cite{Gluck:1999xe,Sutton:1991ay,Owens:1984zj,Gluck:1991ey,
Hecht:2000xa,Aicher:2010cb,Wijesooriya:2005ir} 
(for a review of the pion PDF in the valence-$x$ 
region see also Ref.~\cite{Holt:2010vj}).
The initial-scale, LO-evolved and NLO-evolved distributions are shown 
in Fig.~\ref{Fig-3:f1-evolved}a. 
The LO and NLO evolutions are applied starting from the 
initial scales $\mu_{0, {\rm LO}}^{2}$  and $\mu_{0, {\rm NLO}}^{2}$  in 
Eqs.~\eqref{eq:fixing-initial-scale-LO} and 
\eqref{eq:fixing-initial-scale-NLO}, respectively.
Remarkably, although the initial scales and especially the values 
of $\alpha_s(\mu_0^{2})$ at LO and NLO differ, the evolved results are 
numerically close. This kind of behavior has been interpreted in 
Refs.~\cite{Gluck:1991ey,Gluck:1999xe,Gluck:1998xa,Traini:1997jz} as an 
indication for the ``convergence'' of perturbation theory down to low scales. 

It is important to keep in mind that the LO and NLO parameterizations 
of \cite{Gluck:1991ey,Gluck:1999xe,Gluck:1998xa} differ slightly at 
their respective low scales, such that they allow one to describe data
equally well in the combination with the LO or NLO hard parts in the 
respective LO or NLO treatments. In contrast, our model input at the 
initial scale is identical in LO and NLO.
This inevitably introduces a scheme dependence, when applying
the model results beyond LO. But we feel that such scheme-dependence 
effects are smaller than the generic model accuracy, 
as discussed in Sec.~\ref{Sec-2:initial-scale}. Considering that in 
the context of parton structure studies the generic model accuracy 
is observed to be around (10--30)$\,\%$ \cite{Boffi:2009sh},
we interpret the result in Fig.~\ref{Fig-3:f1-evolved}a,
i.e.\ the ``convergence of the LO and NLO results in the
sense of Refs.~\cite{Gluck:1991ey,Gluck:1999xe,Gluck:1998xa,Traini:1997jz},
as an indication that the issue of applicability of perturbative 
evolution equations down to the low scales in
Eqs.~\eqref{eq:fixing-initial-scale-LO}-\eqref{eq:fixing-initial-scale-NLO}
is not the dominant source of theoretical uncertainty in our approach.

In Fig.~\ref{Fig-3:f1-evolved}b the LFCM results at LO are compared 
with the LO parametrizations of Refs.~\cite{Owens:1984zj,Gluck:1991ey} and 
the calculation using Dyson-Schwinger equations of Ref.~\cite{Hecht:2000xa}.
In Fig.~\ref{Fig-3:f1-evolved}c we compare our NLO results with the NLO 
phenomenological fits of Refs.~\cite{Sutton:1991ay,Gluck:1991ey,Gluck:1999xe} 
and the results from the recent analysis of Ref.~\cite{Aicher:2010cb}.
The evolution effects are important, and change the shape of the distribution 
by leading to the convex-up behavior  near $x=1$, typical of the 
renormalization group equations which populate the sea-quark distribution 
at small $x$ at the expense of the large$-x$ valence-quark contribution.
In particular, the LFCM results are in good agreement with the recent 
analysis of Ref.~\cite{Aicher:2010cb} and the 
calculation~\cite{Hecht:2000xa}, showing a falloff at large $x$ 
much softer than the linear behavior obtained from the other analysis.

We   remark that there is a recent extraction~\cite{Wijesooriya:2005ir} of the pion PDF in the valence region
obtained from an updated NLO analysis of the Fermilab pion DY data. These
results are consistent with the parametrization of Ref.\cite{Gluck:1999xe}
in the valence-$x$ region
and therefore we do not show them explicitly in Fig.\ref{Fig-3:f1-evolved}c.
In summary, we observe that the partonic description of the pion 
works with the same level of accuracy observed for  the  LFCM
of the nucleon \cite{Boffi:2009sh}.

\subsection{\boldmath 
  Results for the Boer-Mulders function at low initial scale}
  \label{Sec-5D:BM-at-low-scale}

Having convinced ourselves that the pion LFCM provides a reasonable
description of the unpolarized TMD, we now focus on what this approach
predicts for  the Boer-Mulders function. 

The overall normalization of the Boer-Mulders function contains
(in leading order of the Wilson line expansion) the parameter $g^2$
in Eqs.~\eqref{eq:bm1},~\eqref{eq:bm-overlap} and \eqref{eq:bm-overlap2}.
At first glance it may appear natural to associate $g^2$ with the strong 
coupling at the low initial scale, $\alpha(\mu_{0}^{2})=g^2/(4\pi)$, and 
eventually we shall do this. But it is worth discussing this choice
in some more detail, because in a nonperturbative calculation this is 
a non-trivial step which should be done with care.
The expansion of the Wilson line is certainly appropriate for 
demonstrating ``matters of principle'' such as the existence of T-odd 
TMDs in QCD \cite{Brodsky:2002cx,Collins:2002kn}. But it is a priori
not clear whether this approach provides an adequate description of 
nonperturbative hadronic physics. From this point of view, one could 
consider the one-gluon-exchange approximation as an effective description. 
Besides the pioneering efforts of Ref.~\cite{Gamberg:2009uk}, nothing is 
known about effects from the Wilson line beyond one-gluon exchange. 
One could therefore understand $g^2$ as a free parameter and choose
its value to ``effectively'' account for higher order effects, 
which would be understood as part of the model. For instance, the value 
of $g^2$ could be adjusted to reproduce data. While in principle perfectly 
legitimate, we feel that here this would be an impractical procedure.

In the context of the pion Boer-Mulders function not much data are
available, and at the present state of the art the analysis of that
data bears uncertainties which are difficult to control. We therefore 
prefer not to introduce a free parameter at this point. Instead we fix 
$\alpha(\mu_{0,\rm NLO}^{2})=g^2/(4\pi)$ in Eq.~(\ref{eq:fixing-initial-scale-NLO}).
One could have also chosen to reproduce the LO value $\alpha(\mu_{0,\rm LO}^{2})$ 
in Eq.~(\ref{eq:fixing-initial-scale-LO}). However, the choice of NLO value 
$\alpha(\mu_{0,\rm NLO}^{2})$ is preferable over the LO value $\alpha(\mu_{0,\rm LO}^{2})$ 
for two reasons. First, the NLO-value can be associated with higher 
stability from the perspective of perturbative convergence
\cite{Pasquini:2004gc,Broniowski:2007si,Davidson:2001cc,Courtoy:2008nf},
and may be interpreted as effectively considering higher order effects
in above explained sense. 
Second, a smaller value of $\alpha(\mu_{0,\rm NLO}^2)$ 
helps to better comply with positivity constraints (see below). 
However, let us stress that fixing the value of $g^2$ 
in the overall normalization of the Boer-Mulders function is part 
of the modeling, and one could revisit this choice, if it 
gave  unsatisfactory phenomenological results. Below we shall see 
that our choice leads to satisfactory results.

\begin{figure}[b!]

\begin{center}\begin{tabular}{ccc}
\epsfig{file=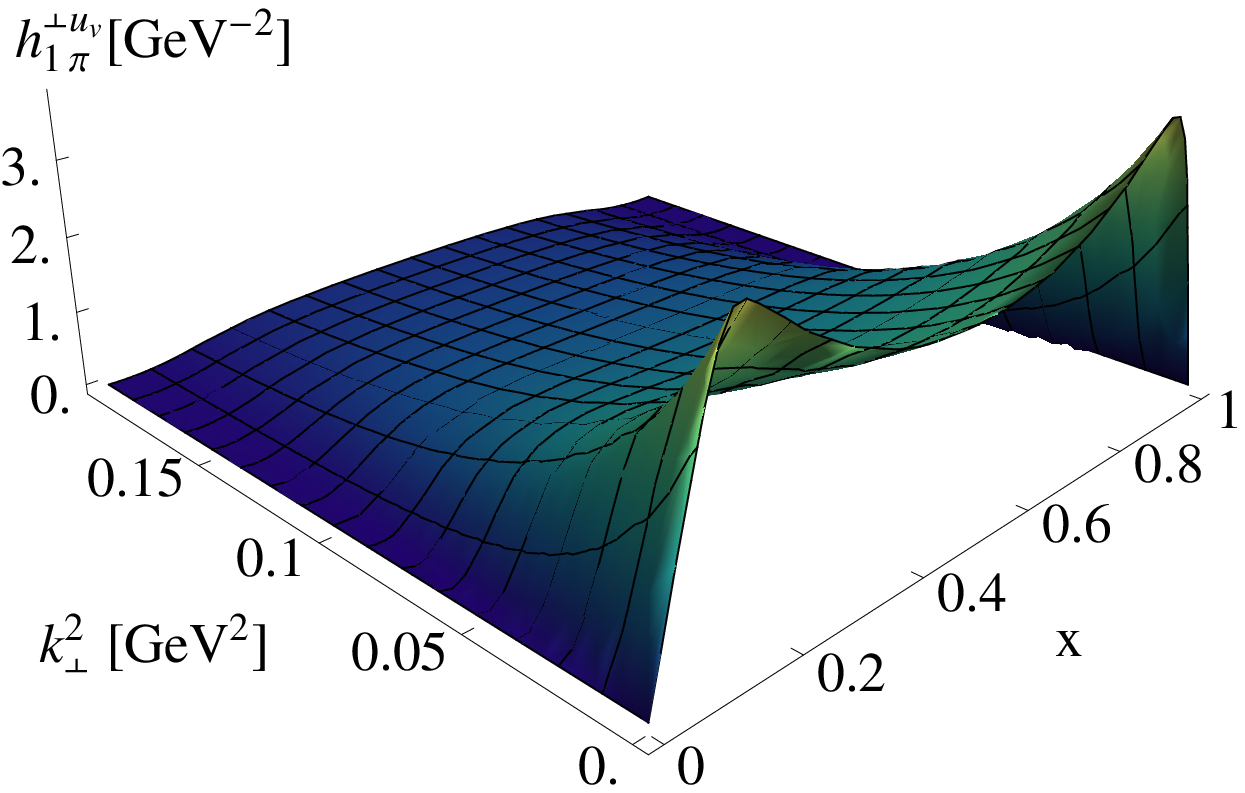,height=4.5cm} \hspace{1cm} \ & 
\epsfig{file=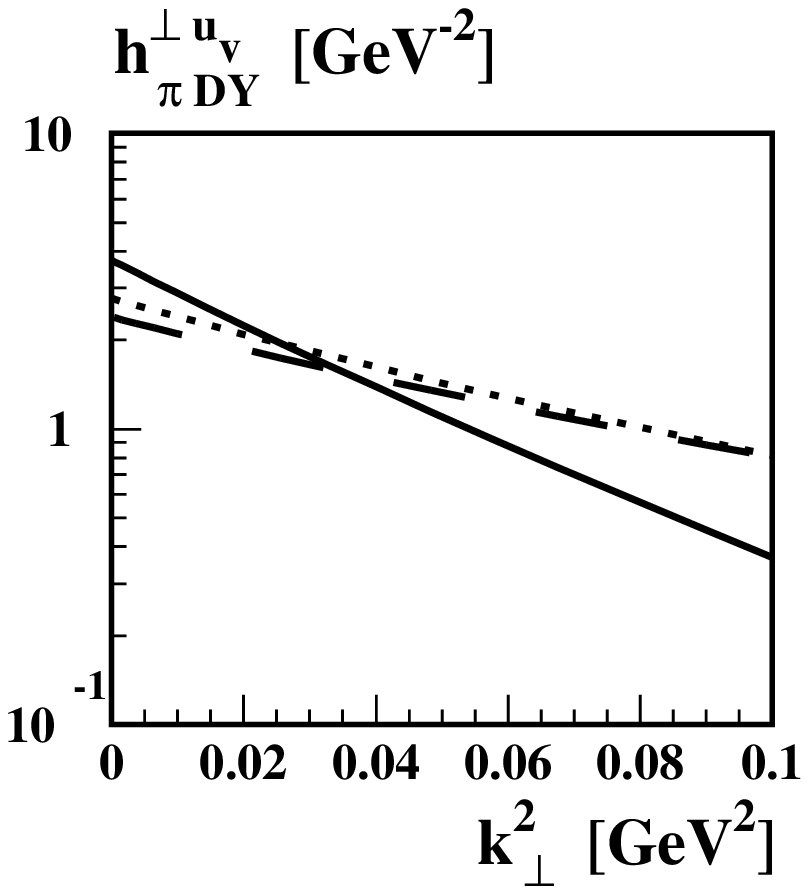,height=5.5cm} \ & 
\epsfig{file=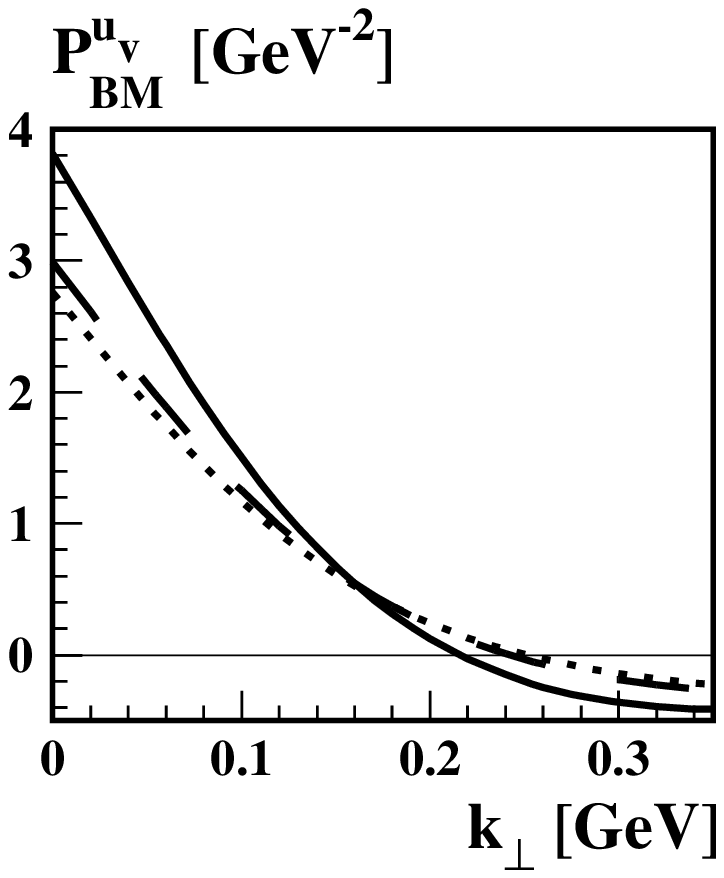,height=5.5cm} \vspace{-4mm} 
\\
\ \hspace{15mm} (a) &  (b) \hspace{15mm} & (c) \hspace{15mm} 
\end{tabular}  \end{center} 	
\caption{
  \label{Fig-4:tmd-bm-x-ksq}
  The Boer-Mulder function of the pion in the DY process from the LFCM
  at initial scale.
  (a) $h_{1,\pi}^{\perp u_v}(x,\boldsymbol k_\perp^2)$     as function of $x$ and $\boldsymbol k^2_\perp$.
  (b) $h_{1,\pi}^{\perp u_v}$ as function of $\boldsymbol k^{2}_{\perp}$ 
      for selected values of $x$ 
      ($x=0.1$ solid curve, $x=0.3$ dotted curve, $x=0.5$ dashed curve).
  (c) The positivity relation~\eqref{eq:positivity} for the valence-$u$ 
      quark in the pion as a function of $k_{\perp}$ at different values 
      of $x$: $x$=0.2 (solid curve), $x=0.35$ (dashed curve) 
      and $x$=0.5 (dotted curve).}
\end{figure}

In Fig.~\ref{Fig-4:tmd-bm-x-ksq}a we show the LFCM results for the 
Boer-Mulders TMD as function of $x$ and $\boldsymbol k^{2}_{\perp}$
with the sign as it is expected to appear in the DY process.
The shape of the distribution is very similar to the unpolarized TMD. 
It is symmetric  with respect to $x=1/2$, with a peak  at $x\sim 0.1$, 
and is rapidly decreasing at larger $\boldsymbol k^{2}_{\perp}$, with a 
fall-off which is not Gaussian but can be approximated reasonably well
by a Gaussian function. 
This is evident from Fig.~\ref{Fig-4:tmd-bm-x-ksq}b which displays the 
$\boldsymbol k^{2}_{\perp}$  dependence at selected values of $x$. The 
slope in $\boldsymbol k^{2}_{\perp}$ of the Boer-Mulders function is 
slightly steeper than that of the unpolarized TMD, in particular 
at larger values of  $x$.

The next important test of the model calculation is posed by positivity 
\cite{Bacchetta:1999kz} which requires that in the pion the unpolarized 
and Boer-Mulders TMD obey the following positivity relation, 
which holds flavor by flavor,
\begin{equation}\label{eq:positivity}
  P_{{\rm BM}}^q(x,\boldsymbol k^{2}_{\perp}) 
  \equiv 
  f_{1,\pi}^{q}(x,\boldsymbol k^{2}_{\perp})
  -\frac{k_\perp}{M_\pi}|h_{1,\pi}^{\perp\, q}(x,\boldsymbol k^{2}_{\perp})|\ge 0 \;.
 \end{equation}
The model results for $P_{{\rm BM}}^q(x,\boldsymbol k^{2}_{\perp})$ at selected 
values of $x$ are plotted in Fig.~\ref{Fig-4:tmd-bm-x-ksq}c.\footnote{
  \label{Footnote-1}
  We remark that if both functions had exactly Gaussian 
  $ k_{\perp}$-behavior (which they have not), 
  the steeper $\boldsymbol k^{2}_{\perp}$-slopes of
  $h_{1,\pi}^{\perp\, q}(x,\boldsymbol k^{2}_{\perp})$ observed in 
  Fig.~\ref{Fig-4:tmd-bm-x-ksq}b as compared to 
  $f_{1,\pi}^{q}(x,\boldsymbol k^{2}_{\perp})$ in 
  Fig.~\ref{Fig-2:tmd-f1-x-ksq}b would be a necessary 
  (though not sufficient) condition to satisfy positivity.} 
We see that the inequality~(\ref{eq:positivity}) is safely satisfied 
for $k_{\perp}\lesssim 0.2\,{\rm GeV}$ but violated for larger $k_\perp$.
Calculations in effective nonperturbative model frameworks may provide
some insights into the properties of TMDs for $k_\perp\ll \mu_0$, but the 
description of the region $k_\perp\sim {\cal O}(\mu_0)$ is out of scope.
Nevertheless, from the point of view of internal consistency, the 
non-compliance with (\ref{eq:positivity}) at large $k_\perp$ is of course 
unsatisfactory.
This happens, to best of our knowledge, also in all presently available 
calculations of T-odd TMDs~\cite{Kotzinian:2008fe}. The general reasons 
for that can be traced back to an inconsistent treatment: T-odd TMDs are
calculated to ``first order of the expansion of the Wilson line,'' 
whereas T-even TMDs like $f_1^q(x,\boldsymbol k^{2}_{\perp})$ are evaluated 
to ``zeroth order'' in that expansion.
To preserve positivity the Wilson link expansion should be truncated 
consistently at the same order for both T-odd and T-even TMDs 
which enter the inequality~(\ref{eq:positivity}) on the same footing 
\cite{Pasquini:2011tk}.

From the point of view of practical applications, it is gratifying 
to observe that the inequality (\ref{eq:positivity}) is violated 
only in the region of small $x$ or large $k_\perp$ 
\cite{Kotzinian:2008fe,Pasquini:2011tk}, i.e.\ in a region of
parameter space that is beyond the range of applicability of effective 
quark models. In particular, we convinced ourselves here that in the 
LFCM of the pion the non-compliance with inequalities in the extreme 
regions of the $(x,k_\perp)$-space has no practical consequences for the 
description of physical processes, provided one uses the model within 
its range of applicability.
The same observation was made in the case of the description of
nucleon T-odd TMDs in the constituent quark model framework
\cite{Pasquini:2011tk}.

\subsection{\boldmath 
  Comparison to results for Boer-Mulders functions from different models}
  \label{Subsec-5E:comparison-bm-models}

It is instructive to compare the Boer-Mulders functions of pion and 
nucleon. Let us define the $(1/2)$- and $(1)$-transverse moments 
of the pion and proton Boer-Mulders functions as
\begin{equation}
\label{eq:moments-bm}
  h_{1,h}^{\perp(1/2)}(x)=\int d^2\boldsymbol k_{\perp}
  \frac{k_{\perp}}{2M_{h}}\, h_{1,h}^{\perp}(x,\boldsymbol k^{2}_{\perp}), \quad
  h_{1,h}^{\perp(1)}(x)=\int d^2\boldsymbol k_{\perp}
  \frac{\boldsymbol{k}^{2}_{\perp}}{2M^{2}_{h}}\, 
  h_{1,h}^{\perp}(x,\boldsymbol k^{2}_{\perp}).
\end{equation}
Owing to the appearance of hadron masses in the correlators 
defining the Boer-Mulders functions in Eq.~\eqref{h1}, the magnitude 
of the $(1)$ moment of the pion Boer-Mulders 
function is artificially enhanced by a factor $\sim M_p/M_\pi$ with 
respect to the nucleon case. Therefore, 
in the following plots, we  will rescale   the results for the 
(1) moment of the proton Boer-Mulders function by that factor,
in such a way that the comparison with the results for the pion   
is not distorted by the numerically very different values of pion 
and nucleon masses.

\begin{figure}[t!]

\begin{center}\begin{tabular}{ccc}
\epsfig{file=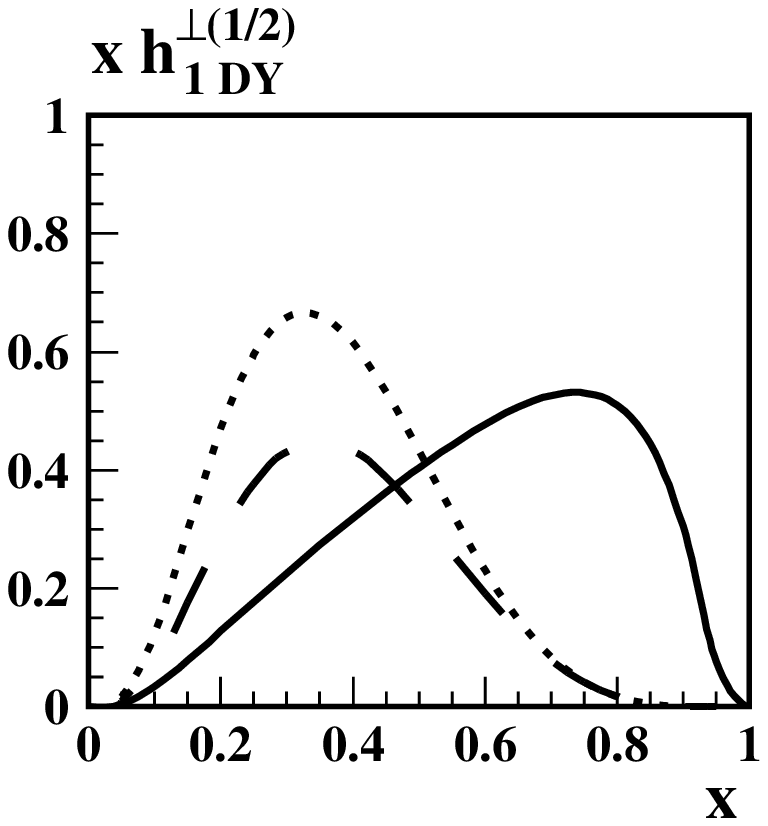,height=6cm} \ &
\epsfig{file=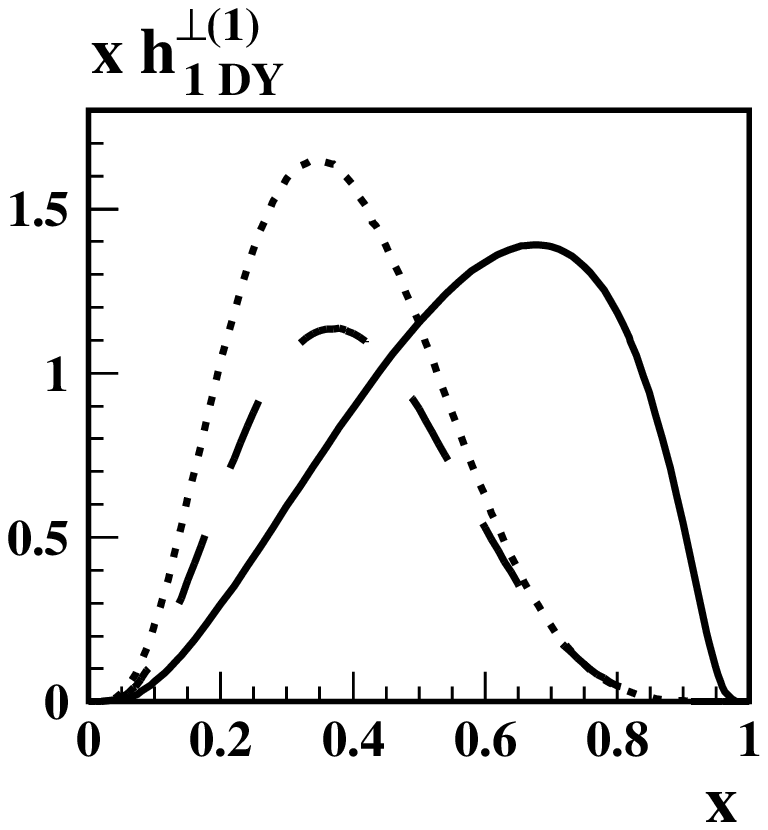,height=6cm} \ &
\epsfig{file=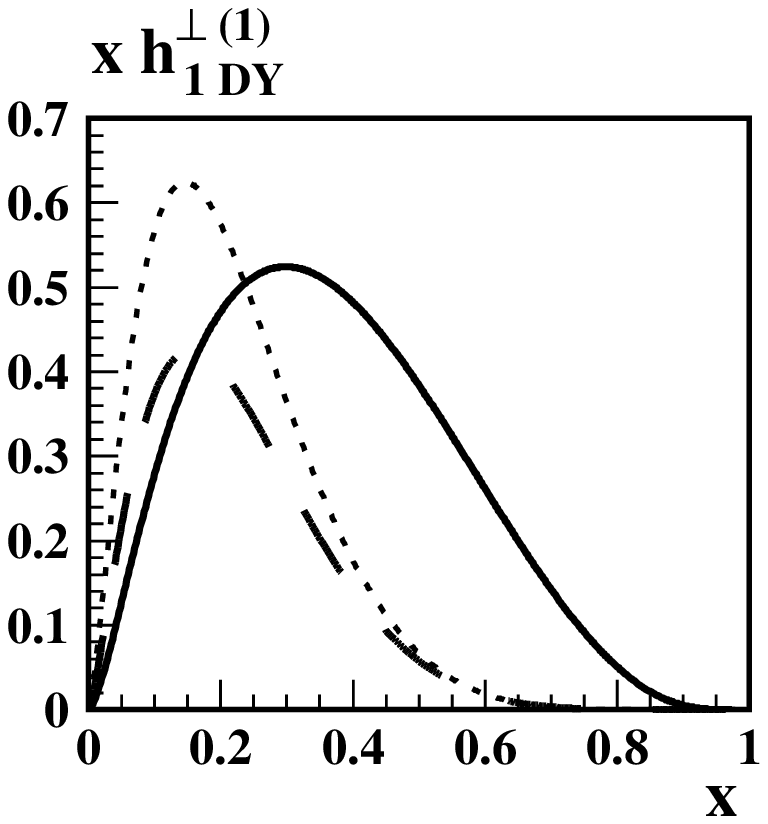,height=6cm} 
\vspace{-9mm} 
\\
(a) & (b) & (c) 
\end{tabular}  \end{center} 	
\caption{
  \label{Fig-5:bm-models-evolution}
  (a) Comparison of $x\,h_{1}^{\perp(1/2)q_v}(x)$ in the DY process as 
      functions of $x$ in pion and nucleon from LFCM approaches 
      at initial scales.
      Solid line: $u_v$-distribution in pion, this work.  
      Dotted (dashed) curve:  $u_v$- ($d_v$-) distribution in proton, 
      Ref.~\cite{Pasquini:2008ax}.
  (b) The results for $x h_{1\, {\rm DY}}^{\perp\, (1)}$ as function of $x$ .
      Solid curve:  $u_v$-distribution in pion.  
      Dotted (dashed) curve: $u_v$- ($d_v$)distribution in proton, 
      from the LFCM of Ref.~\cite{Pasquini:2011tk}. 
      The proton results are rescaled by a factor $M_p/M_\pi$.
  (c) The same as in Fig.~\ref{Fig-5:bm-models-evolution}b 
      but at $Q^2=25$ GeV$^2$, obtained with approximate LO evolution 
      from the LFCM results at the hadronic scale. }
\end{figure}

Fig.~\ref{Fig-5:bm-models-evolution}a compares the results for 
$h_{1,\pi}^{\perp(1/2)u_v}(x)$ obtained here and $h_{1,p}^{\perp(1/2)q_v}(x)$ 
obtained in \cite{Pasquini:2011tk}. 
Similarly, Fig.~\ref{Fig-5:bm-models-evolution}b show the results 
for the (1) moment of the pion Boer-Mulders function in 
comparison with the corresponding results for valence quarks
in the proton, rescaled by a factor $M_p/M_\pi$.
In both cases, the distributions for the valence contribution in the proton and pion have  comparable magnitude, but
similarly to the case of the unpolarized PDF, the $x$ dependence
is quite different. 
The sign of the pion Boer-Mulders function is consistent with the 
sign of the Boer-Mulders function of the proton~\cite{Burkardt:2007xm}, as 
obtained also in lattice calculations~\cite{Engelhardt:2013nba}, 
the MIT-bag model~\cite{Lu:2012hh} 
and spectator models~\cite{Lu:2005rq,Gamberg:2009uk}.
Interestingly, in comparison with other model calculations like 
the spectator model~\cite{Lu:2005rq,Gamberg:2009uk} and MIT-bag 
model~\cite{Lu:2012hh}, the shape and the magnitude of $h_{1,\pi}^{\perp}$
from LFCM are quite different. Similar differences have been found 
also in the comparison of the model results for the proton Boer-Mulders 
function~\cite{Pasquini:2010af,Pasquini:2011tk}.
The LFCM predictions for the nucleon Boer-Mulders function 
 favorably  describe available SIDIS data 
\cite{Pasquini:2011tk}. In Sec.~\ref{Sec-8:BM-in-DY} we will see that the LFCM predictions 
for the pion Boer-Mulders function provide a similarly satisfactory
description of DY data.

\subsection{\boldmath 
  Estimating the $x$-evolution for the Boer-Mulders function}
  \label{Subsec-5F:bm-evolution}

For phenomenological applications we will need the pion Boer-Mulders 
function from the LFCM evolved to experimentally relevant scales.
This requires both, evolution in $x$ and transverse momentum.
In this section we discuss the $x$-evolution
(the evolution of the transverse-momentum dependence will be 
discussed in the next section.)

Recently, substantial progress on the evolution of TMDs has 
been achieved~\cite{Collins-book,Aybat:2011zv,Aybat:2011ge,Cherednikov:2007tw,
Bacchetta:2013pqa,Echevarria:2012pw}.
However, the exact evolution equations for the Boer-Mulders function 
are still under study. At the present stage we have to resort to
approximations in order take into account effects of scale dependence.
To this aim, we will follow the same strategy as we adopted for the 
Boer-Mulders function of proton~\cite{Pasquini:2011tk}, and approximate
the evolution of transverse moments of the Boer-Mulders function by
using the evolution equations of the chiral-odd transversity distribution
function in the nucleon (in a spin-zero hadron like pion there is 
of course no transversity distribution, but the pion Boer-Mulders 
originates from the same unintegrated chiral odd correlator). 

To be more precise, we will evolve the (1)-moments of the 
Boer-Mulders functions. Such transverse moments appear naturally in 
transverse-momentum weighted azimuthal asymmetries, and it was 
argued that asymmetries weighted in this way are less affected by 
Sudakov effects \cite{Boer:2001he}. It will be possible to ultimately
judge the quality of this approximation only after the exact evolution 
equations are known. But we feel confident that the uncertainty
introduced by this step in our theoretical study is not larger than the 
generic accuracy of the LFCM.

Fig.~\ref{Fig-5:bm-models-evolution}c show the results for (1)-moment 
$x h_{1\, {\rm DY}}^{\perp\, (1)}$ after approximate (transversity) LO-evolution 
from the initial scale in Eq.~(\ref{eq:fixing-initial-scale-LO}) to 
$Q^2=25$ GeV$^2$. 
For comparison we include also the results for the nucleon
Boer-Mulders functions, rescaled by a factor $M_p/M_\pi$.
As in the case of the unpolarized PDF, the effects of the evolution 
are sizable, producing a shift of the peak position towards 
smaller $x$ and reducing the magnitude of the distribution. 

In Sec.~\ref{Sec-6:DY} we will use the model predictions to describe
azimuthal asymmetries in DY in a LO treatment. For this purpose, we 
will use the results for $f_{1,\pi}(x)$ and $h_{1,\pi }^{\perp\, (1)}(x)$
LO evolved in $x$ to experimental scales --
exactly and approximately, respectively, as described in 
Sec.~\ref{Subsec-5C:f1-evolution} and the present
Sec.~\ref{Subsec-5F:bm-evolution}.
Before applying the model results to phenomenology, in the following 
section we will estimate the broadening of transverse 
momenta at the large scales typically probed in DY experiments.
\section{The Drell-Yan process with unpolarized hadrons}
\label{Sec-6:DY}

In this Section we introduce the concepts required to describe 
the Drell-Yan process in the parton model taking into account transverse-momentum effects. Our treatment will be pragmatic and phenomenological.

\subsection{Kinematics, variables, conventions}
\label{subsec-6A:DY-general+nuclear-effects}

Let $p_{1,2}$ denote the momenta of the incoming hadrons $h_{1,2}$, 
and let $l$, $l^\prime$ be the momenta of the outgoing lepton pair. 
The kinematics 
of the process is described by the center of mass energy square $s$, 
invariant mass of the lepton pair $Q$, rapidity $y$ or the Feynman 
variable $x_F$, and the variable $\tau$ which are defined 
and related to each other as
\ba
&&	s    =      (p_1+p_2)^2 	\,,\;\;\; 
        q    =       l+l^\prime	\,,\;\;\; 
	Q^2  =      q^2 	\,,\;\;\; 
   	y   =      \frac12\,\ln\frac{p_2\cdot q}{p_1\cdot q} 
	    = \frac12\,\,\ln\frac{x_1}{x_2}\,,\;\;\;\nonumber\\
&&	x_F = x_1 - x_2	\,,\;\;\;
	\tau \equiv \frac{Q^2}{s} = x_1 x_2	\,. \label{Eq:DY-kinematics}
\ea
In the parton model the $x_i$ denote the fractions of the hadron momenta 
$p_i$ carried by (respectively) the annihilating parton or anti-parton,
and are given by (the $+$ signs refer to $x_1$, the $-$ signs $x_2$)
\be\label{Eq:x1-x2}
      x_{1,2} = \pm \frac{x_F}{2} + \sqrt{\frac{x_F^2}{4}+\tau\;} 
            = \sqrt{\tau}\,e^{\pm y}\;.
\ee

\begin{figure}[b!]

\vspace{-0.5cm}

\begin{center}
\epsfig{file=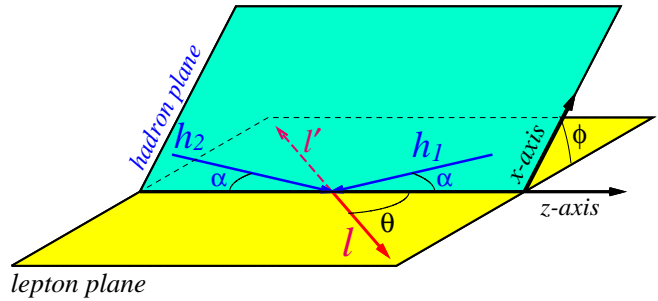,width=9cm}
\end{center}

\vspace{-0.5cm}

\caption{
  \label{Fig-6:CS-frame}
  The definition of the angles $\theta$ and $\phi$ in the 
  Collins-Soper frame. This frame is the center of mass frame of the 
  produced leptons in which the hadrons are incoming symmetrically 
  with respect to the $z$-axis (at an angle $\alpha$ in the figure) 
  with the transverse momentum $q_T$.}
\end{figure}

In the lab frame, where one hadron is a target or where both hadrons 
are beam particles, the produced lepton pair will in general have a
three-momentum $\boldsymbol{q}=\boldsymbol{l}+\boldsymbol{l}^{\;\prime} \neq 0$. 
It is often convenient to analyze the data in a dilepton rest frame.
There are various frames, including several dilepton rest frames, 
that are routinely used for data analyses, see 
Ref.~\cite{McGaughey:1999mq,Reimer:2007iy,Arnold:2008kf,Chang:2013opa,
Peng:2014hta} for an overview. The differences between 
the different frames are of order ${\cal O}(q_T/Q)$.
In the following we will work in the Collins-Soper frame, which is defined 
in Fig.~\ref{Fig-6:CS-frame}, and use only data analyzed in that frame.

In this work we will consider pion-nucleus collisions. The used convention is 
such that $x_1$ describes the momentum fraction of the parton from $\pi^-$, 
while $x_2$ describes the momentum fraction of the parton from the nucleon
bound in the nucleus. In order to describe nuclei with proton
number $Z$ and neutron number $N$ we will neglect nuclear binding 
effects and assume that, for instance, 
$f_{1/\rm nucleus}^u = (Z/A)\,f_{1/\rm proton}^u + (N/A)\,f_{1/\rm neutron}^u$,
where $A=N+Z$ denotes the mass number of the nucleus.
The neglect of nuclear binding effects is a justified step for 
$q_T\lesssim 3\,{\rm GeV}$ \cite{Guanziroli:1987rp,Bordalo:1987cs},
which includes the kinematic region of interest for our study.

\subsection{Structure functions in unpolarized DY}

The angular distribution of the DY lepton pairs originating from 
collisions of unpolarized hadrons is given in the Collins-Soper 
 frame by (see Fig.~\ref{Fig-6:CS-frame} for the definition
of angles),
\begin{equation}\label{Eq:DY-unp-angular-dependence-1}
\frac{dN}{d\Omega} \equiv
\frac{d\sigma}{d^4 q \, d \Omega} \bigg / \frac{d\sigma}{d^4 q}
= \frac{3}{4\pi} \, \frac{1}{\lambda + 3}
 \bigg( 1 + \lambda \cos^2 \theta
      + \mu \sin 2\theta \cos \phi
      + \frac{\nu}{2} \sin^2 \theta \cos 2\phi \bigg) \; .
\end{equation}
In the notation of Ref.~\cite{Arnold:2008kf} the coefficients 
$\lambda$, $\mu$, $\nu$ can be expressed in terms of DY structure functions 
as follows
\begin{equation}\label{Eq:DY-unp-angular-dependence-2}
\lambda = \frac{F_{UU}^{1} - F_{UU}^{2}}{F_{UU}^{1} + F_{UU}^{2}} \,, \qquad
\mu = \frac{F_{UU}^{\cos \phi}}{F_{UU}^{1} + F_{UU}^{2}} \,, \qquad
\nu = \frac{2 \, F_{UU}^{\cos 2\phi}}{F_{UU}^{1} + F_{UU}^{2}} \,.
\end{equation}

The so-called Lam-Tung relation claims $\lambda + 2 \nu = 1$, which reads 
in terms of structure functions $F_{UU}^{2} = 2 \, F_{UU}^{\cos 2\phi}$.
This relation is exact if one treats the DY process to ${\cal O}(\alpha_s)$ 
in the standard collinear factorization QCD framework
\cite{Lam:1978pu,Collins:1978yt}. 
At ${\cal O}(\alpha_s^2)$ 
the  Lam-Tung relation is violated, though at a numerically negligible rate
\cite{Mirkes:1994dp}.
However, DY data from pion-nucleus collisions show that it is strongly 
violated, calling for a nonperturbative leading-twist mechanism 
beyond  collinear factorization. The Boer-Mulders effect provides
such a mechanism \cite{Boer:1999mm}. Alternative nonperturbative
mechanisms to explain this observation have been proposed in
\cite{Brandenburg:1993cj,Brandenburg:1994wf,
Boer:2004mv,Brandenburg:2006xu,Nachtmann:2014qta}.

\subsection{Parton model treatment}
\label{subsec-VIC:parton-model-treatment}

In a tree-level parton model approach including transverse parton 
momenta in the region $q_T\ll Q$ the structure functions $F_{UU}^{1}$ and 
$F_{UU}^{\cos 2\phi}$ are leading twist, $F_{UU}^{\cos \phi}$ is subleading twist,
and $F_{UU}^{2}$ is a power-suppressed higher-twist effect proportional to 
$q_T^2/Q^2$. In such a treatment the transverse dilepton momenta $q_T$ 
arise from the convolutions of (``intrinsic'') transverse momenta of 
the partons as described through TMDs. The leading-twist structure 
functions in the unpolarized DY process are expressed in terms of TMDs 
through the following convolution integrals \cite{Arnold:2008kf}
\ba
     F_{UU}^1(x_1,x_2,q_T)\label{Eq:FUU}
     &=&
     \frac{1}{N_c} \, \sum_a \, e_a^2 \,
     \int d^{2}\boldsymbol{k}_{1\perp} \, d^{2}\boldsymbol{k}_{2\perp} \,
     \delta^{(2)}(\boldsymbol{q}_T - \boldsymbol{k}_{1\perp} - \boldsymbol{k}_{2\perp}) \,
     f_{1,\pi}^{a}(x_1,\boldsymbol k_{1\perp}^{2})\,f_{1,N}^{\bar a}(x_2,\boldsymbol k_{2\perp}^{2})\,,\\
     F_{UU}^{\cos(2\phi)}(x_1,x_2,q_T) \label{Eq:FUUcos2phi}
     &=& \frac{1}{N_c} \, \sum_a \, e_a^2 \,
         \int d^{2}\boldsymbol{k}_{1\perp} \, d^{2}\boldsymbol{k}_{2\perp} \,
         \delta^{(2)}(\boldsymbol{q}_T - \boldsymbol{k}_{1\perp} - \boldsymbol{k}_{2\perp})\;\omega_{\rm BM}\;
         h_{1,\pi}^{\perp     a}(x_1,\boldsymbol k^{2}_{1\perp})_{\rm DY}\,
         h_{1,  N}^{\perp\bar a}(x_2,\boldsymbol k^{2}_{2\perp})_{\rm DY}\,, \;\;\;\\
     \omega_{\rm BM}
     &=& \frac{2\big(\boldsymbol{q}_T\cdot\boldsymbol{k}_{1\perp}\big)
         \big(\boldsymbol{q}_T\cdot\boldsymbol{k}_{2\perp}\big)
         -q_T^2(\boldsymbol{k}_{1\perp}\cdot\boldsymbol{k}_{2\perp})}{M_\pi \,M_N\;q_T^2} \nonumber\;,
     \ea
where the sums go over $a=u,\,\bar u,\,d,\,\bar d$, and, in principle, 
heavier flavors. 

At this point it is important to recall that the parton model description 
is adequate and works reasonably well for some observables, but not for all.
For instance, in order to describe absolute cross sections
(even if averaged over transverse dilepton momenta),
it is necessary to go to the NLO QCD-treatment of the process. 
We will work in a LO (``tree-level'') formalism and consider ratios of 
cross sections where ``overall normalizations'' tend to cancel out.
Indeed, experience in various processes shows that different types 
of corrections may significantly affect absolute cross sections, 
but tend to cancel in cross section ratios. To quote just
a few examples, we mention in this context the weak scale dependence 
of longitudinal spin asymmetries in DIS \cite{Kotikov:1997df}, or
the near cancellation of resummation effects of large double 
logarithmic QCD corrections in longitudinal spin asymmetries 
in SIDIS \cite{Koike:2006fn}. 
In longitudinal and transverse spin asymmetries in DY higher order QCD 
corrections also tend to cancel \cite{Ratcliffe:1982yj,Vogelsang:1992jn,
Boer:2006eq}, 
and the same tendency is found for partonic threshold corrections
\cite{Shimizu:2005fp}. QCD corrections to polarization 
effects in $e^+e^-$ annihilation tend also to cancel \cite{Ravindran:2000rz}.
This is encouraging, but of course does not prove that higher 
order corrections will tend to cancel also for the cross section
ratios considered in this work, and more theoretical work 
is needed to attest this point.
We finally remark, that our parton model treatment does not consider 
the color entanglement effects discussed in Ref.~\cite{Buffing:2013dxa}.

\section{The unpolarized TMD\lowercase{s} in DY}
\label{Sec-7:f1-in-DY}

The LFCM was shown to describe the $x$ dependence of $f_{1,h}$
with an accuracy of (10-30)$\,\%$ within the range of applicability
of the model. (For pion see Sec.~\ref{Sec-5:results-from-LFCM}, 
for nucleon see Ref.~\cite{Lorce:2011dv}.) 
In this section we will therefore focus entirely on the $k_\perp$ dependence.

\subsection{\boldmath 
  Gaussian approximation and estimate of $k_\perp$ broadening 
  for $f^a_{1,h}(x,\boldsymbol{k}^{2}_\perp)$}
  \label{subsec-7A:estimate-broadening-f1}

The LFCM predictions for the $k_\perp$ dependence of TMDs presented
in Sec.~\ref{Sec-5:results-from-LFCM} refer to a low scale of 
$\sim 0.5\,{\rm GeV}$, and cannot be applied directly to describe
DY data which are typically taken in the region 
$Q\sim (4\mbox{--}9)\,{\rm GeV}$ between the $J/\psi$ and 
$\Upsilon$ resonances, or above the $\Upsilon$ resonance region.
In order to estimate the $k_\perp$-evolution effects we shall resort
to the Gaussian Ansatz, and proceed phenomenologically. 
The procedure is motivated and outlined below.

The DY cross section behaves like 
$\di\sigma/\di q_T^2\propto\exp(-q_T^2/\la q_T^2\ra)$ for 
$q_T\ll Q$ \cite{Cox:1982wy,D'Alesio:2007jt,Schweitzer:2010tt}.
This observation is the basis for the popularity of the Gaussian
Ansatz to model the distributions of transverse parton momenta 
in hadrons. Although certainly oversimplifying, the phenomenological 
success of the Gaussian Ansatz indicates that it is a useful working 
assumption. 
We shall therefore recast the model predictions for TMD as follows
\ba\label{Eq:Gauss-Ansatz}
     f_{1,h}^{a}(x,\boldsymbol{k}^{2}_\perp) = f_{1,h}^{a}(x)\,
     \frac{\exp(-k_\perp^2/\pTsqx{x}{unp}{a/h})}{\pi\,\pTsqx{x}{unp}{a/h}}\;,
     &&
     \pTsqx{x}{unp}{a/h} 
     = \frac{\int\di^2\boldsymbol k_\perp\;\boldsymbol k_\perp^2\,f_{1,h}^{a}(x,\boldsymbol k_\perp^{2})}
            {\int\di^2\boldsymbol k_\perp\;       f_{1,h}^{a}(x,\boldsymbol k^{2}_\perp)}\;, 
\ea
where $f_{1,h}^{a}(x)$ is the unpolarized collinear 
parton distribution function. 

Before describing in detail how we estimate $k_\perp$-evolution effects, 
let us comment on a feature concerning Eq.~(\ref{Eq:Gauss-Ansatz}).
In Sec.~\ref{Sec-5:results-from-LFCM} we have seen that the model 
results for pion TMDs exhibit an approximate Gaussian behavior. 
The same was demonstrated in \cite{Boffi:2009sh,Pasquini:2011tk} 
for the nucleon case.
In contrast to Refs.~\cite{Boffi:2009sh,Pasquini:2011tk} 
(where predictions from the LFCM of the nucleon were applied to 
SIDIS phenomenology)
in this work we do not take the Gaussian widths to be $x$-independent 
constants. Rather, in Eq.~(\ref{Eq:Gauss-Ansatz}) we allow a more 
flexible parameterization with $x$-dependent Gaussian widths. 
This has the advantage of further improving the quality of the Gaussian
approximation.

The exact evolution of the $k_\perp$ dependence of the unpolarized 
TMD is known in the Collins-Soper-Sterman (CSS) formalism, which
provides a framework for a quantitative description of 
transverse-momentum broadening effects with increasing energies. 
The underlying physical picture is that with increasing energy
gluon radiation broadens the ``initial'' (or ``intrinsic'') 
parton transverse momentum. There is no practical or theoretical
way to separate ``nonperturbative intrinsic'' and ``perturbative 
gluon-radiation'' effects. However, from a phenomenological point
of view, there is no need for that: both effects are collectively 
parametrized in the effective parameters in Eq.~(\ref{Eq:Gauss-Ansatz}), 
provided one pays due attention to apply this effective description 
to the region of low transverse momenta
$q_T\ll Q$ \cite{D'Alesio:2007jt,Schweitzer:2010tt}.
In order to estimate this effective broadening of the Gaussian 
widths we shall use the results from \cite{Schweitzer:2010tt}.

In principle one could directly work within the CSS formalism. 
However, the CSS-formalism has not yet been established for the 
Boer-Mulders effect. Moreover, even in the unpolarized case, it has not 
yet been studied whether one can use the CSS formalism starting 
from a scale as low as in 
Eqs.~(\ref{eq:fixing-initial-scale-LO}) and~(\ref{eq:fixing-initial-scale-NLO}).
In this work we therefore prefer to use the effective description 
of \cite{Schweitzer:2010tt} to estimate $k_\perp$-evolution effects,
which requires to use the Gaussian Ansatz, as done in Eq.~(\ref{Eq:Gauss-Ansatz}). 
The details of this step will be described below.

Let us now turn our attention to the description of the transverse parton
momenta in DY. We discuss first the mean transverse momenta of the produced 
lepton pairs (see Eqs.~(\ref{Eq:DY-kinematics}) and~(\ref{Eq:x1-x2})
for the relation of $x_F$ with $x_{1,2}$) defined as
\be
     \la q_T^2(x_F,s)\ra = 
     \frac{\int d^2q_T \;q_T^2\,F_{UU}^1(x_1,x_2,q_T)}
          {\int d^2q_T \,F_{UU}^1(x_1,x_2,q_T)}\:.
\ee
It is important to notice that in a LO formalism the energy (or scale) 
dependence is introduced by using parton distributions (LO-) evolved 
to the relevant scale, and using appropriately broadened Gaussian 
widths. We also notice that $\la q_T^2(x_F,s)\ra$ is a ratio of
observables, i.e.\ amenable to the description in a parton model approach
thanks to the approximate cancellation of higher order QCD effects,
as argued in Sec.~\ref{subsec-VIC:parton-model-treatment}.

When using the Gaussian Ansatz in a tree-level parton model approach, 
the $\la q_T^2\ra$ is given by the sum of the Gaussian widths of the 
unpolarized TMDs of the nucleon and pion.
(In general this would hold only if the Gaussian widths were 
flavor independent. In the LFCM, where sea quarks are absent, 
it also holds because only one flavor contributes to the production 
of the lepton pair, namely a valence $\bar u$ from the $\pi^-$ 
and a valence $u$ in the proton annihilate.)

If we used the model results discussed in Sec.~\ref{Sec-5:results-from-LFCM}
at their face value to estimate $\la q_T^2\ra$ we would strongly underestimate 
the data. This is not surprising as the model results have to be evolved. 
In order to estimate evolution effects, we add an energy-dependent constant 
$\la\delta k_\perp^2(s)\ra$ such that
\be\label{Eq:qT2-xF-s}
      \la q_T^2(x_F,s)\ra 
      =\pTsqx{x_1}{unp}{\bar u/\pi^-}
      +\pTsqx{x_2}{unp}{u/N}+\la\delta k_{\perp,\rm unp}^2(s)\ra \;.\;\;\;
\ee
The energy dependence of $\la q_T^2\ra$ enters only through 
$\langle\delta k_{\perp,\rm unp}^2(s)\ra$ which provides the amount 
of transverse-momentum broadening at given $s$. 
The variation of dilepton momenta with $s$ was investigated 
phenomenologically in \cite{Cox:1982wy,Schweitzer:2010tt}. These 
studies allow us to estimate the amount of $k_\perp$ broadening required 
in our approach to be
\be\label{Eq:pT-broadening}
     \la\delta k_{\perp,\rm unp}^2(s)\ra = \delta A_{\rm unp} + B_{\rm unp}\;s \; , \;\;\;
     \delta A_{\rm unp}=0.4\,{\rm GeV}^2 \, , \;\;\;
            B_{\rm unp}=2.6 \times 10^{-3}\,.
\ee
It is important to stress that $\la\delta k_{\perp,\rm unp}^2(s)\ra$ 
constitutes the accumulated $k_\perp$ broadening in both pion and nucleon. 
Notice that $\la\delta k_{\perp,\rm unp}^2(s)\ra$ could also depend on $x_F$ or 
other variables besides $s$, but we disregard this possibility here.
Finally, one should stress that that the linear broadening indicated 
in (\ref{Eq:pT-broadening}) is valid only in a narrow $s$-range 
\cite{Cox:1982wy,Schweitzer:2010tt}. When considering broader energy 
ranges up to collider energies the increase is $\log s$ 
\cite{Landry:2002ix} rather than linear. 

\

\subsection{\boldmath Comparison to data}
\label{subsec-7B:unpol-DY-data}

With the empirical estimate of $k_\perp$-broadening 
effects in Sec.~\ref{subsec-7A:estimate-broadening-f1},
the model results yield a good description of DY data in the region 
$s\sim (50\mbox{--}600)\,{\rm GeV}^2$ 
studied in Refs.~\cite{Cox:1982wy,Schweitzer:2010tt}. 
We present two examples to illustrate this.

\begin{figure}[b!]

\vspace{-0.5cm}

\begin{center}
\epsfig{file=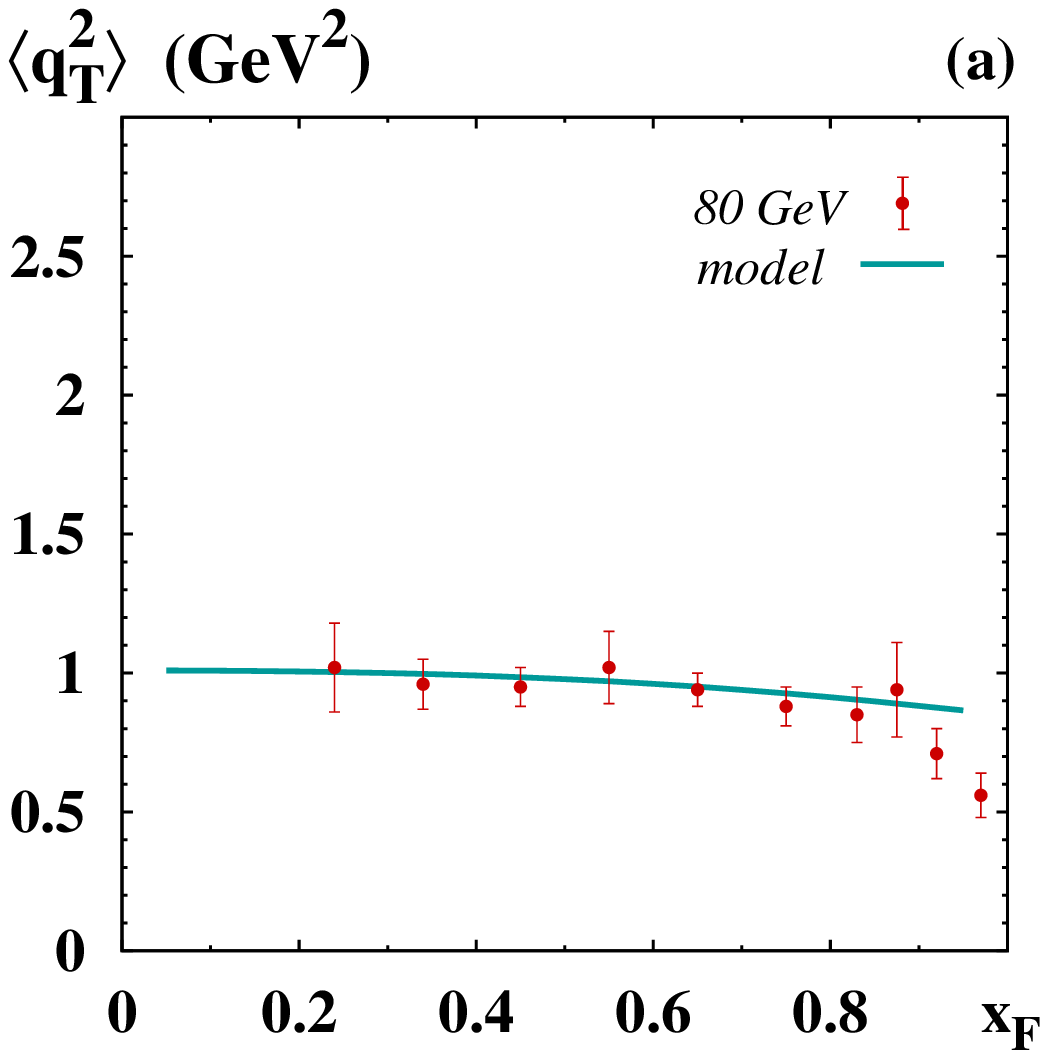,  width=6cm}
\epsfig{file=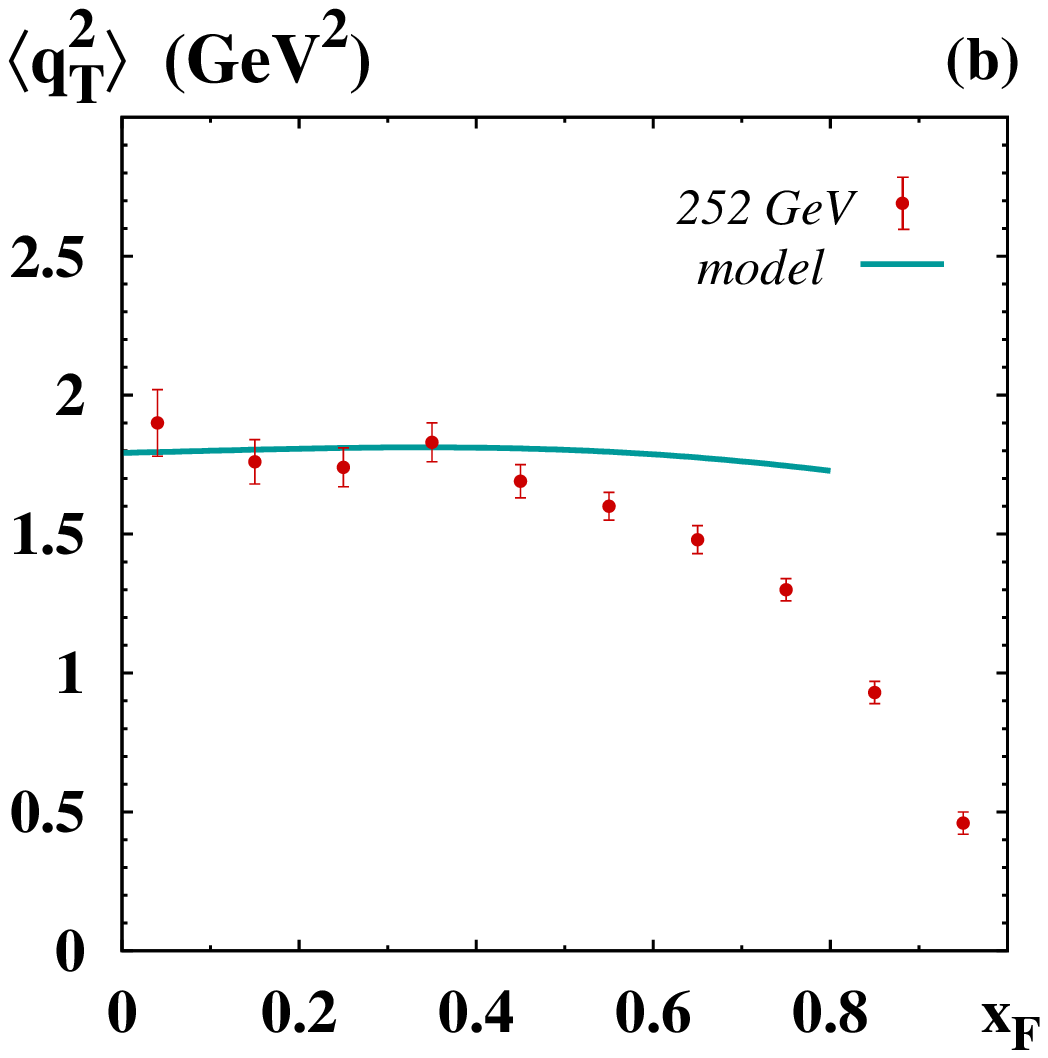,width=6cm}
\end{center}

\vspace{-0.5cm}

\caption{
  \label{Fig-7:pT-in-DY}
  The mean dimuon transverse-momentum square $\la q_T^2\ra$ vs.\
  $x_F$ from the Fermilab E615 experiment taken with respectively
 (a)
  $80\,{\rm GeV}$  \cite{Palestini:1985zc} and 
  (b) $252\,{\rm GeV}$ \cite{Conway:1989fs} 
  $\pi^-$ beams impinging on tungsten targets. 
  The theoretical curves are the result from LFCM obtained in this 
  work with a phenomenological estimate for transverse-momentum broadening,
  see Eqs.~(\ref{Eq:qT2-xF-s}) and (\ref{Eq:pT-broadening}). \vspace{-8mm}}
\end{figure}

Fig.~\ref{Fig-7:pT-in-DY} shows how our approach describes 
Fermilab E615 data on $\la q_T^2(x_F)\ra$ of Drell-Yan lepton pairs 
produced in collisions of $80\,{\rm GeV}$ and $252\,{\rm GeV}$ $\pi^-$ 
beams impinging on tungsten targets \cite{Palestini:1985zc,Conway:1989fs},
which corresponds respectively to $s\simeq 150\,{\rm GeV}^2$ and 
$473\,{\rm GeV}^2$.
We obtain a good description of the  $80\,{\rm GeV}$ data 
\cite{Palestini:1985zc} in the region $0.2\le x_F \lesssim 0.8$.
The $252\,{\rm GeV}$ data are well described for $0 \le x_F \lesssim 0.5$. 
Considering the generic accuracy  $\sim(10\mbox{--}30)\,\%$ of the LFCM,
the description of these data in the region 
$0.5\lesssim x_F \lesssim 0.7$ can be still considered satisfactory. 
However, beyond $x_F \gtrsim 0.7$ the approach breaks down. 
This is not a failure of the model
(which admittedly is not applicable at small- or large-$x$), 
but of the TMD approach in general. The reason is as follows.
At large $x_F$ the breakdown of the description of the DY process
in terms of parton distribution functions is expected.
The limit $x_F\to 1$ corresponds to large $x_1$ in the pion
(and small $x_2$ in the nucleon).
As $x_1\to 1$ the $\bar u$ from the $\pi^-$ is far off-shell, and 
more appropriately described in terms of the pion distribution 
amplitude \cite{Berger:1979du}. While this 
so-called Berger-Brodsky effect provides a unique 
opportunity to access information on the  pion distribution 
amplitude \cite{Bakulev:2007ej}, from the point of view of the 
TMD description of the DY process it is a power correction, 
which dominates as one approaches the limit $x_F\to 1$ of the 
available phase space.
Interestingly the Gaussian Ansatz itself still works 
even for $x_F \gtrsim 0.7$ \cite{Schweitzer:2010tt}. 
In principle one could continue using the TMD description, 
at least in some parts of the large-$x_F$ region.
This would require narrower $\la q_T^2(x_F)\ra$. 
The $x_F$ dependence of $\la q_T^2(x_F)\ra$ implied by the LFCM 
through the $x$ dependence of the Gaussian widths 
in Eq.~(\ref{Eq:Gauss-Ansatz}) is not sufficient for that,
but one could introduce an adequate $x_F$ dependence of the 
transverse-momentum broadening $\la\delta k_{\perp,\rm unp}^2(s)\ra$ 
in addition to its $s$ dependence. 
In this work we shall refrain from such attempts, stick to our  
$x_F$-independent description of transverse-momentum broadening 
in Eqs.~(\ref{Eq:qT2-xF-s}) and ~(\ref{Eq:pT-broadening}), and keep in
mind that this description has limitations at large $x_F$.

\begin{figure}[t!]

\begin{center}\begin{tabular}{cccc} 
\ \hspace{-9mm}
\epsfig{file=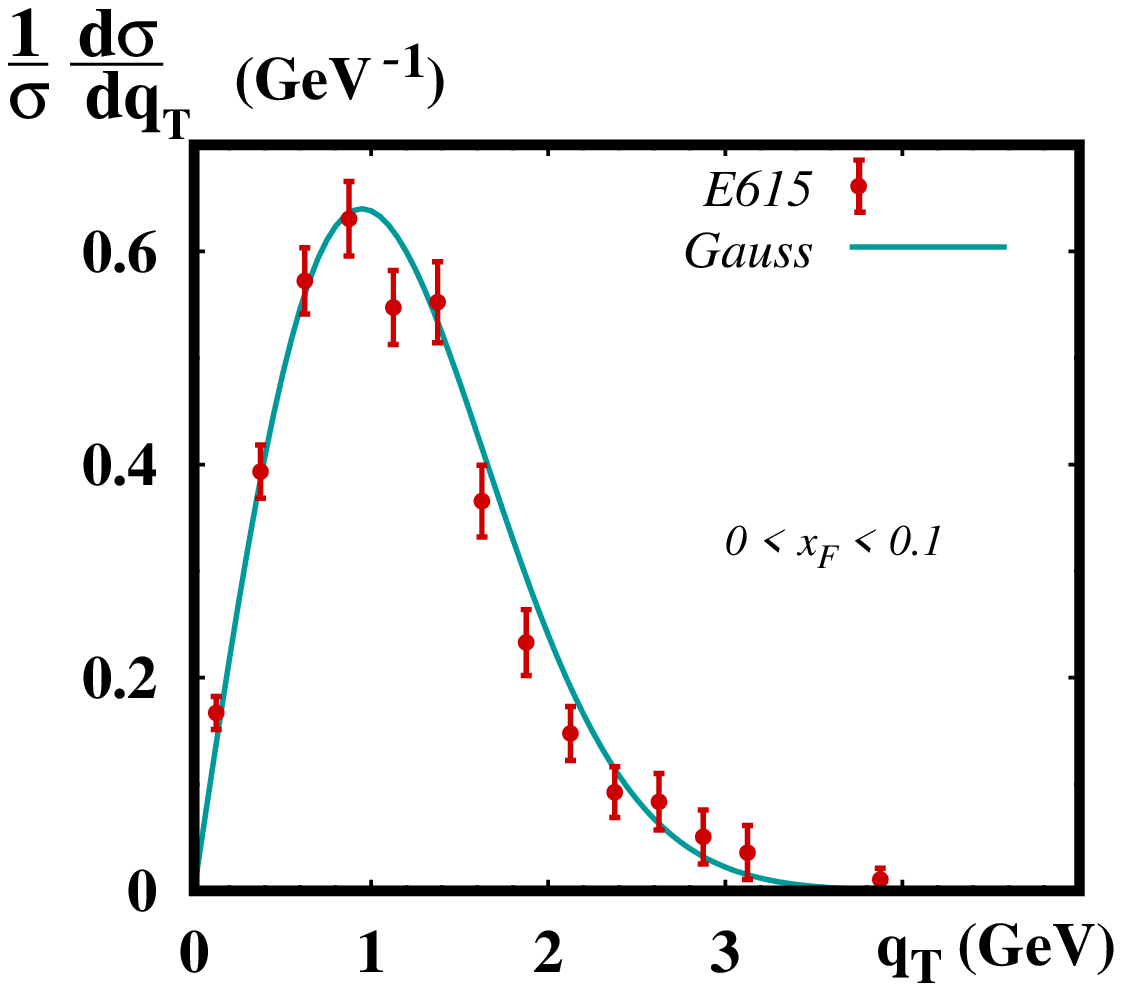,height=3.6cm} \ &
\epsfig{file=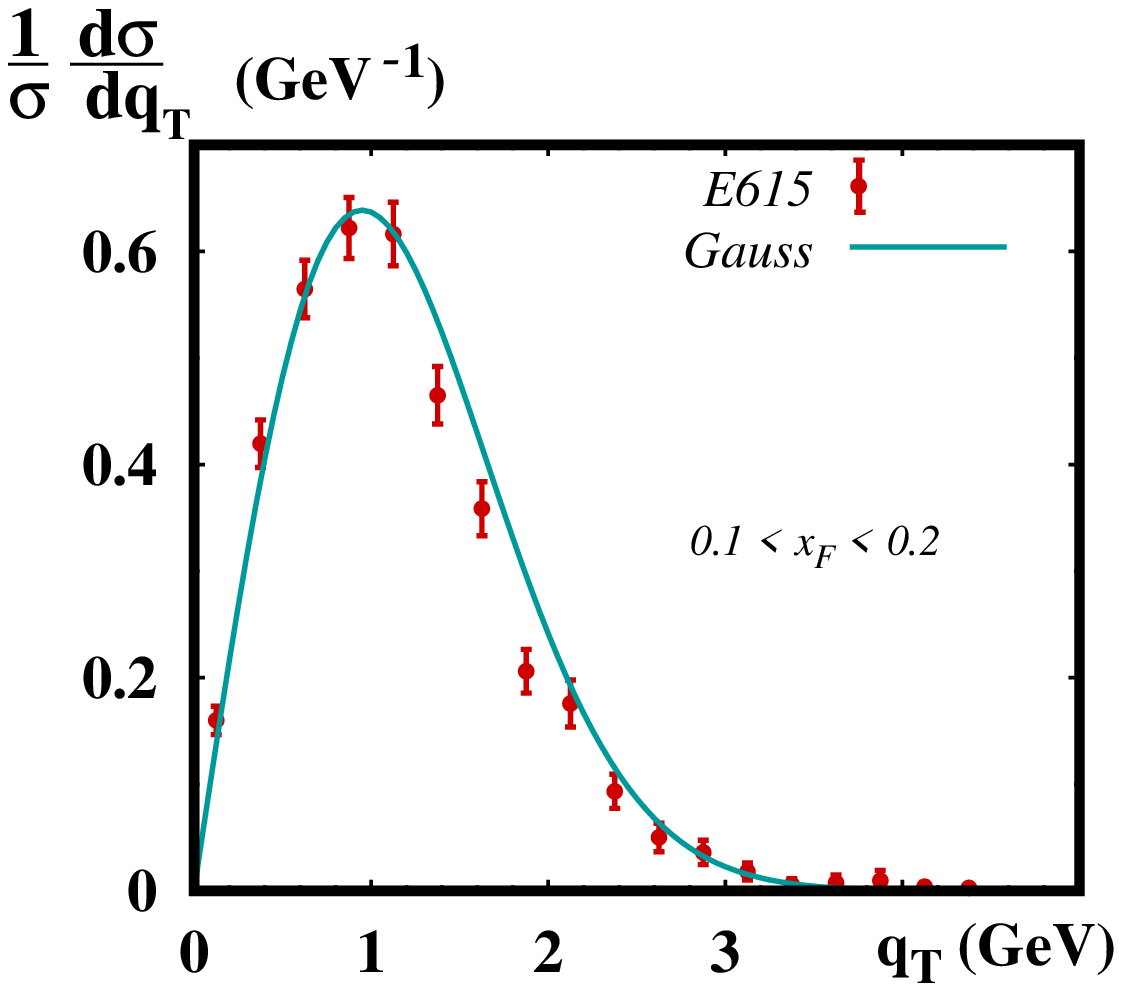,height=3.6cm} \ &
\epsfig{file=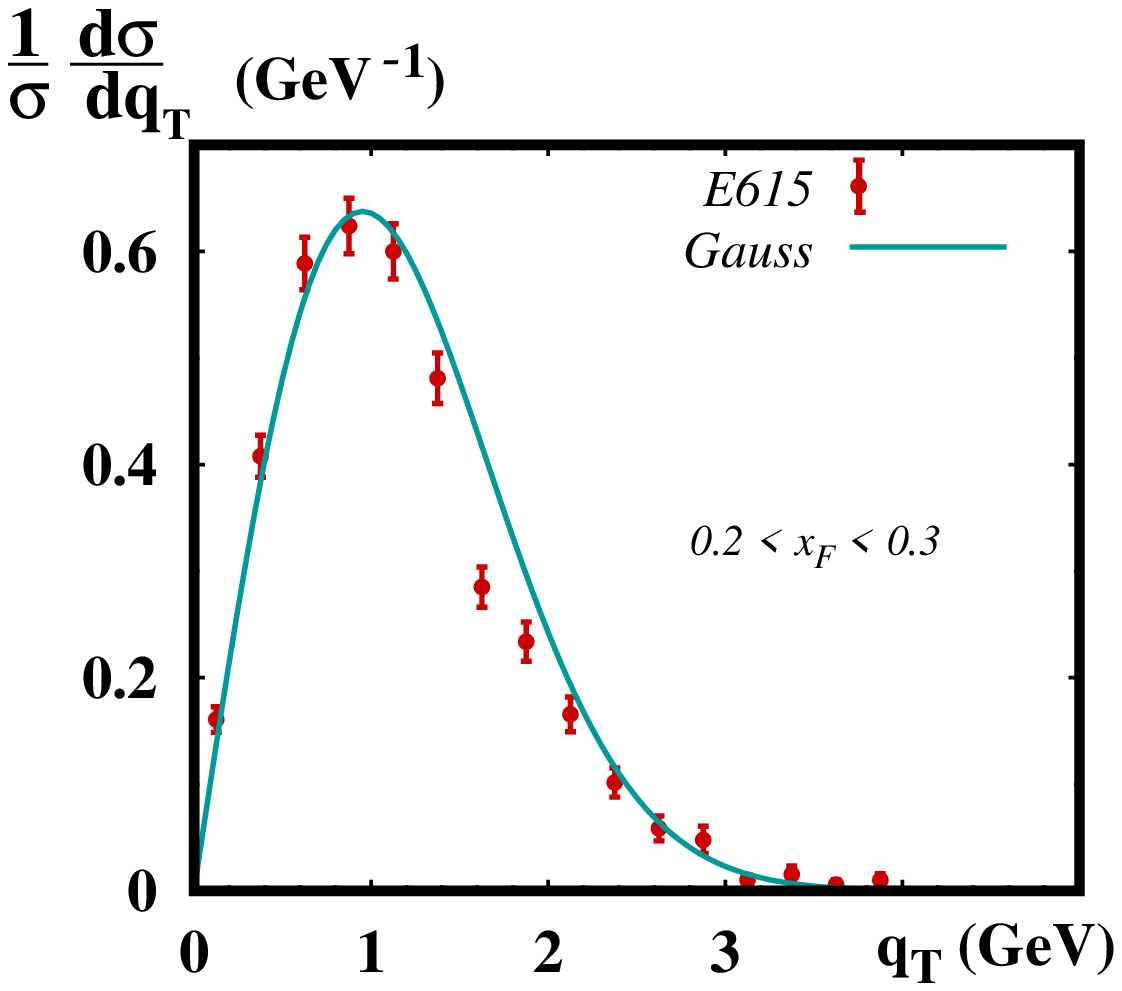,height=3.6cm} \ &
\epsfig{file=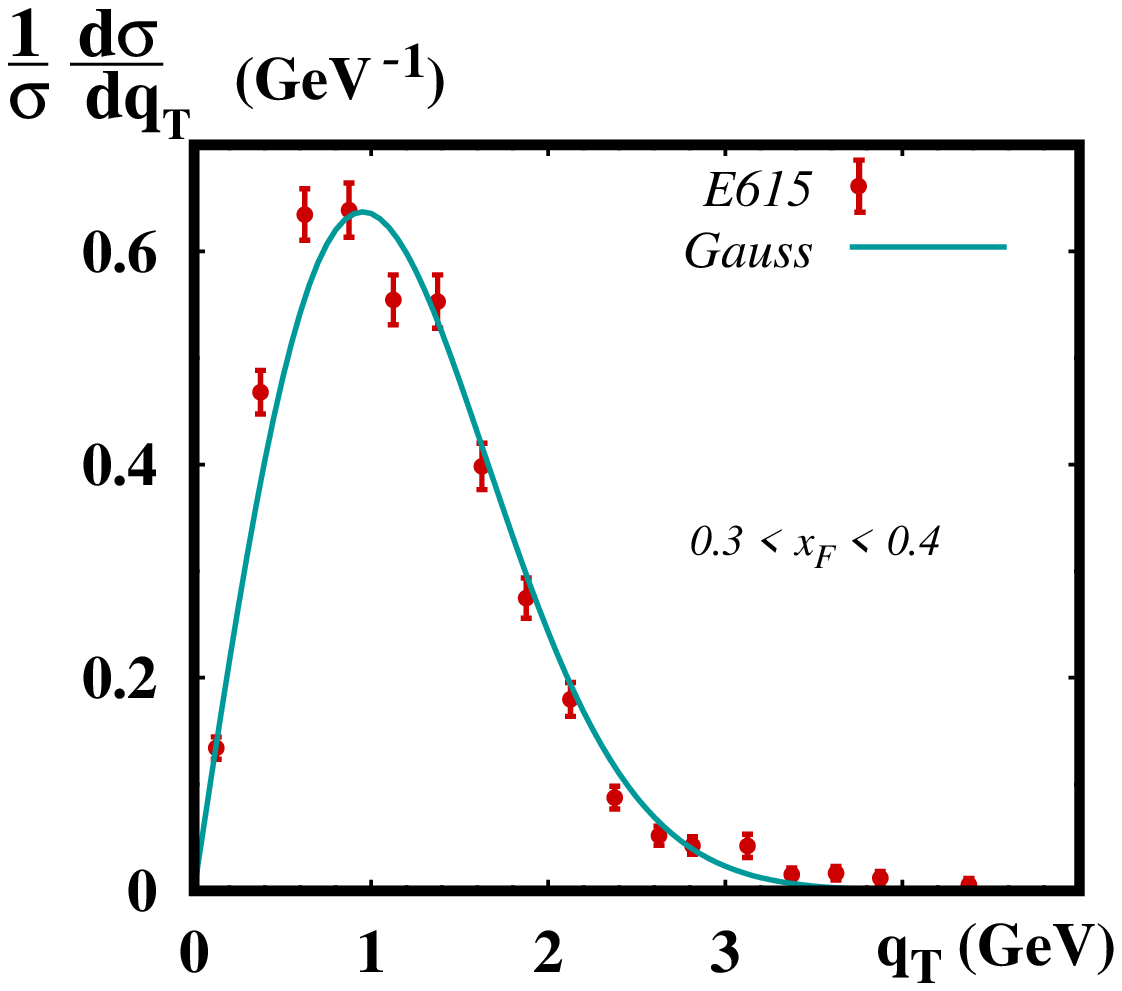,height=3.6cm} \ \\
\ \hspace{-9mm}
\epsfig{file=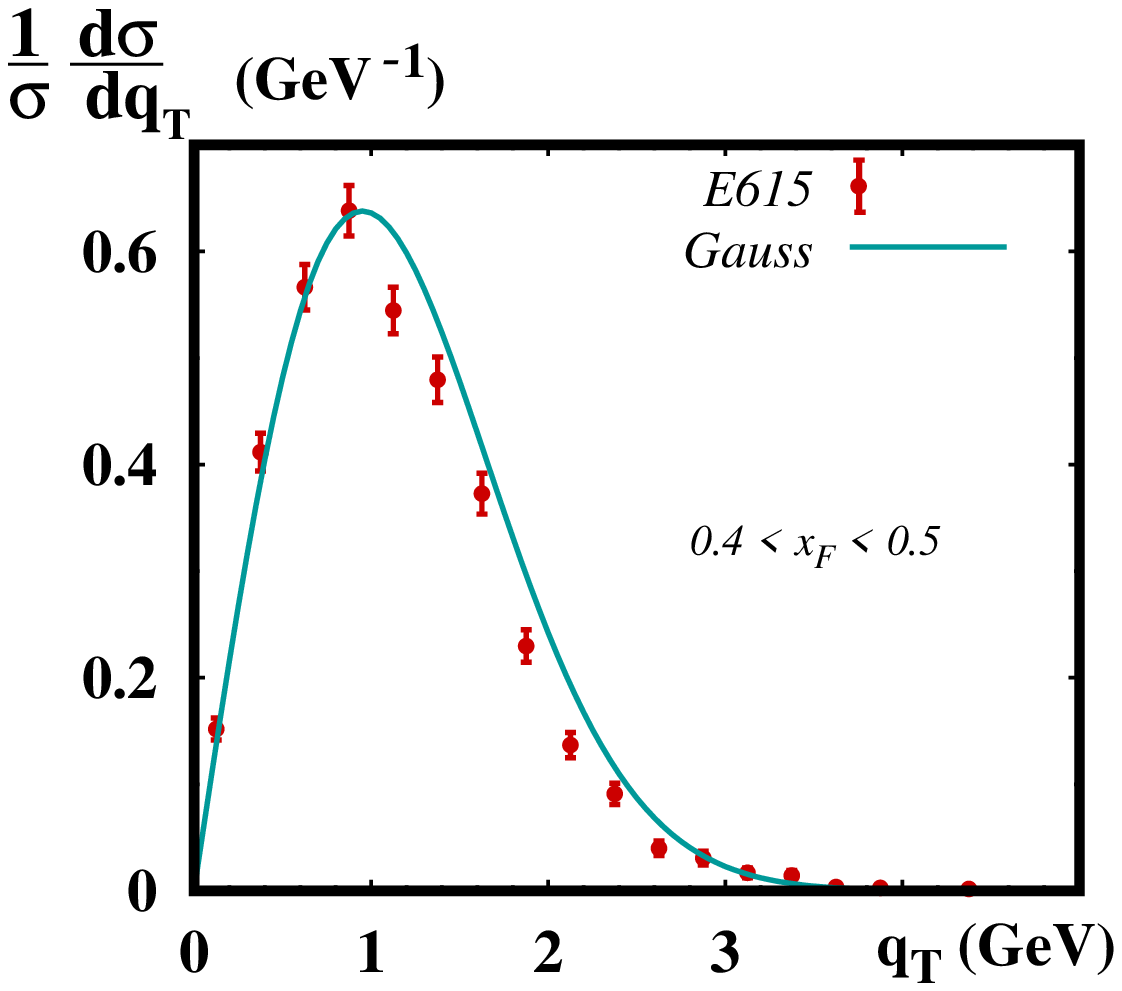,height=3.6cm} \ &
\epsfig{file=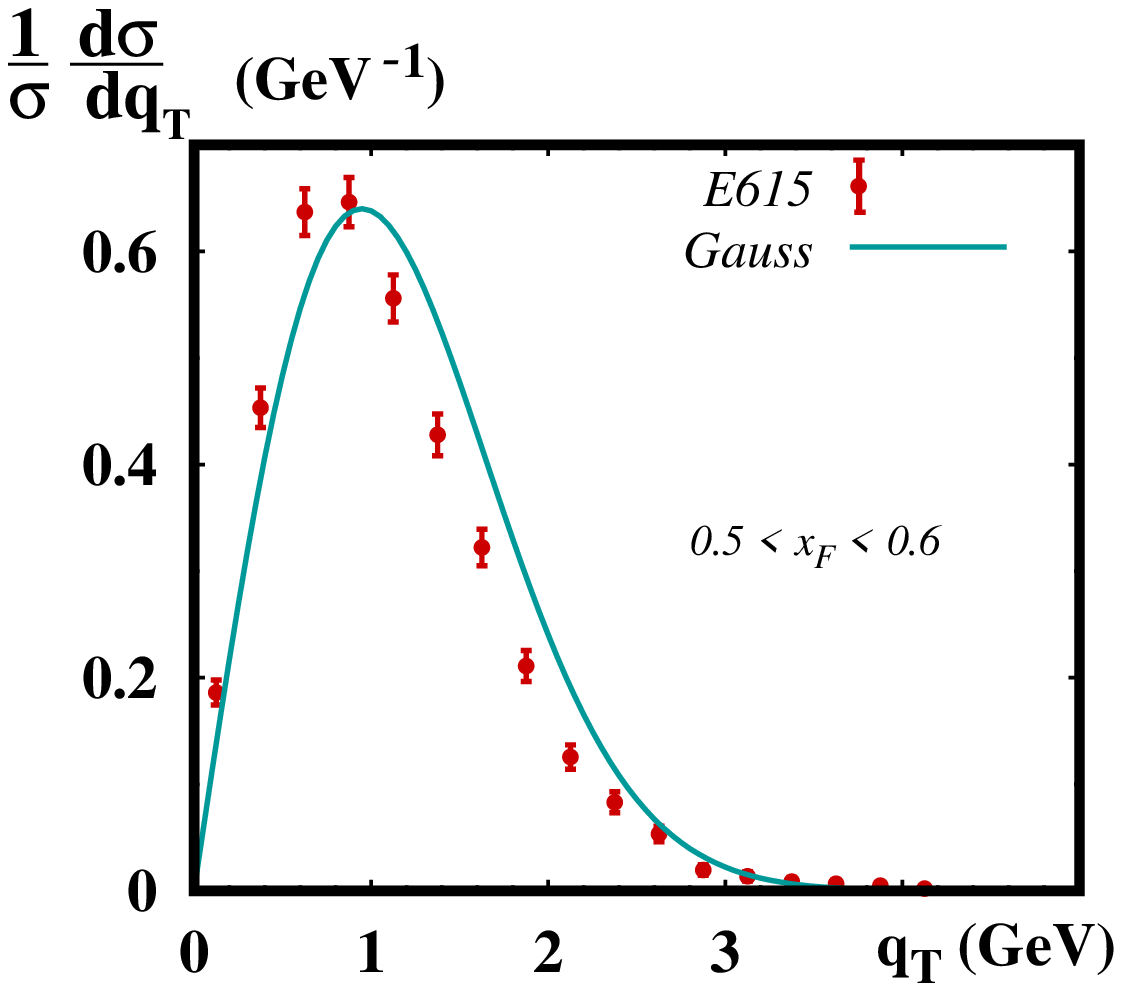,height=3.6cm} \ &
\epsfig{file=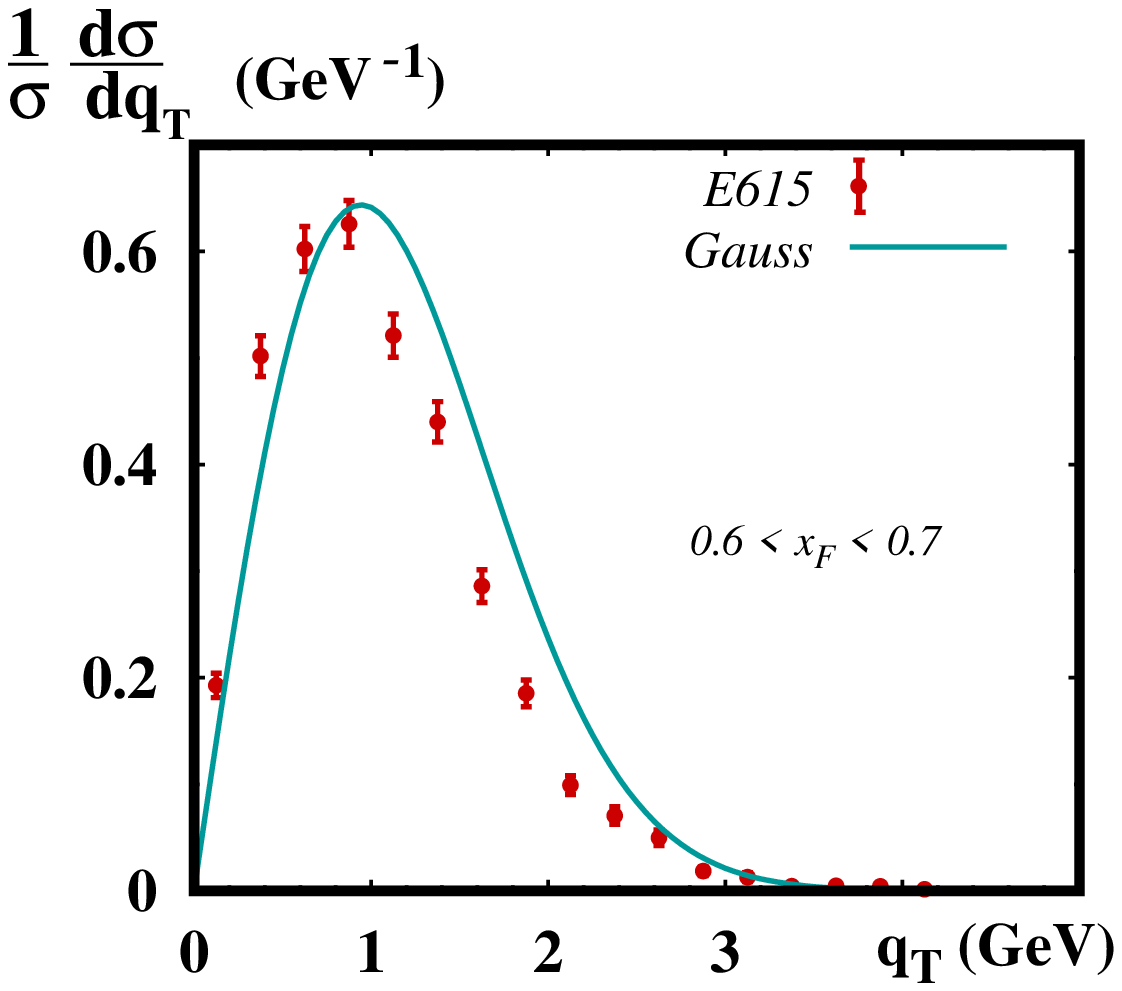,height=3.6cm} \ &
\epsfig{file=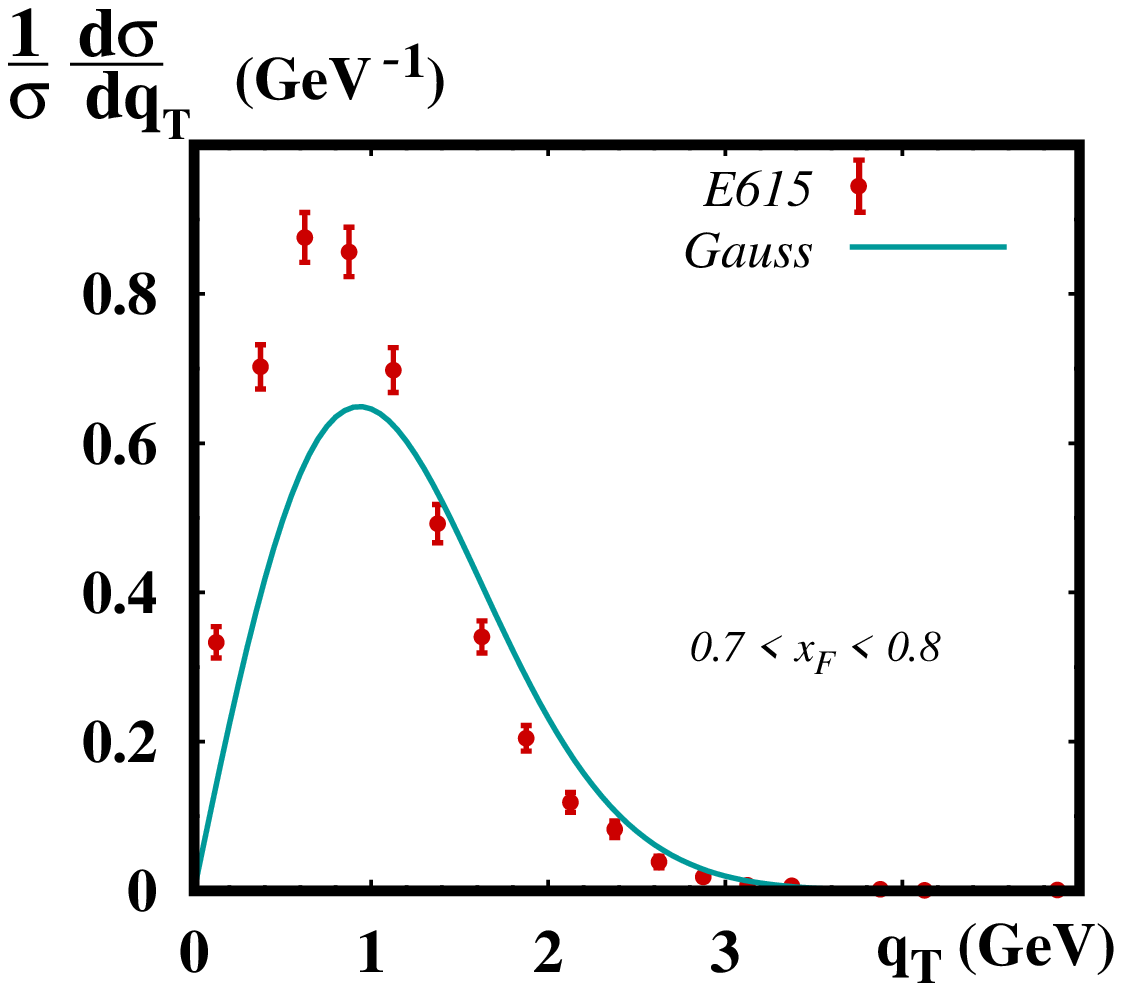,height=3.6cm} \ 
\end{tabular}  \end{center} 	
\caption{
  \label{Fig-8:compare-to-Conway-norm-sigma}
  The normalized cross section $(1/\sigma)\,(\di\sigma(q_T)/\di q_T)$
  as functions of $q_T$ in different $x_F$-bins.
  The data are from Ref.~\cite{Conway:1989fs}.
  The theoretical curves are from the LFCM with transverse-momentum
  broadening effects estimated according to 
  Eqs.~(\ref{Eq:qT2-xF-s}) and ~(\ref{Eq:pT-broadening}).
  The description of the data is very good in the region $0\le x_F\le 0.5$,
  and it is still acceptable for $0.5\le x_F\lesssim 0.7$. For $x_F\gtrsim 0.7$
  the description breaks down, because the TMD approach is not applicable
  and higher twist effects become relevant.}
\end{figure}

The observable $\la q_T^2(x_F)\ra$ shown in Fig.~\ref{Fig-7:pT-in-DY} is 
the result of averaging over DY pair momenta. It is of importance to 
demonstrate that our approach works also for observables depending on $q_T$. 
For that we consider the data from the E615 experiment \cite{Conway:1989fs} 
shown in Fig.~\ref{Fig-8:compare-to-Conway-norm-sigma}
on the normalized cross sections, which we define for brevity as
\be\label{Eq:norm-cross-section}
     \frac{1}{\sigma}\,\frac{\di\sigma(q_T)}{\di q_T} \equiv
     \frac{\di^2\sigma(q_T,x_F)}{\di q_T \,\di x_F} \Biggl/ 
     \frac{\di\sigma(x_F)}{\di x_F} =
     \frac{2\pi\,q_T\,\la F_{UU}^1(x_1,x_2,q_T)\ra }{\la F_{UU}^1(x_1,x_2)\ra },
\ee
where  $\la\,\cdots\,\ra$ denote averages over $x_F$ in certain bins,
and $\sigma$  in the first term of Eq.~(\ref{Eq:norm-cross-section}) is a 
short-cut notation for the differential cross section $\di\sigma/\di x_F$.
The normalization is such that one obtains unity after 
integrating over $q_T$ in Eq.~(\ref{Eq:norm-cross-section}). 
Using the Gaussian Ansatz, the structure functions are given by
\ba\label{Eq:FUU1-xF-qT}
   F_{UU}^1(x_1,x_2,q_T)
   &=& \frac{1}{N_c}\,\sum_a e_a^2 f_{1,\pi}^{     a}(x_1)\,f_{1,N  }^{\bar a}(x_2)\,
   \frac{\exp(-{q_T^2}/\la q_T^2\ra)}{\pi\la q_T^2\ra}
   \;,\\
   F_{UU}^1(x_1,x_2) 
   &=& \frac{1}{N_c}\,\sum_a e_a^2 f_{1,\pi}^a(x_1)\,f_{1,N}^{\bar a}(x_2)\;.
   \label{Eq:FUU1-xF}
\ea
Notice that  $F_{UU}^1(x_1,x_2,q_T)$ in Eq.~(\ref{Eq:FUU1-xF-qT}) 
depends on the Gaussian model, but after integrating out transverse
momenta one obtains the model-independent structure function
$F_{UU}^1(x_1,x_2)=\int\di^2q_T\,F_{UU}^1(x_1,x_2,q_T)$ in Eq.~(\ref{Eq:FUU1-xF}).
The data refer to $4.05 \le Q/{\rm GeV}\le 8.55$ and were taken
with a 252$\,$GeV $\pi^-$ beam impinging on a nuclear (tungsten) target
\cite{Conway:1989fs}. Thus $s\simeq 473\,{\rm GeV}^2$ in this experiment.
Strictly speaking we could only retrieve E615 data on 
$\di^2\sigma/(\di q_T\,\di x_F)$ from Ref.~\cite{Stirling:1993gc}, 
and estimated the differential cross sections $\di\sigma/\di x_F$ 
ourselves, to obtain the normalized data in
Fig.~\ref{Fig-8:compare-to-Conway-norm-sigma}. We are confident 
that the data shown in Fig.~\ref{Fig-8:compare-to-Conway-norm-sigma} 
are normalized with an accuracy of $10\,\%$, which is comparable 
or better than the accuracy of the~LFCM. 
(We recall that we work in a LO approach. Thus, we could have 
alternatively studied the $q_T$ dependence of the differential 
cross sections $\di^2\sigma/(\di q_T\,\di x_F)$  fixing the overall 
normalizations ``by hand,''  or estimating ``K-factors.'' 
Both alternatives are not more rigorous than our treatment.)

\begin{figure}[b!]

\vspace{-0.5cm}

\begin{center}
\epsfig{file=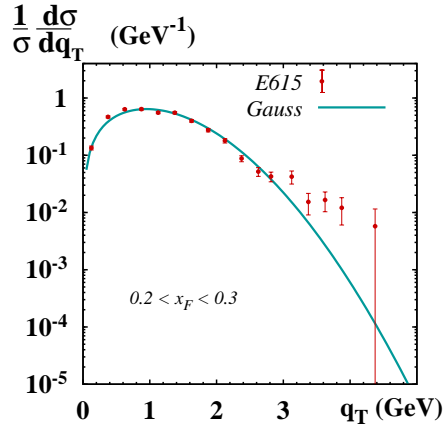,height=6cm}
\end{center}

\vspace{-0.5cm}

\caption{
  \label{Fig-9:compare-to-Conway-norm-sigma-log}
  The normalized cross section $(1/\sigma)\,(\di\sigma(q_T)/\di q_T)$
  as function of $q_T$ for $0.2\le x_F\le 0.3$.
  The data are from Ref.~\cite{Conway:1989fs}.
  The theoretical curves are from the LFCM with transverse-momentum 
  broadening effects estimated according to 
  Eqs.~(\ref{Eq:qT2-xF-s},~\ref{Eq:pT-broadening}).
  The logarithmic plot shows the limitation of the Gaussian 
  description which works well for $q_T\lesssim(2\mbox{--}3)\,{\rm GeV}$.}
\end{figure}

Fig.~\ref{Fig-8:compare-to-Conway-norm-sigma} shows that the
description of the $q_T$ dependence of the normalized cross
sections works very well in the region $0\le x_F\le 0.5$, 
is still reasonably good for $0.5\le x_F\lesssim 0.7$, but 
for $x_F\gtrsim 0.7$ it clearly breaks down, which is not
surprising given our earlier findings concluded from 
Fig.~\ref{Fig-7:pT-in-DY} and the expectations from 
QCD for $x_F\to 1$ \cite{Berger:1979du}.
It is important to stress that we do not only expect limitations
of the approach at large $x_F$, but in particular also at large
$q_T$, where the Gaussian Ansatz is at variance with QCD which
predicts a power-like decay \cite{Bacchetta:2008xw}.
These limitations cannot be seen in 
Fig.~\ref{Fig-8:compare-to-Conway-norm-sigma}. We therefore 
present a logarithmic plot of the E615 data \cite{Conway:1989fs}
on the normalized cross section in 
Fig.~\ref{Fig-9:compare-to-Conway-norm-sigma-log}
which demonstrates that the Gaussian description is applicable 
for $q_T\lesssim(2\mbox{--}3)\,{\rm GeV}$ but not beyond that.
Since $4.05 \le Q/{\rm GeV}\le 8.55$ and we need $q_T\ll Q$ for 
the TMD factorization to be applicable, one certainly cannot expect 
the approach to work beyond $q_T\lesssim(2\mbox{--}3)\,{\rm GeV}$. 
In Fig.~\ref{Fig-9:compare-to-Conway-norm-sigma-log} we limit
ourselves to showing the data in the bin $0.2\le x_F\le 0.3$ only,
because this $x_F$-bin shows the limitations of the $q_T$-description 
most clearly. 
The data sets from \cite{Conway:1989fs} in the other $x_F$-bins 
shown in the Fig.~\ref{Fig-8:compare-to-Conway-norm-sigma}
happen to be less accurate at larger $q_T$ and show deviations from
the Gaussian  Ansatz less clearly. Depending on the energy, 
the Gaussian model was shown to work satisfactorily in DY up to 
$q_T\lesssim(2\mbox{--}3)\,{\rm GeV}$ also in \cite{Schweitzer:2010tt}.

At this point it is worth recalling that we neglect nuclear 
binding effects, which is justified for $q_T\lesssim 3\,{\rm GeV}$ 
\cite{Guanziroli:1987rp,Bordalo:1987cs}. Thus, nuclear effects
become important only beyond the range of $q_T$ we are interested in.
Moreover, since in the LFCM the $\pTsqx{x}{unp}{}$ are equal for 
$u$- and $d$-quarks in protons and neutrons, we do not need to
distinguish protons and neutrons in the tungsten target.

To summarize, a parton model description of cross section ratios 
in DY with the LFCM predictions for pion and nucleon unpolarized
TMDs with the phenomenological estimate of transverse-momentum
broadening effects in Eqs.~(\ref{Eq:qT2-xF-s}) and ~(\ref{Eq:pT-broadening})
works well for $s\sim (50\mbox{--}600)\,{\rm GeV}^2$ in the regions of 
$q_T\lesssim(2\mbox{--}3)\,{\rm GeV}$ and $x_F\lesssim(0.7\mbox{--}0.8)$.
Although the LFCM has its own limitations, this is the range of 
applicability of the TMD approach expected on general grounds,
and we shall keep it in mind when embarking on the description 
of the Boer-Mulders effect in DY in the next section.

\section{Boer-Mulders effect in DY}
\label{Sec-8:BM-in-DY}

In this section we describe the Boer-Mulders effect in the DY process.
The treatment is in large part parallel to the discussion of the
unpolarized TMDs in Sec.~\ref{Sec-7:f1-in-DY}.

\subsection{\boldmath 
  Gaussian approximation and estimate of $k_\perp$ broadening for 
  $h^{\perp a}_{1,h}(x,\boldsymbol{k}^{2}_\perp)$}
  \label{subsec-6A:estimate-broadening-BM}

In analogy to the unpolarized TMDs in Eq.~(\ref{Eq:Gauss-Ansatz-BM}),
also in the case of the Boer-Mulders functions it is convenient to 
recast the model predictions in terms of a Gaussian Ansatz as follows
\ba
\label{Eq:Gauss-Ansatz-BM}
     h_{1,h}^{\perp a}(x,\boldsymbol{k}^{2}_\perp) = h_{1,h}^{\perp a}(x)\,
     \frac{\exp(-k_\perp^2/\pTsqx{x}{BM}{a/h})}{\pi\,\pTsqx{x}{BM}{a/h}}\;,
     &&
     \pTsqx{x}{BM}{a/h} 
     = \frac{\int\di^2\boldsymbol k_\perp\;\boldsymbol k_\perp^{2}\,h_{1,h}^{\perp a}(x,\boldsymbol{k}^{2}_\perp)}
            {\int\di^2\boldsymbol k_\perp\;       h_{1,h}^{\perp a}(x,\boldsymbol{k}^{2}_\perp)}\;.
\ea
The model results for $h_{1,h}^{\perp a}(x,\boldsymbol{k}^{2}_\perp)$ at the initial 
scale exhibit an approximate Gaussian $k_\perp$-behavior, see
Fig.~\ref{Fig-4:tmd-bm-x-ksq}b in this work for pion and
\cite{Boffi:2009sh} for nucleon. This is well approximated by 
Eq.~(\ref{Eq:Gauss-Ansatz-BM}) thanks to the flexible $x$-dependent 
Gaussian width. Moreover, also in the case of the Boer-Mulders 
functions the Gaussian Ansatz will facilitate the estimate of 
transverse-momentum broadening effects, as described below.

We remark that $h_{1,h}^{\perp a}(x)= \int\di^2\boldsymbol k_\perp\,h_{1,h}^{\perp a}(x,\boldsymbol{k}^{2}_\perp)$, 
though well-defined in models, would have an involved QCD definition
because one should ``divide out'' a power of transverse momentum from 
the correlator in Eq.~(\ref{h1}). However, this quantity appears here 
merely as an ``intermediate-step construct'' and will be eliminated in 
favor of the (1)-moment of the Boer-Mulders function in the final expression.
We remark that treatments of the Boer-Mulders effect in DY 
in the Gaussian Ansatz were reported e.g.\ in 
Refs.~\cite{Zhang:2008nu,Barone:2010gk}, though from our point of 
view the used Gaussian widths were sometimes chosen unacceptably small.

Using the Gaussian Ansatz (\ref{Eq:Gauss-Ansatz-BM}), one can analytically 
evaluate the convolution integral in the structure function 
(\ref{Eq:FUUcos2phi}). There are 
``infinitely many'' possible ways to express the result. We choose to 
write it in terms of (1)-moments of the Boer-Mulders function as follows
\ba
  F_{UU}^{\cos(2\phi)}(x_1,x_2,q_T)
  &=& \frac{1}{N_c}\,\sum_a e_a^2 
      h_{1,\pi}^{\perp(1)     a}(x_1)_{\rm DY}\,
      h_{1,N  }^{\perp(1)\bar a}(x_2)_{\rm DY}\;
      \frac{4M_\pi M_N}{\la q_T^2\ra_{\rm BM}^{}}\;\,
      \frac{q_T^2\,\exp\big(-{q_T^2}/\la q_T^2\ra_{\rm BM}^{})}
           {\pi\la q_T^2\ra^2_{\rm BM}} \;,\label{Eq:FUUcos2phi-xF-qT}\\
  F_{UU}^{\cos(2\phi)}(x_1,x_2) 
  &=& \frac{1}{N_c}\,\sum_a e_a^2 
      h_{1,\pi}^{\perp(1)     a}(x_1)_{\rm DY}\,
      h_{1,N  }^{\perp(1)\bar a}(x_2)_{\rm DY}\;
      \frac{4M_\pi M_N}{\la q_T^2\ra_{\rm BM}^{}}\;,\label{Eq:FUUcos2phi-xF}\\
  \la q_T^2(x_1,x_2,s)\ra_{\rm BM }
  &=& \pTsqx{x_1}{BM }{a/\pi}+\pTsqx{x_2}{BM }{\bar a/N}
      + \la\delta k_{\perp,\rm BM}^2(s)\ra \;.
\ea
One could also use $h_{1,h}^{\perp a}(x)$ or 
$h_{1,h}^{\perp(1/2)a}(x)$, or any other moment $h_{1,h}^{\perp(n)a}(x)$ 
defined analogously to Eq.~(\ref{eq:moments-bm}),
in order to express the structure functions in 
Eqs.~(\ref{Eq:FUUcos2phi-xF-qT}) and~(\ref{Eq:FUUcos2phi-xF}).
From the point of view of the Gaussian model, all such expressions
would be equally acceptable. From phenomenological point of view,
our choice in Eqs. (\ref{Eq:FUUcos2phi-xF-qT}) and (\ref{Eq:FUUcos2phi-xF}) 
is preferred in the sense that this is the only case, where one deals with 
a single parameter, $\la\delta k_{\perp,\rm BM}^2(s)\ra$, describing the accumulated 
$k_\perp$ broadening of the pion and nucleon Boer-Mulders functions. All other 
choices would require to explicitly estimate the $k_\perp$ broadenings of 
the separate pion and nucleon Gaussian widths $\pTsqx{x}{BM }{a/h}$.

We find a good description of data \cite{Guanziroli:1987rp,Conway:1989fs}
on the $q_T$ dependence of the Boer-Mulders effect in DY with
\be\label{Eq:BM-broadening}
    \la\delta k_{\perp,\rm BM}^2(s)\ra = 1.3\,{\rm GeV}^2 \;\;\; \mbox{at} \;\;\;
    s \approx \mbox{(470--540)\,GeV$^2$}
\ee
in the range of $q_T$ up to (2--3) GeV in which the Gaussian Ansatz 
was shown to be applicable for unpolarized TMDs in 
Sec.~\ref{subsec-7B:unpol-DY-data}. DY data on the  Boer-Mulders 
effect are available also for smaller center-of-mass energies $s$ 
\cite{Badier:1981ti,Palestini:1985zc,Falciano:1986wk,Guanziroli:1987rp,
Conway:1989fs}. But we observe that we cannot describe these data 
using  Eqs.~(\ref{Eq:FUUcos2phi-xF-qT}) and (\ref{Eq:FUUcos2phi-xF}). 
More precisely, descriptions of the data at smaller $s$ are possible, 
but in a more limited range $q_T\lesssim 1$ GeV. We also found 
that different prescriptions to describe the structure function, say 
in terms of $h_{1,h}^{\perp a}(x)$ or $h_{1,h}^{\perp(1/2)a}(x)$, do not yield 
better descriptions.

These observations should not come as a surprise. None of such Gaussian 
Ansatz descriptions can be expected to adequately describe the true 
QCD scale dependence of the Boer-Mulders functions. However, as we 
will show in the next section, the Gaussian Ansatz 
is useful in a specific range of $s$ and $q_T$ with the 
understanding that $q_T\ll Q$. Only after the full CSS-evolution for 
the the Boer-Mulders functions will be available, it will be possible 
to undertake an attempt to describe Boer-Mulders data at all energies. 
Furthermore, 
it is important to compare the value of $\la\delta k_{\perp,\rm BM}^2(s)\ra$
in Eq.~(\ref{Eq:BM-broadening}) with the broadening
$\la\delta k_{\perp,\rm unp}^2(s)\ra = $ (1.6--1.8) GeV$^2$ of unpolarized TMDs 
in the same range of $s$.
The accumulated $k_\perp$ broadening of the unpolarized TMDs is 
larger than that of the Boer-Mulders functions. This is a necessary
(cf.\ Footnote~\ref{Footnote-1}) and, in our case, numerically also 
sufficient condition to comply with positivity.

\subsection{Comparison to data}

With the descriptions of the unpolarized structure function in 
Eqs.~(\ref{Eq:qT2-xF-s}) and (\ref{Eq:FUU1-xF}) and the Boer-Mulders 
structure function in 
Eqs.~(\ref{Eq:FUUcos2phi-xF-qT})--(\ref{Eq:BM-broadening})
we are now in the position to evaluate the coefficient $\nu$ in the 
angular distribution of the DY cross section in the Collins-Soper 
frame as defined through Eqs.~(\ref{Eq:DY-unp-angular-dependence-1}) 
and (\ref{Eq:DY-unp-angular-dependence-2}).

We will compare to the data from the NA10 CERN experiment 
\cite{Guanziroli:1987rp} and the E615 Fermi Lab experiment 
\cite{Conway:1989fs}. In both experiments secondary $\pi^-$ 
beams were collided with nuclear targets. 
In the NA10 experiment \cite{Guanziroli:1987rp} several beam energies
were used. We will focus on the NA10 data taken with 286 GeV $\pi^-$ 
beams impinging on a tungsten or deuterium targets. The covered range 
of $Q$ was $4.0<Q<8.5\,{\rm GeV}$ and $Q > 11 \,{\rm GeV}$ to remove 
the influence of the $J/\psi$- and $\Upsilon$-resonance regions. 
In order to discard the Berger-Brodsky higher twist effect 
\cite{Berger:1979du} the cut $x_1<0.7$ was imposed.
In the E615 Fermi Lab experiment \cite{Conway:1989fs} a 252 GeV $\pi^-$ 
beam was collided with a tungsten target, and the kinematic region
$4.05<Q<8.55\,{\rm GeV}$ between the $J/\psi$- and $\Upsilon$-resonances  
was covered with $0.2<x_1<1$.
For our theoretical calculation we assume for simplicity
$\la Q^2\ra = 25\,{\rm GeV}^2$ as typical hard scale in both experiments.

\begin{figure}[b!]
\vspace{-0.2cm}
\begin{center}
\epsfig{file=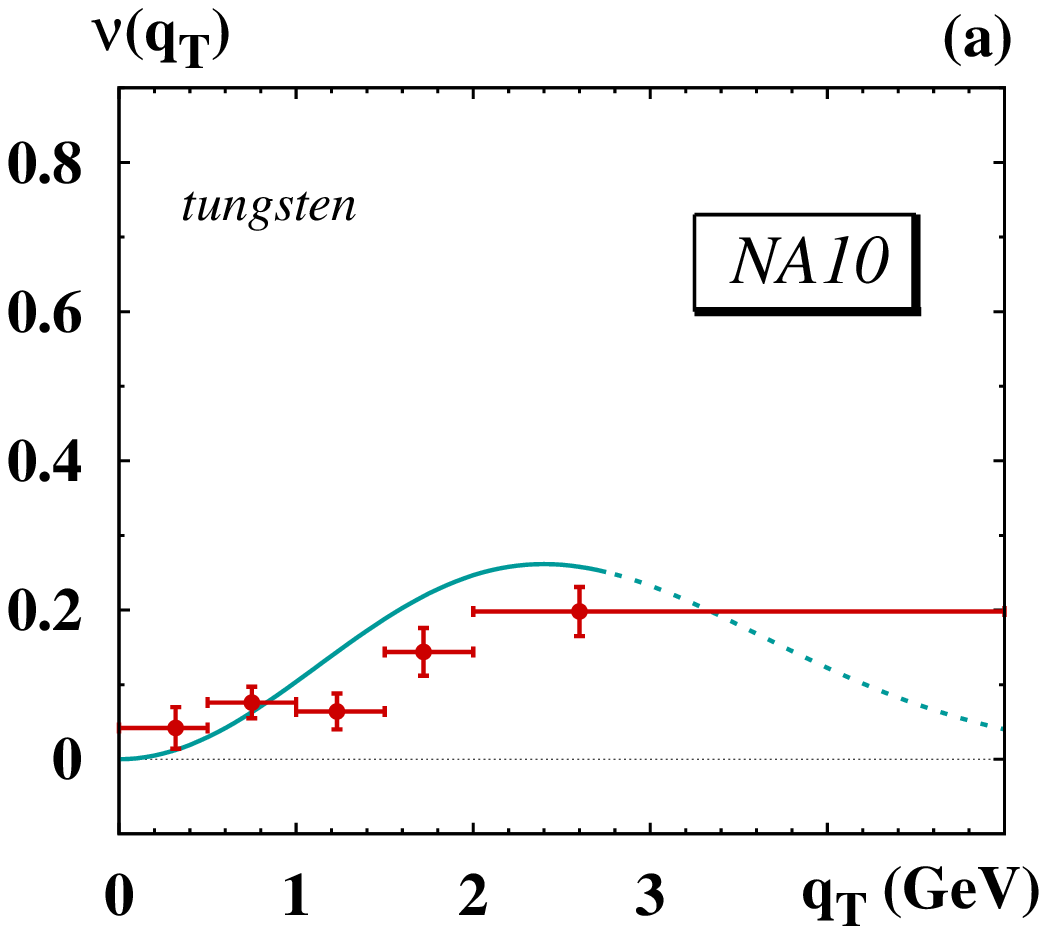,height=5cm} \
\epsfig{file=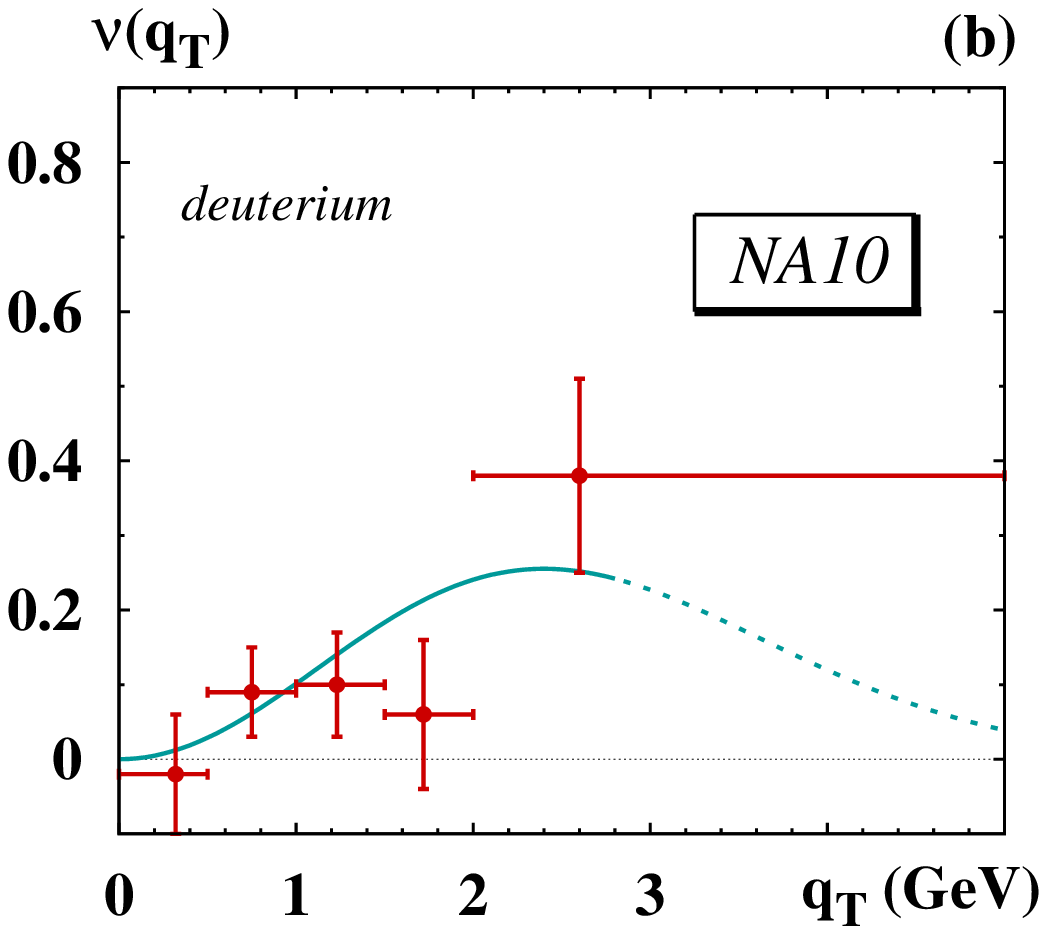,height=5cm} \
\epsfig{file=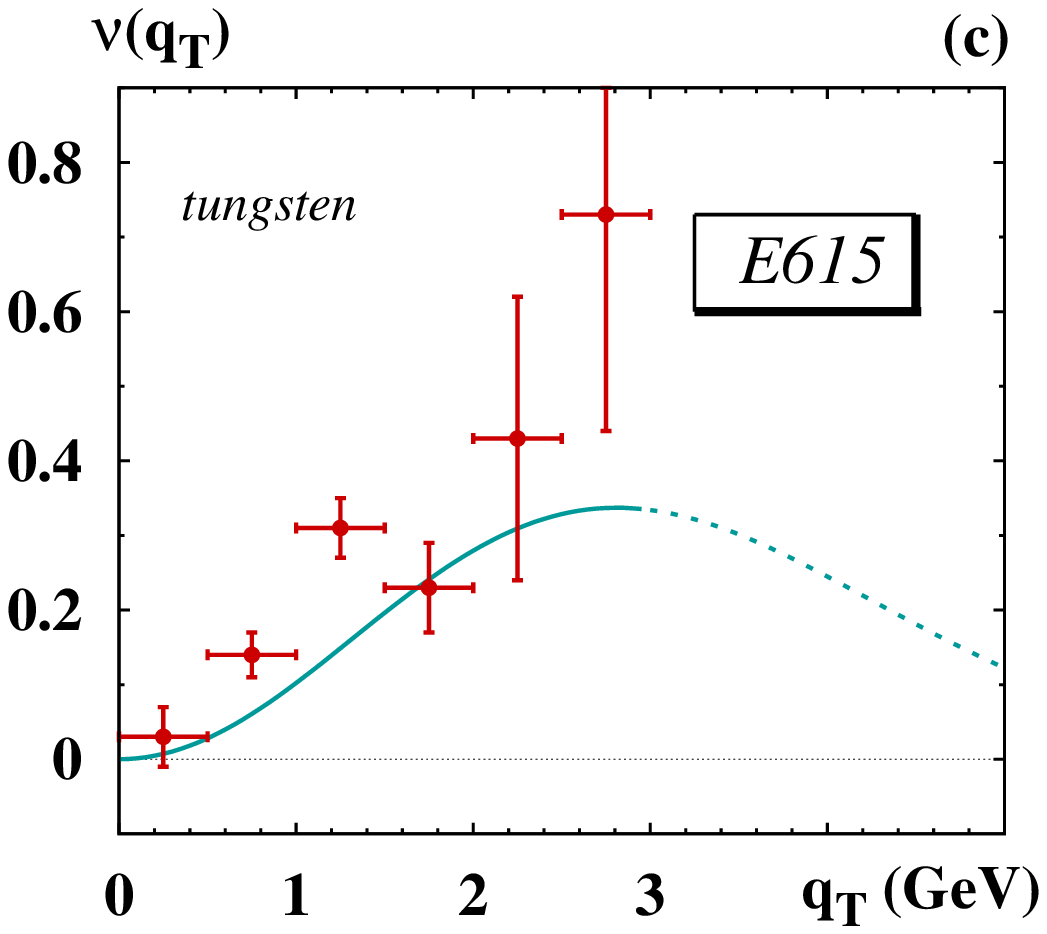,height=5cm}
\end{center}

\vspace{-0.6cm}

\caption{
  \label{Fig-10:nu-qT}
  The coefficient $\nu$ in the $\pi^-$-nucleus DY angular distribution 
  as function of $q_T$. The data are from the
  NA10 CERN experiment with $E_{\rm beam} = 286\,{\rm GeV}$ 
  using tungsten (a) and deuterium (b) targets \cite{Guanziroli:1987rp}, 
  and the E615 Fermi Lab experiment with $E_{\rm beam} = 252\,{\rm GeV}$ 
  using a tungsten target \cite{Conway:1989fs}. 
  The theoretical curves are obtained using the LFCM predictions
  for the pion Boer-Mulders function obtained here, and the analog
  nucleon predictions from \cite{Pasquini:2011tk}. The solid (dotted)
  lines indicate where the TMD approach is applicable (not applicable).}
\end{figure}

Let us first discuss the $q_T$ dependence of the coefficient $\nu$.
In the observable $\nu(q_T)$ the model input determines the overall
normalization, while the $q_T$ dependence is dictated by the Gaussian 
Ansatz with the estimated $k_\perp$ broadening of the Boer-Mulder functions
in Eqs.~(\ref{Eq:FUUcos2phi-xF-qT}) and (\ref{Eq:BM-broadening}).
In fact, more than testing the LFCM predictions, this comparison
shows
that the use of the Gaussian Ansatz for the Boer-Mulders
function with the estimated broadening (\ref{Eq:BM-broadening})
is compatible with data, as can be seen in Fig.~\ref{Fig-10:nu-qT}.

Several comments are in order. First, the NA10 tungsten data shown 
in Fig.~\ref{Fig-10:nu-qT}a have a 10-times larger statistics than
the NA10 deuterium data in Fig.~\ref{Fig-10:nu-qT}b. Within
the statistical uncertainty of the data, no significant nuclear 
dependence was observed \cite{Guanziroli:1987rp}. We exploited 
this observation when we defined our simplistic approach to 
estimate nuclear TMDs in Sec.~\ref{subsec-6A:DY-general+nuclear-effects}.
Second, there seems to be a tendency in our approach to slightly 
overestimate the tungsten data from NA10 in Fig.~\ref{Fig-10:nu-qT}a, 
and to slightly underestimate the tungsten data from E615 in 
Fig.~\ref{Fig-10:nu-qT}c.
The effect is not statistically significant. If it was, an explanation 
for that could be the fact that in the E615 data the Berger-Brodsky
effect was included ($x_1 < 1$) but not in the NA10 data ($x_1 < 0.7$).
Indications for the Berger-Brodsky effect were seen in the E615 
experiment \cite{Conway:1989fs}.
The slightly different energies in the two experiments could also
play a role. 
Third, in Sec.~\ref{subsec-7B:unpol-DY-data} we learned that a
Gaussian Ansatz for unpolarized TMDs works well in the region 
$q_T\lesssim\mbox{(2--3)}\,{\rm GeV}$, but breaks down beyond that. 
Our descriptions of $\nu(q_T)$ in Fig.~\ref{Fig-10:nu-qT} are
therefore certainly not valid for $q_T\gtrsim 3\,{\rm GeV}$ 
and we have emphasized this region with dotted lines.
Clearly, in the region $q_T\lesssim\mbox{(2--3)}\,{\rm GeV}$
(indicated by solid lines) our description of $\nu(q_T)$ is
compatible with data. 
Forth, it should be noted that the TMD approach in general 
requires $q_T\ll Q$. Thus, our results in Fig.~\ref{Fig-10:nu-qT} 
indicate that in the range $s\approx\mbox{(470--540)\,GeV$^2$}$
$\nu(q_T)$ can be well described in the TMD approach with
the Gaussian Ansatz. 
Finally, we remark that our results safely comply with the 
model-independent positivity bound $|\frac{\nu}{2}| \le 1$.

\begin{figure}[t!]
\vspace{-0.2cm}
\begin{center}
\epsfig{file=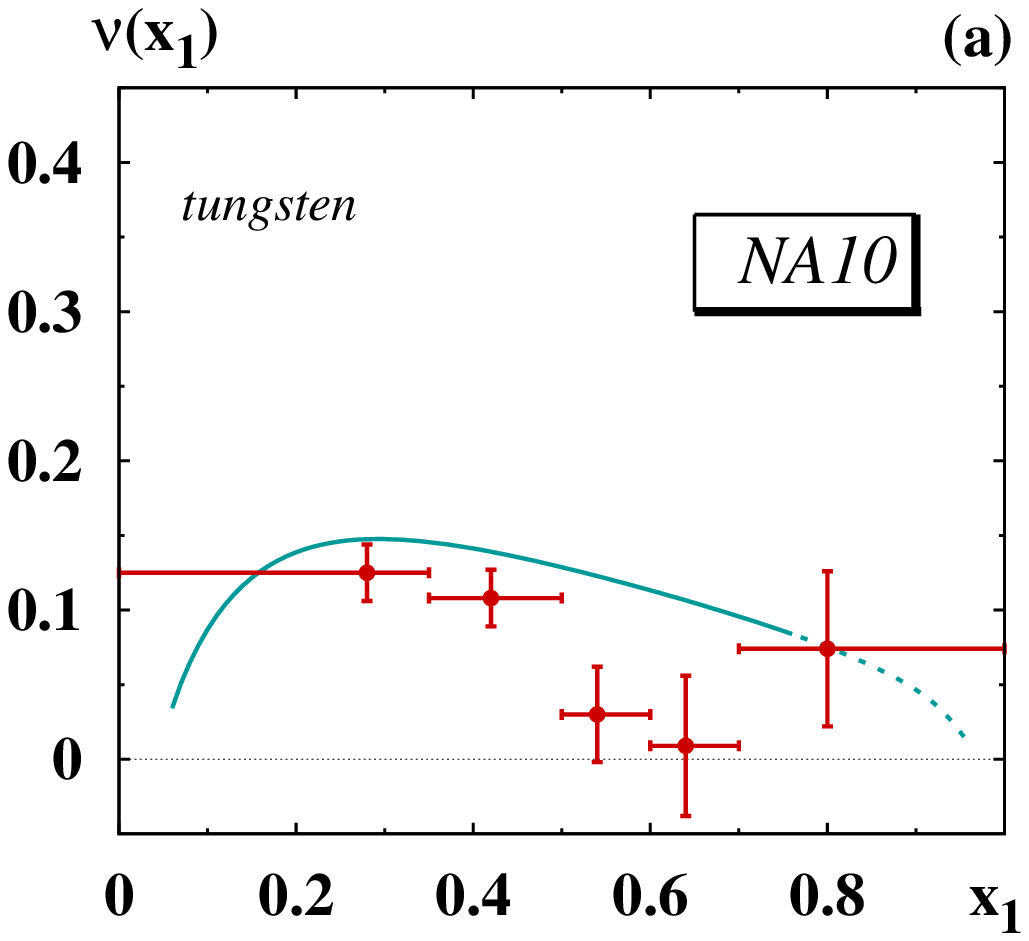,height=5cm} \
\epsfig{file=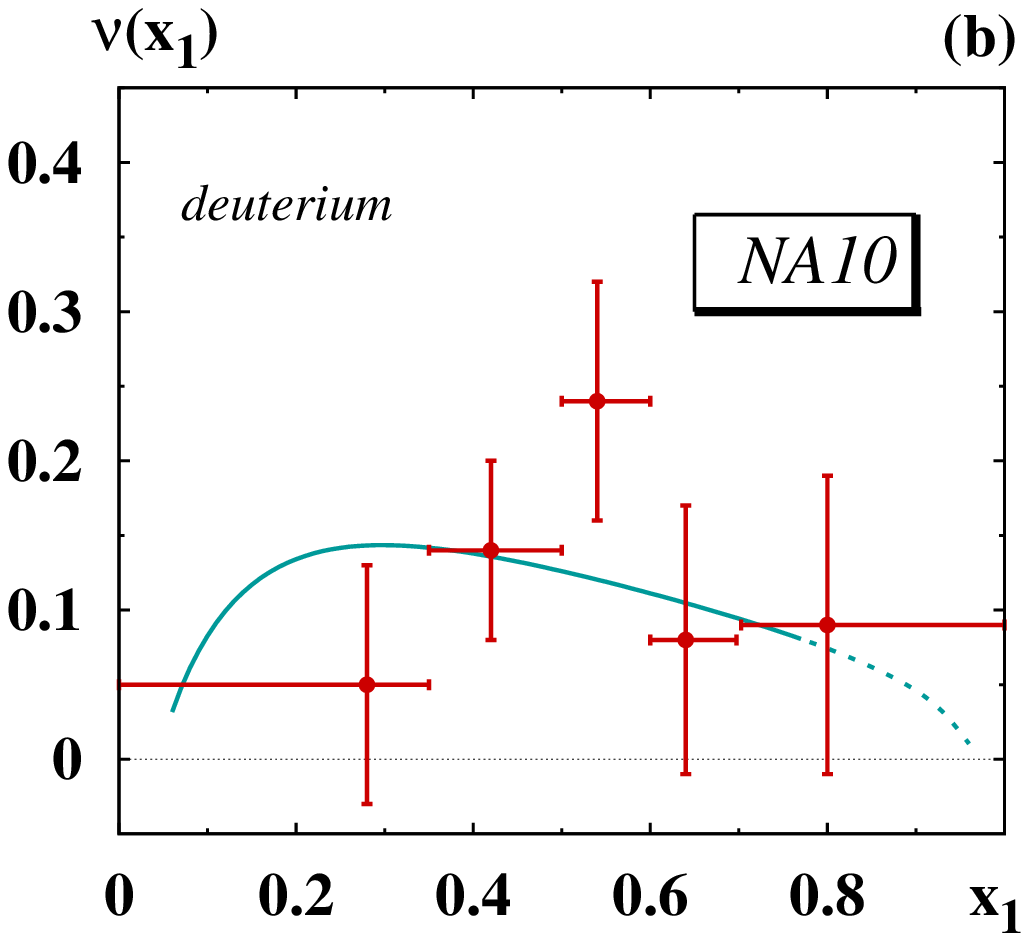,height=5cm} \
\epsfig{file=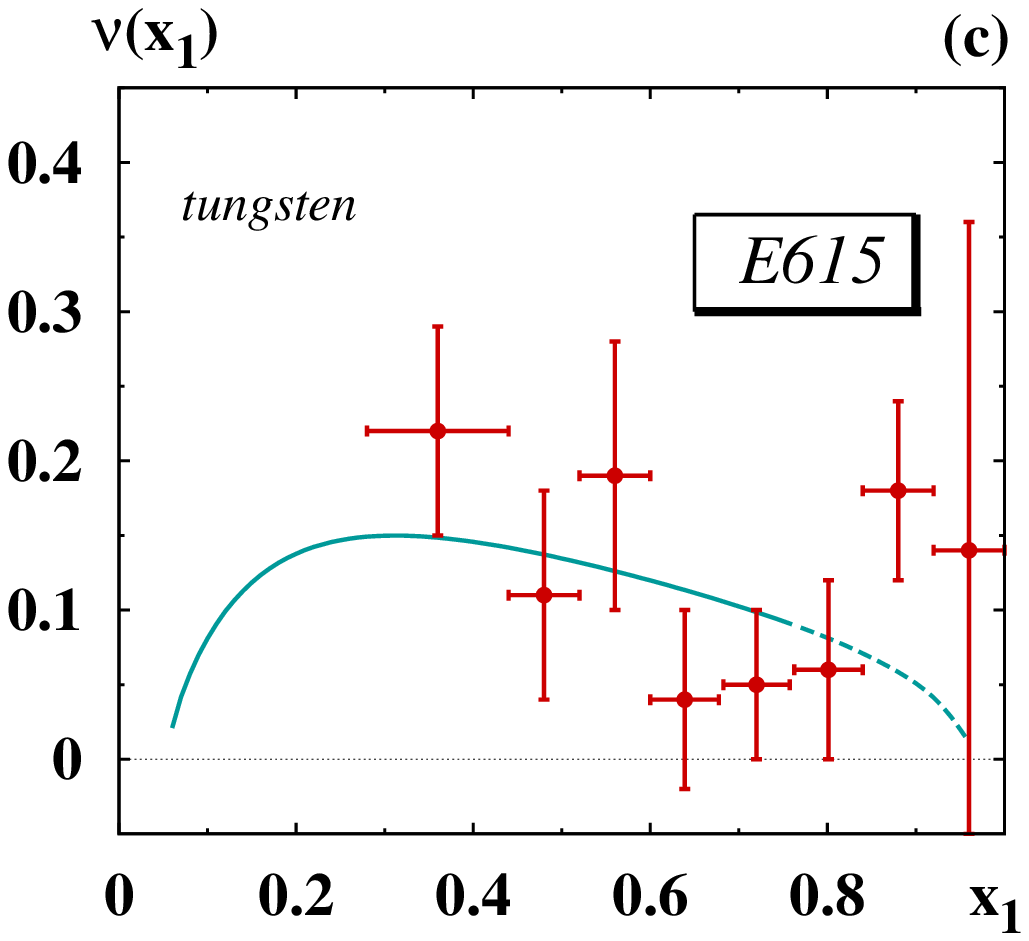,height=5cm}
\end{center}

\vspace{-0.6cm}

\caption{
  \label{Fig-11:nu-x1}
  The coefficient $\nu$ in the $\pi^-$-nucleus DY angular distribution 
  as function of $x_1$. The data are from the
  NA10 CERN experiment with $E_{\rm beam} = 286\,{\rm GeV}$ 
  using tungsten (a) and deuterium (b) targets \cite{Guanziroli:1987rp}, 
  and the E615 Fermi Lab experiment with $E_{\rm beam} = 252\,{\rm GeV}$ 
  using a tungsten target \cite{Conway:1989fs}. 
  The theoretical curves are obtained using the LFCM predictions
  for the pion Boer-Mulders function obtained here, and the corresponding
  nucleon predictions from \cite{Pasquini:2011tk}. The solid (dotted)
  lines indicate where the TMD approach is applicable (not applicable).}
\end{figure}

Next, we turn our attention to the $x_1$ dependence of 
the coefficient $\nu$ shown in Fig.~\ref{Fig-11:nu-x1}. 
We recall that $x_1$ corresponds to the momentum fraction 
carried by the parton which originates from the pion. 
We use this variable here, because it is the only common 
kinematical variable (besides $q_T$) used to analyze data 
in both experiments \cite{Guanziroli:1987rp,Conway:1989fs}.
The observable $\nu(x_1)$ provides a more stringent test of the model
results, in the sense that the shapes of the theoretical curves in 
Fig.~\ref{Fig-11:nu-x1} are directly dictated by the LFCM predictions,
although their overall normalizations are influenced through 
Eq.~(\ref{Eq:FUUcos2phi-xF}) by the choice of the parameter 
$\la\delta k_{\perp,\rm BM}^2(s)\ra$ in Eq.~(\ref{Eq:BM-broadening}).

The comparison with the data in  Fig.~\ref{Fig-11:nu-x1} is satisfactory. 
The most precise data set, namely the 
NA10 tungsten data in Fig.~\ref{Fig-11:nu-x1}a, may indicate that 
our model results somewhat overshoot the data in the region around 
$x_1\sim 0.6$, but the effect is not significant. 
Even if it was, one should recall that the typical accuracy of the 
LFCM in applications to TMD phenomenology is  (10--30)$\,\%$ 
\cite{Boffi:2009sh,Pasquini:2011tk}.
The NA10 deuterium data \cite{Guanziroli:1987rp} 
in Fig.~\ref{Fig-11:nu-x1}b and the E615 tungsten data 
\cite{Guanziroli:1987rp,Conway:1989fs} in Fig.~\ref{Fig-11:nu-x1}c 
have larger error bars, and our model results are compatible with 
them in the entire region of $x_1$.

It is important to keep in mind that the TMD approach is not 
applicable in the full range of $x_1$. 
In Sec.~\ref{subsec-7B:unpol-DY-data} we have seen that we can describe 
well the E615 data \cite{Conway:1989fs} on the (normalized) DY cross 
sections for $x_F\lesssim 0.7$, but not in the region $x_F\gtrsim 0.7$, 
where the Berger-Brodsky effect becomes increasingly significant.
In the kinematics of the NA10 and E615 experiments this $x_F$ region
corresponds to $x_1\gtrsim 0.76$, and we have indicated this region by dotted
lines in Fig.~\ref{Fig-11:nu-x1}. The Berger-Brodsky effect is not 
prominent in the NA10 data shown in Figs.~\ref{Fig-11:nu-x1}a and 
\ref{Fig-11:nu-x1}b. 
(Notice that the region of $x_1>0.7$ was excluded in the NA10 analysis of
$\nu$ as function of $q_T$ which we discussed in Fig.~\ref{Fig-10:nu-qT}.)
However, there is an indication of this effect in E615 data
shown in Fig.~\ref{Fig-11:nu-x1}c. 

To conclude, we observe that the predictions from the LFCM for the
pion Boer-Mulders functions, from this work, and nucleon,
from \cite{Pasquini:2011tk}, are in good agreement with the 
NA10 and E615 data taken at $s\approx\mbox{(470--540)\,GeV$^2$}$ 
\cite{Guanziroli:1987rp,Conway:1989fs}. The good agreement
is based also on our use of the Gaussian Ansatz in the TMD factorization
approach, and the chosen method to estimate $k_\perp$-broadening effects,
which corresponds to estimating CSS-evolution effects.

\section{Summary and Outlook}
\label{Sec-9:summary}

In this work we studied the structure of the pion, as described in terms
of the leading twist TMDs $f_{1,\pi}(x,\boldsymbol{k}^{2}_\perp)$ and
$h_{1,\pi}^{\perp }(x,\boldsymbol{k}^{2}_\perp)$,
using a light-front constituent model where the pion is described in
terms of the minimal Fock-state component consisting of a
quark and antiquark.
In a first step we determined the initial scale of this
constituent approach to the pion, following a similar procedure commonly
used in  hadronic models with effective valence degrees of freedom.
The resulting initial scale is about $\mu_0\sim 0.5\,{\rm GeV}$
and numerically similar to the initial scale in the case of the
of constituent approach of the nucleon \cite{Boffi:2009sh},
supporting the validity of the constituent approach.

The $q\bar q$ LFWF of the pion was shown to involve two
independent amplitudes describing the different orbital angular
momentum components of the constituent quark and antiquark in the
pion state \cite{Burkardt:2002uc,Ji:2003yj}.
In this work we derived a model-independent representation of
leading-twist pion TMDs in terms of overlaps of
light-front amplitudes which reveals
the role of the different
orbital angular momentum components for the structure of the pion.
We applied these expressions to a specific model, which
has been successfully employed to describe the pion electromagnetic
form factor~\cite{Schlumpf:1994bc,Chung:1988mu}.
Our predictions  for the pion TMDs are in qualitative agreement
with results from spectator and bag models
\cite{Lu:2004hu,Burkardt:2007xm,Gamberg:2009uk,Lu:2012hh}
and lattice QCD \cite{Engelhardt:2013nba}.
We then evolved the model result for the collinear valence
pion distribution function from the low hadronic scale to
experimentally relevant scales, and demonstrated that it is 
in good agreement with available parametrizations.
We observed that the $k_\perp$ dependence of the model TMDs
is not exactly Gaussian, but can be usefully approximated by
a Gaussian Ansatz. In comparison with the model results for the
nucleon Boer-Mulders function \cite{Pasquini:2010af,Pasquini:2011tk},
we confirm that in LFCM approaches ``all Boer-Mulders functions are alike,''
in the qualitative sense of Ref.~\cite{Burkardt:2007xm}.

As a phenomenological application, we studied the pion-nucleus
induced Drell-Yan process.
We re-expressed the model results in terms of an
effective Gaussian Ansatz for the $k_\perp$ dependence of TMDs, which is
well supported (in the model and by data), and incorporated
phenomenologically the energy-dependent transverse-momentum
broadening effects.
We have shown that the model predictions obtained in this way for the 
(normalized) cross sections, given in terms of the unpolarized
pion and nucleon  TMDs, compare very well
with the data up to $q_T\lesssim\,$(2--3) GeV
for $x_F\lesssim 0.7$, which is basically the general range
of applicability of the TMD factorization approach in DY.

We studied also the coefficient $\nu$ in the dilepton angular
distribution in the Collins-Soper frame, which is described
in the parton model \cite{Boer:1999mm,Arnold:2008kf}
in terms of the pion and nucleon Boer-Mulders functions.
We obtained  a satisfactory description of available experimental
data for $s\approx\mbox{(470--540)\,GeV$^2$}$ and
in the range of applicability of the TMD factorization approach
established in our study of (normalized) cross sections.

The primary goal of this work was to extend the successful
LFCM phenomenology of the nucleon to the pion case. The
LFCM of the nucleon was shown to describe effects related
to nucleon TMDs in SIDIS in the valence-$x$ region within an
accuracy of (10--30)$\,\%$ \cite{Boffi:2009sh,Pasquini:2011tk}.
In this work we demonstrated that the pion LFCM (in combination
with nucleon LFCM results) yields a similarly good description
of pion-induced DY.

There are also several model-independent conclusions of our study.
First, it is a remarkable fact that valence degrees of freedom
are capable of successfully catching the main features of the
pion-induced DY process, including (normalized) cross sections
differential in $q_T\lesssim\,\,$(2--3)$\,$GeV and $x_F\lesssim 0.7$
and the coefficient $\nu$.
This may indicate that the color entanglement effects discussed in
\cite{Buffing:2013dxa} are not large, though more work is needed
to shed further light in this respect.
Second, the Gaussian Ansatz is well capable of describing the
Boer-Mulders effect in DY in the region $q_T\ll Q$, at least
if one works in a limited range of energies. This point will
be further clarified, when the CSS-evolution equations for the
Boer-Mulders functions will be available and make possible a more 
comprehensive analysis of data at all energies.

Forthcoming or proposed pion induced DY experiments
will open new windows. The forthcoming COMPASS DY experiment
\cite{Quintans:2011zz,Gautheron:2010wva}, 
where a 190$\,{\rm GeV}$ pion beam is available,
is scheduled to start data taking this year and will include also
polarized targets. The SPASCHARM experiment \cite{Abramov:2011zza}, 
where (10--70) GeV pion beams would be available, is in preparation 
at the IHEP facility in Protvino.
The main focus of these experiments is to measure the single
spin asymmetry in DY due to the other (besides Boer-Mulders function)
T-odd TMD of the nucleon, namely the Sivers function \cite{Sivers:1989cc},
and test the predicted sign-change between DIS and DY of this TMD
\cite{Collins:2002kn} which was estimated to be feasible,
see \cite{Efremov:2004tp} for an early estimate.
The sign-change for the nucleon Boer-Mulders function can
also be tested, but this requires measurements of several
single spin asymmetries in DY, and it is less clear whether
the measurements are feasible.
In any case, these experiments will also give new
insights into the structure of the pion.
The present study in the LFCM will be extended to provide 
model predictions for these experiments.

\begin{acknowledgments}
This work has been partially supported by the European Community 
Joint Research Activity ``Study of Strongly Interacting Matter'' 
(acronym HadronPhysics3, Grant Agreement No.~283286) under the 
Seventh Framework Programme of the European Community. The Feynman 
diagram in this paper was drawn using JaxoDraw~\cite{Binosi:2008ig}.
This work was supported partially through GAUSTEQ (Germany and U.S.
Nuclear Theory Exchange Program for QCD Studies of Hadrons and Nuclei)
under contract number DE-SC0006758.
The authors acknowledge the hospitality at the Physics Institute of the 
Gutenberg University Mainz where a part of this work was performed,
and thank Marc Vanderhaeghen for fruitful discussions.

\end{acknowledgments}


\begin{thebibliography}{99}

\bibitem{Collins:1981uw} 
  J.~C.~Collins and D.~E.~Soper,
  Nucl.\ Phys.\ B {\bf 194}, 445 (1982).

\bibitem{Collins:2003fm} 
  J.~C.~Collins,
  Acta Phys.\ Polon.\ B {\bf 34}, 3103 (2003).

\bibitem{Collins-book}
  J.~C.~Collins, ``Foundations of Perturbative QCD'' 
  (Cambridge University Press, Cambridge, 2011).

\bibitem{Tangerman:1994eh} 
  R.~D.~Tangerman and P.~J.~Mulders,
  Phys.\ Rev.\ D {\bf 51}, 3357 (1995).

\bibitem{Kotzinian:1994dv}
  A.~Kotzinian,
  Nucl.\ Phys.\  B {\bf 441}, 234 (1995).

\bibitem{Mulders:1995dh}
  P.~J.~Mulders and R.~D.~Tangerman,
  Nucl.\ Phys.\ B {\bf 461}, 197 (1996)
  [Erratum-ibid.\ B {\bf 484}, 538 (1997)].

\bibitem{Boer:1997nt}
  D.~Boer and P.~J.~Mulders,
  Phys.\ Rev.\  D {\bf 57}, 5780 (1998).

\bibitem{Bacchetta:2006tn}
  A.~Bacchetta, M.~Diehl, K.~Goeke, A.~Metz, P.~J.~Mulders and M.~Schlegel,
  JHEP {\bf 0702}, 093 (2007).

\bibitem{Christenson:1970um}
  J.~H.~Christenson, G.~S.~Hicks, L.~M.~Lederman, P.~J.~Limon, B.~G.~Pope, 
  E.~Zavattini,
  Phys.\ Rev.\ Lett.\  {\bf 25}, 1523 (1970).

\bibitem{Drell:1970wh}
  S.~D.~Drell and T.~M.~Yan,
  Phys.\ Rev.\ Lett.\  {\bf 25}, 316 (1970)
  [Erratum-ibid.\  {\bf 25}, 902 (1970)];
  Annals Phys.\  {\bf 66}, 578 (1971).

\bibitem{Badier:1981ti}
  J.~Badier {\it et al.}  [NA3 Collaboration],
  Z.\ Phys.\  C {\bf 11}, 195 (1981).

\bibitem{Palestini:1985zc}
  S.~Palestini {\it et al.},
  Phys.\ Rev.\ Lett.\  {\bf 55}, 2649 (1985).

\bibitem{Falciano:1986wk}
  S.~Falciano {\it et al.}  [NA10 Collaboration],
  Z.\ Phys.\  C {\bf 31}, 513 (1986).

\bibitem{Guanziroli:1987rp}
  M.~Guanziroli {\it et al.}  [NA10 Collaboration],
  Z.\ Phys.\  C {\bf 37}, 545 (1988).

\bibitem{Conway:1989fs}
  J.~S.~Conway {\it et al.},
  Phys.\ Rev.\  D {\bf 39}, 92 (1989).

\bibitem{Bordalo:1987cs}
  P.~Bordalo {\it et al.}  [NA10 Collaboration],
  Phys.\ Lett.\  B {\bf 193}, 368 (1987);
  Phys.\ Lett.\  B {\bf 193}, 373 (1987).

\bibitem{Stirling:1993gc}
  W.~J.~Stirling and M.~R.~Whalley,
  J.\ Phys.\ G {\bf 19}, D1 (1993).

\bibitem{McGaughey:1999mq}
  P.~L.~McGaughey, J.~M.~Moss and J.~C.~Peng,
  Ann.\ Rev.\ Nucl.\ Part.\ Sci.\  {\bf 49}, 217 (1999).

\bibitem{Reimer:2007iy}
  P.~E.~Reimer,
  J.\ Phys.\ G {\bf 34}, S107 (2007).

\bibitem{Arnold:2008kf}
  S.~Arnold, A.~Metz and M.~Schlegel,
  Phys.\ Rev.\  D {\bf 79}, 034005 (2009).

\bibitem{Chang:2013opa} 
  W.-C.~Chang and D.~Dutta,
  Int.\ J.\ Mod.\ Phys.\ E {\bf 22}, 1330020 (2013).

\bibitem{Peng:2014hta} 
  J.-C.~Peng and J.-W.~Qiu,
  Prog.\ Part.\ Nucl.\ Phys.\  {\bf 76}, 43 (2014).

\bibitem{Collins:1981uk}
  J.~C.~Collins and D.~E. Soper,
  Nucl.\ Phys.\ B {\bf 193}, 381 (1981)
  [Erratum-ibid.\ B {\bf 213}, 545 (1983)].

\bibitem{Ji:2004wu}
  X.~D.~Ji, J.~P.~Ma and F.~Yuan,
  Phys.\ Rev.\ D {\bf 71}, 034005 (2005);
  Phys.\ Lett.\ B {\bf 597}, 299 (2004).

\bibitem{Collins:2004nx}
  J.~C.~Collins and A.~Metz,
  Phys.\ Rev.\ Lett.\  {\bf 93}, 252001 (2004).

\bibitem{Echevarria:2011rb} 
  M.~G.~Echevarria, A.~Idilbi and I.~Scimemi,
  JHEP {\bf 1207}, 002 (2012).

\bibitem{Collins:1984kg}
  J.~C.~Collins, D.~E.~Soper and G.~Sterman,
  Nucl.\ Phys.\ B {\bf 250}, 199 (1985).

\bibitem{Aybat:2011zv} 
  S.~M.~Aybat and T.~C.~Rogers,
  Phys.\ Rev.\ D {\bf 83}, 114042 (2011).

\bibitem{Aybat:2011ge} 
  S.~M.~Aybat, J.~C.~Collins, J.-W.~Qiu, T.~C.~Rogers and ,
  Phys.\ Rev.\ D {\bf 85}, 034043 (2012).
  
\bibitem{Cherednikov:2007tw} 
  I.~O.~Cherednikov and N.~G.~Stefanis,
  Phys.\ Rev.\ D {\bf 77}, 094001 (2008);
  Nucl.\ Phys.\ B {\bf 802}, 146 (2008);
  Phys.\ Rev.\ D {\bf 80}, 054008 (2009).
  I.~O.~Cherednikov, A.~I.~Karanikas and N.~G.~Stefanis,
  Nucl.\ Phys.\ B {\bf 840}, 379 (2010).

\bibitem{Bacchetta:2013pqa} 
  A.~Bacchetta and A.~Prokudin,
  Nucl.\ Phys.\ B {\bf 875}, 536 (2013).

\bibitem{Echevarria:2012pw} 
  M.~G.~Echevarr\'ia, A.~Idilbi, A.~Sch\"afer and I.~Scimemi,
  Eur.\ Phys.\ J.\ C {\bf 73}, 2636 (2013).
  M.~G.~Echevarr\'ia, A.~Idilbi and I.~Scimemi,
  Phys.\ Lett.\ B {\bf 726}, 795 (2013);
  arXiv:1402.0869 [hep-ph].
  M.~G.~Echevarria, A.~Idilbi, Z.-B.~Kang and I.~Vitev,
  Phys.\ Rev.\ D {\bf 89}, 074013 (2014).

\bibitem{Vladimirov:2014aja} 
  A.~A.~Vladimirov,
  arXiv:1402.3182 [hep-ph].

\bibitem{Boer:1999mm}
  D.~Boer,
  Phys.\ Rev.\  D {\bf 60}, 014012 (1999).

\bibitem{Brodsky:2002cx}
  S.~J.~Brodsky, D.~S.~Hwang and I.~Schmidt,
  Phys.\ Lett.\ B {\bf 530}, 99 (2002).

\bibitem{Collins:2002kn}
  J.~C.~Collins,
  Phys.\ Lett.\ B {\bf 536}, 43 (2002).

\bibitem{Ji:2002aa}
  X.~D.~Ji and F.~Yuan,
  Phys.\ Lett.\ B {\bf 543}, 66 (2002).

\bibitem{Brodsky:2002rv}
  S.~J.~Brodsky, D.~S.~Hwang and I.~Schmidt,
  Nucl.\ Phys.\ B {\bf 642}, 344 (2002).

\bibitem{Boer:2002ju} 
  D.~Boer, S.~J.~Brodsky and D.~S.~Hwang,
  Phys.\ Rev.\ D {\bf 67}, 054003 (2003).

\bibitem{Belitsky:2002sm}
  A.~V.~Belitsky, X.~Ji and F.~Yuan,
  Nucl.\ Phys.\ B {\bf 656}, 165 (2003).

\bibitem{Boer:2003cm}
  D.~Boer, P.~J.~Mulders and F.~Pijlman,
  Nucl.\ Phys.\ B {\bf 667}, 201 (2003).

\bibitem{Lam:1978pu}
  C.~S.~Lam and W.~K.~Tung,
  Phys.\ Rev.\  D {\bf 18}, 2447 (1978);
  Phys.\ Rev.\  D {\bf 21}, 2712 (1980).

\bibitem{Collins:1978yt} 
  J.~C.~Collins,
  Phys.\ Rev.\ Lett.\  {\bf 42}, 291 (1979).

\bibitem{Mirkes:1994dp} 
  E.~Mirkes and J.~Ohnemus,
  Phys.\ Rev.\ D {\bf 51}, 4891 (1995).

\bibitem{Brandenburg:1993cj} 
  A.~Brandenburg, O.~Nachtmann and E.~Mirkes,
  Z.\ Phys.\ C {\bf 60}, 697 (1993).

\bibitem{Brandenburg:1994wf} 
  A.~Brandenburg, S.~J.~Brodsky, V.~V.~Khoze and D.~Mueller,
  Phys.\ Rev.\ Lett.\  {\bf 73}, 939 (1994).

\bibitem{Boer:2004mv} 
  D.~Boer, A.~Brandenburg, O.~Nachtmann and A.~Utermann,
  Eur.\ Phys.\ J.\ C {\bf 40}, 55 (2005).

\bibitem{Brandenburg:2006xu} 
  A.~Brandenburg, A.~Ringwald and A.~Utermann,
  Nucl.\ Phys.\ B {\bf 754}, 107 (2006).

\bibitem{Nachtmann:2014qta} 
  O.~Nachtmann,
  arXiv:1401.7587 [hep-ph].

\bibitem{Zhu:2006gx} 
  L.~Y.~Zhu {\it et al.}  [NuSea Collaboration],
  Phys.\ Rev.\ Lett.\  {\bf 99}, 082301 (2007);
  Phys.\ Rev.\ Lett.\  {\bf 102}, 182001 (2009).

\bibitem{Boffi:2002yy}
  S.~Boffi, B.~Pasquini and M.~Traini,
  Nucl.\ Phys.\  B {\bf 649}, 243 (2003).

\bibitem{Boffi:2003yj}
  S.~Boffi, B.~Pasquini and M.~Traini,
  Nucl.\ Phys.\  B {\bf 680}, 147 (2004).

\bibitem{Pasquini:2004gc}
  B.~Pasquini, M.~Traini and S.~Boffi,
  Phys.\ Rev.\  D {\bf 71}, 034022 (2005).

\bibitem{Pasquini:2005dk}
  B.~Pasquini, M.~Pincetti and S.~Boffi,
  Phys.\ Rev.\  D {\bf 72}, 094029 (2005).

\bibitem{Pasquini:2006iv}
  B.~Pasquini, M.~Pincetti and S.~Boffi,
  Phys.\ Rev.\  D {\bf 76}, 034020 (2007).

\bibitem{Pasquini:2007xz}
  B.~Pasquini and S.~Boffi,
  Phys.\ Lett.\  B {\bf 653}, 23 (2007).

\bibitem{Pasquini:2007iz} 
  B.~Pasquini and S.~Boffi,
  Phys.\ Rev.\ D {\bf 76}, 074011 (2007).

\bibitem{Boffi:2007yc}
  S.~Boffi and B.~Pasquini,
  Riv. Nuovo Cim. {\bf 30}, 387 (2007).

\bibitem{Pasquini:2009ki}
  B.~Pasquini, M.~Pincetti and S.~Boffi,
  Phys.\ Rev.\  D {\bf 80}, 014017 (2009);
  S.~Boffi and B.~Pasquini,
  Mod.\ Phys.\ Lett.\  A {\bf 24}, 2882 (2009).

\bibitem{Lorce:2011dv} 
  C.~Lorc\'e, B.~Pasquini and M.~Vanderhaeghen,
  JHEP {\bf 1105}, 041 (2011).

\bibitem{Pasquini:2008ax}
  B.~Pasquini, S.~Cazzaniga and S.~Boffi,
  Phys. Rev. D {\bf 78}, 034025 (2008).

\bibitem{Boffi:2009sh}
  S.~Boffi, A.~V.~Efremov, B.~Pasquini and P.~Schweitzer,
  Phys.\ Rev.\  D {\bf 79} (2009) 094012;
  B.~Pasquini, S.~Boffi and P.~Schweitzer,
  Mod.\ Phys.\ Lett.\  A {\bf 24}, 2903 (2009).

\bibitem{Pasquini:2010af} 
  B.~Pasquini and F.~Yuan,
  Phys.\ Rev.\ D {\bf 81}, 114013 (2010).

\bibitem{Pasquini:2011tk} 
  B.~Pasquini and P.~Schweitzer,
  Phys.\ Rev.\ D {\bf 83}, 114044 (2011).

\bibitem{Schlumpf:1994bc} 
  F.~Schlumpf,
  Phys.\ Rev.\ D {\bf 50}, 6895 (1994).

\bibitem{Chung:1988mu} 
  P.~L.~Chung, F.~Coester and W.~N.~Polyzou,
  Phys.\ Lett.\ B {\bf 205}, 545 (1988).

\bibitem{Frederico:2009fk}
  T.~Frederico, E.~Pace, B.~Pasquini and G.~Salm\`e,
  Phys.\ Rev.\  D {\bf 80}, 054021 (2009);
  Nucl.\ Phys.\ B Proc. Suppl. {\bf 199}, 264 (2010).

\bibitem{Salme':2012rv} 
  G.~Salm\'e, E.~Pace and G.~Romanelli,
  Few Body Syst.\  {\bf 54}, 769 (2013);
  Few Body Syst.\  {\bf 52}, 301 (2012).

\bibitem{Lu:2004hu} 
  Z.~Lu and B.-Q.~Ma,
  Phys.\ Rev.\ D {\bf 70}, 094044 (2004).

\bibitem{Burkardt:2007xm} 
  M.~Burkardt and B.~Hannafious,
  Phys.\ Lett.\ B {\bf 658}, 130 (2008)

\bibitem{Gamberg:2009uk}
  L.~Gamberg and M.~Schlegel,
  Phys.\ Lett.\  B {\bf 685}, 95 (2010).

\bibitem{Lu:2012hh} 
  Z.~Lu, B.-Q.~Ma and J.~Zhu,
  Phys.\ Rev.\ D {\bf 86}, 094023 (2012)

\bibitem{Engelhardt:2013nba}
  M.~Engelhardt, B.~Musch, P.~H\"agler, J.~Negele and A.~Sch\"afer,
  arXiv:1310.8335 [hep-lat].\\
  B.~U.~Musch, P.~H\"agler, M.~Engelhardt, J.~W.~Negele and A.~Sch\"afer,
  Phys.\ Rev.\ D {\bf 85}, 094510 (2012).

\bibitem{Bianconi:2006hc} 
  A.~Bianconi and M.~Radici,
  Phys.\ Rev.\ D {\bf 73}, 114002 (2006);
  Phys.\ Rev.\ D {\bf 71}, 074014 (2005).\\
  
\bibitem{Lu:2005rq} 
  Z.~Lu and B.-Q.~Ma,
  Phys.\ Lett.\ B {\bf 615}, 200 (2005).

\bibitem{Gamberg:2005ip} 
  L.~P.~Gamberg and G.~R.~Goldstein,
  Phys.\ Lett.\ B {\bf 650}, 362 (2007).

\bibitem{Sissakian:2005yp} 
  A.~Sissakian, O.~Shevchenko, A.~Nagaytsev, O.~Denisov and O.~Ivanov,
  Eur.\ Phys.\ J.\ C {\bf 46}, 147 (2006).\\
  A.~Sissakian, O.~Shevchenko, A.~Nagaytsev and O.~Ivanov,
  Eur.\ Phys.\ J.\ C {\bf 59}, 659 (2009).

\bibitem{Barone:2006ws} 
  V.~Barone, Z.~Lu and B.-Q.~Ma,
  Eur.\ Phys.\ J.\ C {\bf 49}, 967 (2007).

\bibitem{Zhang:2008nu}
  B.~Zhang, Z.~Lu, B.~Q.~Ma and I.~Schmidt,
  Phys.\ Rev.\  D {\bf 77} (2008) 054011.

\bibitem{Lu:2009ip} 
  Z.~Lu and I.~Schmidt,
  Phys.\ Rev.\ D {\bf 81}, 034023 (2010).

\bibitem{Barone:2010gk} 
  V.~Barone, S.~Melis and A.~Prokudin,
  Phys.\ Rev.\ D {\bf 82}, 114025 (2010).

\bibitem{Lu:2011mz} 
  Z.~Lu and I.~Schmidt,
  Phys.\ Rev.\ D {\bf 84}, 094002 (2011).

\bibitem{Liu:2012fha} 
  T.~Liu and B.-Q.~Ma,
  Eur.\ Phys.\ J.\ C {\bf 73}, 2291 (2013).

\bibitem{Liu:2012vn} 
  T.~Liu and B.-Q.~Ma,
  Eur.\ Phys.\ J.\ C {\bf 72}, 2037 (2012).

\bibitem{Chen:2013zpy} 
  L.~Chen, J.-h.~Gao and Z.-T.~Liang,
  Phys.\ Rev.\ C {\bf 89}, 035204 (2014).

\bibitem{Chang:2013pba} 
  C.-P.~Chang and H.-N.~Li,
  Phys.\ Lett.\ B {\bf 726}, 262 (2013).

\bibitem{Broniowski:2007si}
  W.~Broniowski, E.~R.~Arriola and K.~Golec-Biernat,
  Phys.\ Rev.\  D {\bf 77}, 034023 (2008).

\bibitem{Davidson:2001cc}
  R.~M.~Davidson and E.~Ruiz Arriola,
  Acta Phys.\ Polon.\  B {\bf 33}, 1791 (2002).

\bibitem{Courtoy:2008nf}
  A.~Courtoy and S.~Noguera,
  Phys.\ Lett.\  B {\bf 675}, 38 (2009).

\bibitem{Gluck:1999xe}
  M.~Gluck, E.~Reya and I.~Schienbein,
  Eur.\ Phys.\ J.\  C {\bf 10}, 313 (1999).

\bibitem{Sutton:1991ay}
  P.~J.~Sutton, A.~D.~Martin, R.~G.~Roberts and W.~J.~Stirling,
  Phys.\ Rev.\  D {\bf 45}, 2349 (1992).

\bibitem{Martin:2009iq}
  A.~D.~Martin, W.~J.~Stirling, R.~S.~Thorne and G.~Watt,
  Eur.\ Phys.\ J.\  C {\bf 63} (2009) 189.

\bibitem{Burkardt:2002uc}
  M.~Burkardt, X.~Ji and F.~Yuan,
  Phys.\ Lett.\  B {\bf 545}, 345 (2002).     

\bibitem{Ji:2003yj} 
  X.-D.~Ji, J.-P.~Ma and F.~Yuan,
  Eur.\ Phys.\ J.\ C {\bf 33}, 75 (2004).

\bibitem{Melosh:1974cu}
  H.~J.~Melosh,
  Phys.\ Rev.\  D {\bf 9}, 1095 (1974).

\bibitem{Gluck:1991ey} 
  M.~Gl\"uck, E.~Reya and A.~Vogt,
  Z.\ Phys.\ C {\bf 53}, 651 (1992).

\bibitem{Owens:1984zj} 
  J.~F.~Owens,
  Phys.\ Rev.\ D {\bf 30}, 943 (1984).

\bibitem{Hecht:2000xa}
M.~B.~Hecht, C.~D.~Roberts and S.~M.~Schmidt,
Phys.\ Rev.\  C {\bf 63}, 025213 (2001).

\bibitem{Aicher:2010cb} 
  M.~Aicher, A.~Schafer and W.~Vogelsang,
  Phys.\ Rev.\ Lett.\  {\bf 105}, 252003 (2010).

\bibitem{Wijesooriya:2005ir}
  K.~Wijesooriya, P.~E.~Reimer and R.~J.~Holt,
  Phys.\ Rev.\  C {\bf 72}, 065203 (2005)
  [arXiv:nucl-ex/0509012].

\bibitem{Holt:2010vj} 
  R.~J.~Holt and C.~D.~Roberts,
  Rev.\ Mod.\ Phys.\  {\bf 82}, 2991 (2010).


\bibitem{Gluck:1998xa}
  M.~Gl\"uck, E.~Reya and A.~Vogt,
  Eur.\ Phys.\ J.\ C {\bf 5}, 461 (1998).

\bibitem{Traini:1997jz} 
  M.~Traini, A.~Mair, A.~Zambarda and V.~Vento,
  Nucl.\ Phys.\ A {\bf 614}, 472 (1997).
  
  \bibitem{Bacchetta:1999kz}
  A.~Bacchetta, M.~Boglione, A.~Henneman and P.~J.~Mulders,
  Phys.\ Rev.\ Lett.\  {\bf 85}, 712 (2000).

\bibitem{Kotzinian:2008fe} 
  A.~Kotzinian,
  arXiv:0806.3804 [hep-ph].

\bibitem{Boer:2001he} 
  D.~Boer,
  Nucl.\ Phys.\ B {\bf 603}, 195 (2001).

\bibitem{Kotikov:1997df} 
  A.~V.~Kotikov and D.~V.~Peshekhonov,
  Phys.\ Atom.\ Nucl.\  {\bf 60}, 653 (1997);
  Eur.\ Phys.\ J.\ C {\bf 9}, 55 (1999).

\bibitem{Koike:2006fn} 
  Y.~Koike, J.~Nagashima and W.~Vogelsang,
  Nucl.\ Phys.\ B {\bf 744}, 59 (2006).

\bibitem{Ratcliffe:1982yj} 
  P.~Ratcliffe,
  Nucl.\ Phys.\ B {\bf 223}, 45 (1983).\\ 
  V.~Barone, A.~Cafarella, C.~Coriano, M.~Guzzi and P.~Ratcliffe,
  Phys.\ Lett.\ B {\bf 639}, 483 (2006).

\bibitem{Vogelsang:1992jn} 
  W.~Vogelsang and A.~Weber,
  Phys.\ Rev.\ D {\bf 48}, 2073 (1993).\\
  A.~P.~Contogouris, B.~Kamal and Z.~Merebashvili,
  Phys.\ Lett.\ B {\bf 337}, 169 (1994).\\
  B.~Kamal,
  Phys.\ Rev.\ D {\bf 53}, 1142 (1996).

\bibitem{Boer:2006eq} 
  D.~Boer and W.~Vogelsang,
  Phys.\ Rev.\ D {\bf 74}, 014004 (2006).

\bibitem{Shimizu:2005fp} 
  H.~Shimizu, G.~F.~Sterman, W.~Vogelsang and H.~Yokoya,
  Phys.\ Rev.\ D {\bf 71}, 114007 (2005).\\
  A.~Mukherjee and W.~Vogelsang,
  Phys.\ Rev.\ D {\bf 73}, 074005 (2006).

\bibitem{Ravindran:2000rz} 
  V.~Ravindran and W.~L.~van Neerven,
  Nucl.\ Phys.\ B {\bf 589}, 507 (2000).

\bibitem{Buffing:2013dxa} 
  M.~G.~A.~Buffing and P.~J.~Mulders,
  Phys.\ Rev.\ Lett.\  {\bf 112}, 092002 (2014).

\bibitem{Cox:1982wy}
  B.~Cox and P.~K.~Malhotra,
  Phys.\ Rev.\  D {\bf 29}, 63 (1984).

\bibitem{D'Alesio:2007jt}
  U.~D'Alesio and F.~Murgia,
  Prog.\ Part.\ Nucl.\ Phys.\  {\bf 61}, 394 (2008).

\bibitem{Schweitzer:2010tt} 
  P.~Schweitzer, T.~Teckentrup and A.~Metz,
  Phys.\ Rev.\ D {\bf 81}, 094019 (2010).

\bibitem{Landry:2002ix} 
  F.~Landry, R.~Brock, P.~M.~Nadolsky and C.~P.~Yuan,
  Phys.\ Rev.\ D {\bf 67}, 073016 (2003).

\bibitem{Berger:1979du} 
  E.~L.~Berger and S.~J.~Brodsky,
  Phys.\ Rev.\ Lett.\  {\bf 42}, 940 (1979).
  E.~L.~Berger,
  Z.\ Phys.\ C {\bf 4}, 289 (1980).

\bibitem{Bakulev:2007ej} 
  A.~P.~Bakulev, N.~G.~Stefanis and O.~V.~Teryaev,
  Phys.\ Rev.\ D {\bf 76}, 074032 (2007).

\bibitem{Bacchetta:2008xw} 
  A.~Bacchetta, D.~Boer, M.~Diehl and P.~J.~Mulders,
  JHEP {\bf 0808}, 023 (2008).

\bibitem{Gautheron:2010wva} 
  F.~Gautheron {\it et al.}  [COMPASS Collaboration],
  SPSC-P-340. 

\bibitem{Quintans:2011zz} 
  C.~Quintans [COMPASS Collaboration],
  J.\ Phys.\ Conf.\ Ser.\  {\bf 295}, 012163 (2011).  
  
\bibitem{Abramov:2011zza} 
  V.V.~Abramov, N.I.~Belikov, Y.M.~Goncharenko, V.N.~Grishin, 
  A.M.~Davidenko, A.A.~Derevshchikov, V.A.~Kachanov 
  and D.~A.~Konstantinov {\it et al.},
  J.\ Phys.\ Conf.\ Ser.\  {\bf 295}, 012018 (2011).  

\bibitem{Sivers:1989cc}
  D.~W.~Sivers,
  Phys.\ Rev.\ D {\bf 41}, 83 (1990);
  Phys.\ Rev.\ D {\bf 43}, 261 (1991).

\bibitem{Efremov:2004tp}
  A.~V.~Efremov, K.~Goeke, S.~Menzel, A.~Metz and P.~Schweitzer,
  Phys.\ Lett.\  B {\bf 612}, 233 (2005).

\bibitem{Binosi:2008ig} 
  D.~Binosi, J.~Collins, C.~Kaufhold and L.~Theussl,
  Comput.\ Phys.\ Commun.\  {\bf 180}, 1709 (2009).

\end{thebibliography}
\end{document}